\documentclass[%
reprint,
nofootinbib,
 amsmath,amssymb,
 aps,
pra,
floatfix,
]{revtex4-2}

\usepackage{graphicx}% Include figure files
\usepackage{dcolumn}% Align table columns on decimal point
\usepackage{bm}% bold math
\usepackage{subcaption}
\usepackage{hyperref}% add hypertext capabilities
\usepackage{amsthm}
\newtheorem{theorem}{Theorem}[section]

\begin{document}

\title{Ghost Projection}

\author{David Ceddia}
\email{David.Ceddia@gmail.com}
\author{David M.~Paganin}
\affiliation{School of Physics and Astronomy, Monash University, Victoria 3800, Australia}

\date{\today}

\begin{abstract}
Ghost imaging is a developing imaging technique that employs random masks to image a sample. Ghost projection utilizes ghost-imaging concepts to perform the complementary procedure of projection of a desired image.  The key idea underpinning ghost projection is that any desired spatial distribution of radiant exposure may be produced, up to an additive constant, by spatially-uniformly illuminating a set of random masks in succession. We explore three means of achieving ghost projection: (i) weighting each random mask, namely selecting its exposure time, according to its correlation with a desired image, (ii)  selecting a subset of random masks according to their correlation with a desired image, and (iii) numerically optimizing a projection for a given set of random masks and desired image. The first two protocols are analytically tractable and conceptually transparent. The third is more efficient but less amenable to closed-form analytical expressions. A comparison with existing image-projection techniques is drawn and possible applications are discussed. These potential applications include: (i) a data projector for matter and radiation fields for which no current data projectors exist, (ii) a universal-mask approach to lithography, (iii) tomographic volumetric additive manufacturing, and (iv) a ghost-projection photocopier. 
\end{abstract}

\maketitle

%\tableofcontents

\section{Introduction}

Imaging, namely the direct or indirect measurement of a spatial distribution of radiation, has a long and diverse history ranging from biological manifestations with the evolution of the eye, to the first cameras in the early nineteenth century \cite{HechtOpticsBook}. Projection, namely the creating of a known radiation distribution, has a coupled yet distinct history \cite{johnson1960}. Examples of image projection include human cave-drawings, which indirectly create a desired radiation distribution when sunlight is reflected from them, as well as shadow formation using specified shapes, pinhole projectors  and  optical projectors. 

A common image-projection strategy is the serial approach of projecting one resolution element at a time, for example via raster scanning a finely focused beam.  This `one point at a time' analogue of scanning-probe imaging \cite{Pennycook2011} corresponds to the synthesis of a desired function, such as a specified two-dimensional distribution of radiant exposure, via a linear combination of localized Dirac-delta basis elements.  Another analogy, here, is creating a desired drawing on a sheet of paper using a single pencil.  Examples of this serial approach to image projection include the creation of an analogue television image by raster scanning an electron beam impinging on a fluorescent screen \cite{Trundle2001}, and electron-beam lithography \cite{Chen2015}. 

A complementary projection strategy is the parallel approach of projecting all resolution elements simultaneously.  This `all points at once' strategy has a close analogue in the lithographic printing of artworks and books by smearing paint or ink over a pre-fabricated mask and then subsequently printing an entire artwork or page of text in a single shot. Examples of this parallelized projection strategy include the optical projection of images using a photographic slide or optical data projector, analogue optical photography, x-ray medical radiography using photographic film or image plates, and mask-based photolithography \cite{Bourdillon2000,Bourdillon2001}.  

A third image-projection strategy creates a desired distribution of radiant exposure in a serial manner using a set of `extended brushes', by having a series of structured basis elements that are each significantly larger in spatial extent than the resolution of the desired projection image.  Here, each structured basis element has detail that is finer in scale than the diameter of each `extended brush' \cite{Svalbe2020}. Extrapolating the previous pencil-and-paper analogy, our third image-projection strategy corresponds to writing using a sheaf of pencils, all of which write in unison. One example of this third approach is the multi-layer lithographic printing of artworks, with several separate masks being used to create a series of overlapping images.  Another example is the synthesis of a two-dimensional distribution of radiant exposure by superposition of two-dimensional Fourier harmonics \cite{Bracewellbook}, or other members of a complete set of two-dimensional basis functions.  In addition to the previously-mentioned Fourier basis, for this third image-projection approach one could also employ a wavelet basis \cite{WaveletBook}, a Zernike-polynomial basis \cite{BornWolf}, a Hadamard basis \cite{HadamardBook}, a Huffman basis \cite{Svalbe2020}, or a random-function basis \cite{Gorban2016}. A three-dimensional example is the `tomography in reverse' approach to volumetric manufacturing \cite{Beer2019,TomographyInReverse2019}, which works by illuminating a photoresist volume from a variety of angles.  Conceptually, the third approach to projection blends the serial `one point at a time' nature of the first approach with the `all points at once' concept of the second approach. Stated differently, this third approach uses a weighted superposition of non-localized basis elements, each of which cover at least several resolution cells.

\begin{figure*}[ht!]
     \centering
     \begin{subfigure}{\textwidth}
         \centering
         \includegraphics[width=0.73\textwidth]{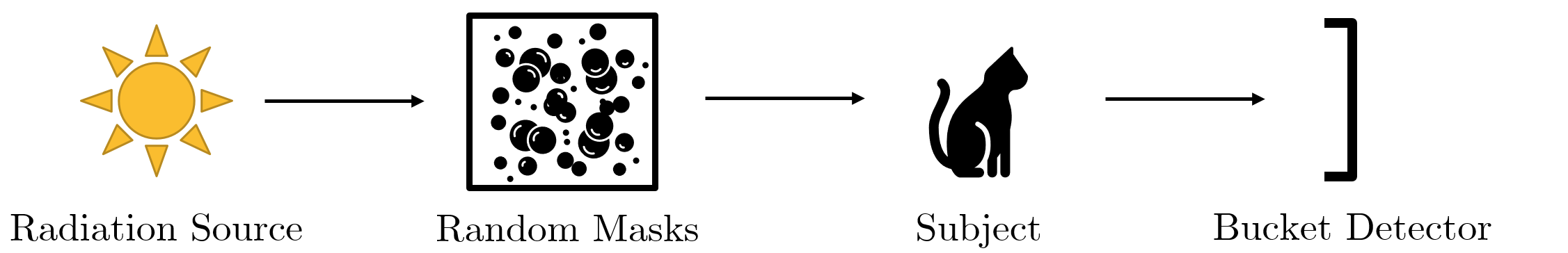}
         \caption{ }
         \label{subfig:schematic 1}
     \end{subfigure}
     \begin{subfigure}{\textwidth}
         \centering
         \includegraphics[width=0.58\textwidth]{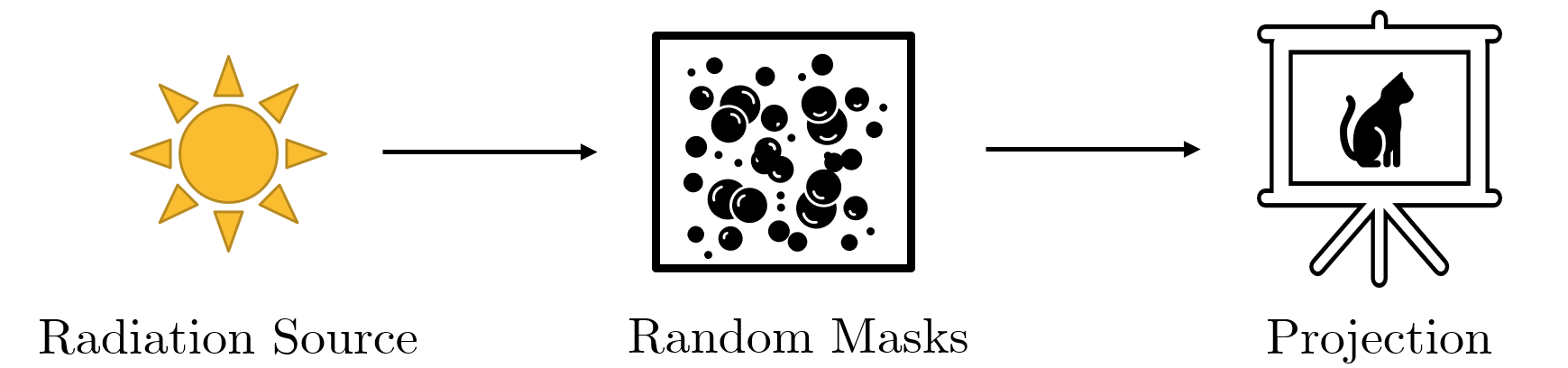}
         \caption{ }
         \label{subfig:schematic 2}
     \end{subfigure}
     \begin{subfigure}{\textwidth}
         \centering
         \includegraphics[width=0.78\textwidth,clip,trim={0cm 1cm 0cm 1cm}]{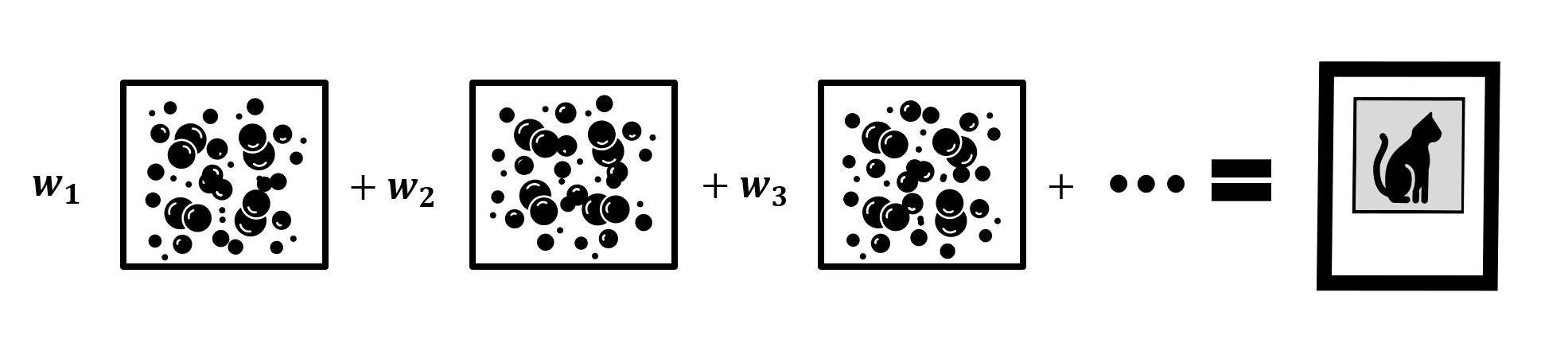}
         \caption{ }
         \label{subfig:schematic 3}
     \end{subfigure}
        \caption{(a) Schematic ghost imaging set-up. (b) Schematic ghost projection set-up. (c) Illustration of ghost reconstruction technique. If we allow the weights (mask exposure times) $w_1,w_2,\cdots$ to be both positive and negative, and perform the reconstruction computationally, then we have ghost imaging. If we restrict the weights to be non-negative, however, and perform the reconstruction experimentally, then we have ghost projection. }
        \label{fig:Ghost Schematics}
\end{figure*}

Two particularly topical imaging-related concepts are classical computational ghost imaging \cite{shapiro2008computational,erkmen2010ghost,Shapiro2012,Padgett2017} and the closely-related field of random-mask compressive sensing \cite{Donoho2006,Rani2017}. Computational ghost imaging takes the correlation measurement between a set of known random masks and a desired subject using a single-pixel `bucket' detector, as shown in Fig.~\ref{subfig:schematic 1}. A spatially resolved image is then subsequently computationally reconstructed \cite{Bromberg2009}. Assuming a known sparse basis, this reconstruction is efficiently achieved through compressive sensing algorithms. Other reconstruction approaches include matrix inversion \cite{NumericalRecipes} and iterative refinement \cite{KingstonIEEE2019}.

Borrowing from both fields, we propose the complementary process of ghost projection. In this, we employ the random basis and correlation measurements from ghost imaging and compressive sensing, but apply them in the reverse direction of image projection \cite{paganin2019writing}.  Rather than building a desired projection in an entirely-serial pencil-beam manner by projecting one resolution element at a time, or in an entirely-parallel manner via a single-shot exposure of a single mask, ghost projection seeks to synthesize a desired distribution of radiant exposure via a linear combination of spatially-extended random masks, as shown in Fig.~\ref{subfig:schematic 2}. The core concept is that a desired image may be \textit{computationally} expressed as a weighted sum of illuminated spatially-random masks (ghost imaging), or \textit{experimentally} expressed up to an additive constant via a non-negative weighted sum of illuminated spatially-random masks (ghost projection).  See Fig.~\ref{subfig:schematic 3}. The ensemble of illuminated spatially-random masks, which can be represented in continuum terms as random basis functions or discrete terms as a random-matrix basis, may be generated by transversely scanning a single illuminated spatially-random mask \cite{paganin2019writing}.  

While our primary interest and motivation is in the fundamental optical physics underpinning the ghost-projection concept, several possible applications motivate this work. (i) Ghost projection requires neither a specifically-fabricated mask nor a tightly transversely localized beam, but rather may be performed using a single well-characterized transversely-scanned spatially random mask. Once it has been characterized, the spatially-random mask is universal insofar as it is able to write any desired distribution of radiant exposure, up to both an additive offset and a limiting spatial resolution that is dictated by the finest spatial features that are present in the projected image of the mask. Such a random mask might be employed in a lithographic setting using x-rays \cite{Paganin2006} and gamma rays, together with electron \cite{CowleyBook}, neutron \cite{NeutronOpticsHandbook} or muon \cite{MuonRadiography} radiation, as well as atomic \cite{AtomBeamBook}, molecular \cite{MolcularBeamsBook} and ion beams \cite{IonBeamBook}. (ii) Ghost projection may be useful for generalizing the concept of a spatial light modulator to radiation and matter wave fields for which no current mask-based technique exists to project an arbitrary image, or for which existing data projectors have limited spatial resolution.  (iii) Tomographic volumetric additive manufacturing \cite{Beer2019,TomographyInReverse2019}, in which a radiation-sensitive three-dimensional resist is illuminated from a variety of angles in order to deposit a desired position-dependent distribution of dose, may also benefit from the ghost-projection concept.  Again, it is in regimes where high-resolution image projectors or generalized spatial light modulators do not exist---such as, for example, in the hard-x-ray or ion-beam domains---where the concept of ghost projection might be especially useful.  %(iv)  Intensity-modulated medical radiotherapy \cite{Cho2018}, whereby a desired distribution of radiation dose is delivered to a three-dimensional volume of biological soft tissue such as the volume occupied by a cancerous growth, may also benefit from a ghost-projection approach. 

\begin{figure*}[ht!]
     \centering
     \begin{subfigure}{0.3\textwidth}
         \centering
         \includegraphics[width=0.75\textwidth,clip,trim={0cm 8cm 0cm 0cm}]{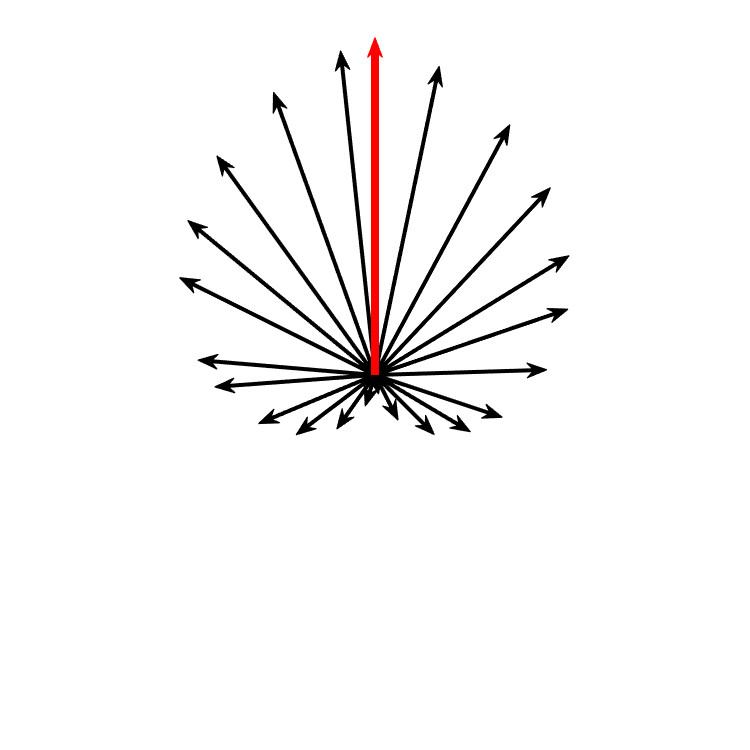}
         \caption{ }
         \label{subfig: Intro Fig 1}
     \end{subfigure}
     \begin{subfigure}{0.3\textwidth}
         \centering
         \includegraphics[width=0.75\textwidth,clip,trim={0cm 8cm 0cm 0cm}]{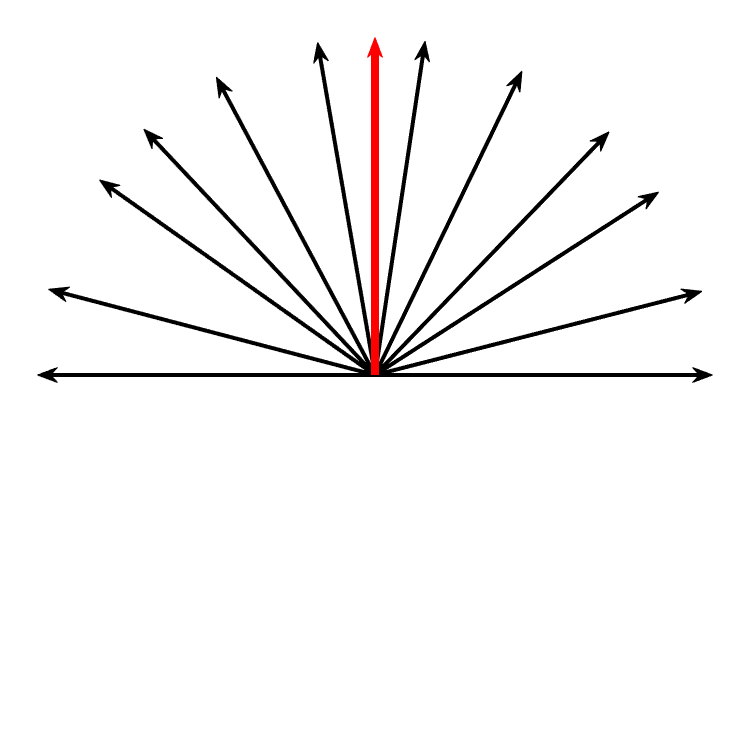}
         \caption{ }
         \label{subfig: Intro Fig 2}
     \end{subfigure}
     \begin{subfigure}{0.3\textwidth}
         \centering
         \includegraphics[width=0.75\textwidth,clip,trim={0cm 8cm 0cm 0cm}]{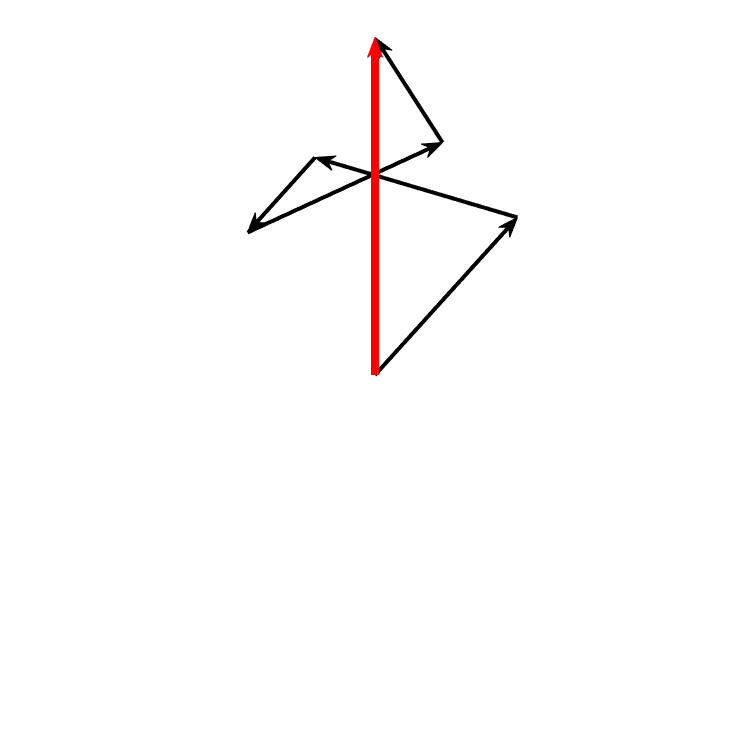}
         \caption{ }
         \label{subfig: Intro Fig 3}
     \end{subfigure}
        \caption{Function-space visualization of the ghost-projection concept.  Black arrows indicate random-basis elements in the function space associated with the random-matrix basis, with red vectors representing the desired image. (a) Weighting each random mask by its correlation with a desired image and averaging. (b) Retaining only random masks that are positively correlated with the desired image and averaging.  (c) Numerically optimized non-negative weights and random masks that best reconstruct the desired image.  Panels (a), (b) and (c) correspond to the ghost-projection schemes developed in Secs.~III, IV and VII, respectively.  }
        \label{fig: Vectors Intro Fig}
\end{figure*}

The paper is structured as follows.  Section II summarizes some key results on the use of random matrices as a basis, with respect to which specified images may be decomposed. Section III develops the direct ghost-projection analogue of classical ghost imaging using a spatially-random basis. Section IV improves on this scheme, via what we term pseudo-correlation filtered ghost projection. Pseudo-correlation filtered ghost projection is then examined under the influence of Poisson noise in Sec.~V.  Section VI presents a color variant of ghost projection, whereby a specified energy spectrum may be deposited within each resolution element.  Section VII employs iterative numerical refinement schemes to increase the efficacy of ghost projection, thereby improving on the closed-form analytical expressions developed in earlier sections, at the cost of forsaking the conceptual clarity that is provided by such closed-form solutions. An illustration and comparison of the different approaches to ghost projection is presented in Fig.~\ref{fig: Vectors Intro Fig}, in which panels (a), (b) and (c) correspond to Secs.~III, IV and VII, respectively. Further considerations on the influence of two forms of noise are developed in Sec.~VII: (i) the Poisson noise associated with the finite number of imaging quanta, and (ii) the normally-distributed exposure noise associated with experimental uncertainty in the duration for which each ghost-projection random mask is exposed.  Section VIII discusses some broader implications of our work, and makes several suggestions for future research.  Some concluding remarks are made in Sec.~IX. 

\section{Random-Matrix Basis}

Suppose we want to project a known image $I_{ij}$ over a specified planar surface.  Here, $I$ is a matrix, indexed by the integers $ij$ which range from $[1,m]$ and $[1,n]$, respectively. This image projection can be performed in units of the number of photons\footnote{In this paper, we employ the word `photon' to mean one detector count of electromagnetic quanta.} or other imaging quanta such as the number of neutrons or ions etc.  We can also work in terms of intensity or exposure time. To cover all of these cases, we will work with $I_{ij}$ expressed as a dimensionless transmission coefficient, which can be multiplied by any choice of the number of incoming photons (or other imaging quanta), illumination intensity or exposure time to obtain the final units of interest.

We now wish to express this image as a linear combination of `noise maps' or, synonymously, as a linear combination of elements of a `random-matrix basis'. Every basis element must be non-negative, and every weighting coefficient in the linear combination must be positive.  Let the non-negative random-matrix basis be denoted by the tensor $R_{ijk}$, where $k$ indexes the set of basis members ranging from $[1,N]$ and $ij$ indexes the matrix entries of each basis member. Our random-matrices model experimentally acquired spatially-random masks, at the resolution of the `speckles' in such random masks. Examples of such spatially-random masks, for the x-ray and neutron domains respectively, are given in Pelliccia {\em et al.}~\cite{PellicciaIUCrJ} and Kingston {\em et al.}~\cite{Kingston2020}.
 Explicitly, $R$ is also a transmission coefficient, and is a random number deviate drawn from the interval  $[0,1]$. The elements of the random tensor $R$ can have any distribution within these bounds (e.g.~uniform, Binomial, Gaussian, Poissonian, etc.) and all results will be left in terms of the parameters of these distributions (i.e.~expectation value E$[R]$, variance Var$[R]$, etc.). We employ tensor notation with the Einstein summation convention assumed, e.g.~$R_{ijk} I^{ij} = \sum_{i=1}^{m} \sum_{j=1}^{n} R_{ijk} I^{ij} $ is the spatial inner product between the $k^{\text{th}}$ random-matrix basis member and the desired image. In this context, there is no meaning attached to the subscript or superscript tensor indices (e.g.~$I_{ij} = I^{ij}$ or $R_{ijk}I^{ij}= R^{ij}_{\ \ k} I_{ij}$) and they are used to denote summation only. Following \citeauthor{ceddia2018random} \cite{ceddia2018random}, we can adapt the orthogonality and completeness expressions to include our non-zero centered basis which will now produce an additive offset.

\subsection{Orthogonality}

We define our offset orthogonality relationship as the appropriately normalized, spatial-product of two random matrices,
\begin{align} \nonumber
\lim_{n,m \to \infty} & \frac{1}{nm \ \text{Var}\left[ R \right]}  R_{ijk} R_{\ \ k'}^{ij} 
\\ \nonumber &= \delta_{kk'} + \frac{\text{E}[R]^2}{\text{Var}[R]}J_{kk'} \\
&= \frac{1}{\text{Var}\left[ R \right]}
\begin{cases}
\text{E} \left[  R \right]^2  \text{for} \ k \neq k' \\
\text{E} \left[  R^2 \right]  \text{for} \ k = k',
\end{cases}
\end{align}
where $\delta_{kk'}$ is the standard Kronecker delta, i.e. if $k=k'$, then $\delta_{kk} = \delta_{k'k'} = 1$, and if $k\neq k'$, then $\delta_{kk'} = 0$, and $J$ is a tensor of all ones. The above expression is a standard orthogonality relationship that is offset by E$[R]^2$/($nm$ Var$[R]$), i.e.
\begin{align}
\lim_{n,m \to \infty} \frac{1}{nm \ \text{Var}\left[ R \right]} \left( R_{ijk} R_{\ \ k'}^{ij}  - \text{E} [R]^2 J_{kk'} \right)
= \delta_{kk'}.
\end{align}
Recall that this is a random-matrix basis and instead of considering orthogonality in an absolute sense, we instead speak of orthogonality in an expected sense, with allowable probabilistic variations. In the above expression, we have an infinite spatial random-matrix basis to ensure the expectation value is reached. Supposing we had a finite spatial random-matrix basis set (which is always the case in practice), we would anticipate probabilistic variations from the expected result. This can be calculated explicitly via the variance with a finite $nm$:
\begin{align} \nonumber
\left( \frac{1}{nm \ \text{Var}\left[ R \right]} \right)^2   \text{Var} \left[ R_{ijk} R_{\ \ k'}^{ij} \right] \quad\quad\quad\quad\quad\quad\quad\quad
\\  = \frac{1}{nm \ \text{Var}\left[ R \right]^2} 
\begin{cases}
\text{E} \left[ R^2 \right]^2 - \text{E} [R]^4 \ \text{for} \ k \neq k', \\
\text{Var} \left[ R^2 \right] \text{for} \ k = k'.
\end{cases}
\end{align}

\subsection{Completeness}
We define our offset completeness relationship to be:
\begin{align} \label{eq:Offset Completeness relationship}
\lim_{N\to \infty} \frac{1}{N \ \text{Var}\left[ R \right] } R_{ijk}  R_{i'j'}^{\ \ \ \ k} = \delta_{ii'} \delta_{jj'} + \frac{\text{E}[R]^2}{\text{Var}\left[ R \right]} J_{ii'} J_{jj'},
\end{align}
which can be written in the more conventional form:
\begin{align}
\nonumber \lim_{N\to \infty} \frac{1}{N \ \text{Var}\left[ R \right] } \left(  R_{ijk} R_{i'j'}^{\ \ \ \ k} - \text{E}[R]^2  J_{ijk} J_{i'j'}^{\ \ \ \ k} \right) \\ = \delta_{ii'} \delta_{jj'}.
\end{align}
Examining the variance of our offset completeness relationship, which will give us an indication of the error we might expect for a truncated sum, we have:
\begin{align} \nonumber
\frac{1}{N^2 \ \text{Var}\left[ R \right]^2 } & \text{Var} \left[ R_{ijk}  R_{i'j'}^{\ \ \ \ k} \right] \\ \nonumber
&= \frac{1}{N \ \text{Var}\left[ R \right]^2 } \text{Var} \left[ R_{ij} R_{i'j'} \right] \\
&= \frac{1}{N \ \text{Var}\left[ R \right]^2 } 
\begin{cases}
\text{Var}[R_1 R_2] \ \text{for} \ ij \neq i'j', \\
\text{Var} \left[ R^2 \right] \text{for} \ ij = i'j'.
\end{cases}
\end{align}
Here, $\text{Var}[R_1 R_2]$ is the variance of two independent deviates drawn from $P(R)$.

\section{Pseudo-correlation Weighted Ghost Projection with a Random-Matrix Basis} \label{sec:Naive Ghost Projection with a Random Matrix Basis}

\subsection{Pseudo-correlation Weighted Ghost Projection} \label{subsec:Naive ghost projection}

Suppose we wish to project a desired image $I_{ij}$, up to an additive constant, using a non-negative random-matrix basis $R_{ijk}$ weighted by non-negative coefficients. To do this, we can mirror the procedure set out by ghost imaging 
\cite{erkmen2010ghost,ceddia2018random,Bromberg2009,Katz2009}. That is, take our offset completeness relationship (Eq.~(\ref{eq:Offset Completeness relationship})), multiply by $I_{i'j'}$ and sum over the primed indices:
\begin{align}
\nonumber \lim_{N\to \infty} \frac{1}{N \ \text{Var}\left[ R \right] } R_{ijk}  R_{i'j'}^{\ \ \ \ k} I^{i'j'} \quad\quad\quad\quad \\ = \delta_{ii'} \delta_{jj'} I^{i'j'}  +  \frac{\text{E}[R]^2}{\text{Var}\left[ R \right]} J_{ii'} J_{jj'} I^{i'j'}.
\end{align}
Employing the simplifications:
\begin{equation}
\delta_{ii'} \delta_{jj'} I^{i'j'} = I_{ij}
\end{equation}
and 
\begin{equation}
 J_{ii'} J_{jj'} I^{i'j'} = J_{ij} nm \ \text{E}[I],  
\end{equation}
we have:
\begin{align}
\lim_{N\to \infty} \frac{1}{N \ \text{Var}\left[ R \right] } R_{ijk}  R_{i'j'}^{\ \ \ \ k} I^{i'j'} = I_{ij} + nm \frac{\text{E}[R]^2 \text{E} [I]}{\text{Var}\left[ R \right]} J_{ij}.
\end{align}
We call the following quantity our `exposure':
\begin{equation}
t_k \equiv \frac{1}{N \ \text{Var}\left[ R \right] } R_{i'j'k} I^{i'j'}.
\label{eq:t_k}
\end{equation}
The above expression is the ghost-projection construction analogous to a `bucket measurement' from a ghost imaging context.  We then have the theoretical result that a desired image can be exactly projected from a random-matrix basis up to an additive constant, via:
\begin{align}
\lim_{N\to \infty} t_k R_{ij}^{\ \ k} = I_{ij} + nm \frac{\text{E}[R]^2 \text{E} [I]}{\text{Var}\left[ R \right]} J_{ij}.
\end{align}
In practice, we would truncate this infinite sum at some finite $N$ and thereby obtain a prescription for practically achievable ghost projection $P_{ij}$ of a desired image:
\begin{align} \label{eq: Standard Ghost Projection}
P_{ij} \equiv t_k R_{ij}^{\ \ k} \approx I_{ij} + nm \frac{\text{E}[R]^2 \text{E} [I]}{\text{Var}\left[ R \right]} J_{ij}.
\end{align}
Note that $P_{ij}$ denotes our ghost projection scheme, throughout this paper. We consequently adopt a different definition in each section and, in some cases, subsections. Rather than having a different variant of the symbol $P_{ij}$ for every case, we instead alert the reader to the context-dependent definition of $P_{ij}$. 

Examining the parameters that make up the additive constant $(\text{E}[R]^2/\text{Var}\left[ R \right]) nm \text{E} [I]$, we see that this can be minimized by picking a random-matrix basis that has a small E$[R]$ with a maximal Var$[R]$. Additionally, if $I_{ij}$ has an additive constant, removing this will reduce $\text{E}[I]$ and reduce the additive constant inherent in the projection scheme. 

Turning to the random-matrix basis reconstruction noise that arises from truncating the sum, we can quantify this by the variance of the ghost projection:
\begin{align}
\nonumber
\text{Var}[P_{ij}] 
&= \text{Var} [t_k R_{ij}^{\ \ k}] \\ \nonumber
&= \text{Var} \left[ \frac{1}{N \ \text{Var}\left[ R \right] } R_{ijk} R^{i'j' k} I_{i'j'} \right] \\ \nonumber
&= \frac{1}{N^2 \ \text{Var}\left[ R \right]^2 } \text{Var} \left[ R_{ijk} R^{i'j' k} I_{i'j'} \right] \\
&= \frac{1}{N \ \text{Var}\left[ R \right]^2 } \text{Var} \left[ R_{ij} R^{i'j'} I_{i'j'} \right]. 
\end{align}
We can split the above sum into the dominant case of $i'j' \neq ij$ and the sub-dominant case of $i'j'=ij$. Pursing the dominant case, we can ignore the sub-dominant case and subsequent complication of covariances. That is: 
\begin{align}
\nonumber \text{Var} & \left[ R_{ij} R^{i'j'} I_{i'j'} \right] \\ 
\nonumber &\approx \text{E} \left[ \left(  R_{ij} R^{i'j'}I_{i'j'} \right)^2 \right]  - \left( \text{E} \left[   R_{ij} R^{i'j'}I_{i'j'}  \right] \right)^2 \\
\nonumber & \approx \text{E}[R^2 ] \text{E} \left[ \left(R^{i'j'}I_{i'j'} \right)^2 \right]  - \text{E} [R]^4 (nm \text{E} [I])^2 \\
& \approx \text{E}[R^2 ] \text{Var}[R] (nm \text{E} [I])^2, \label{eq: one I am after}
\end{align}
where, in the second last line, we have employed:
\begin{align}
    \left( R^{\alpha \beta }I_{\alpha \beta } \right)^2 = J^{\alpha \beta } (R_{\alpha \beta }I_{\alpha \beta })^2 +  \sum_{\alpha \beta \neq \mu \nu } R_{\alpha \beta }I_{\alpha \beta } R_{\mu \nu}I_{\mu \nu },
\end{align}
disregarded the first term as another sub-dominant contribution, and then subsequently used the further simplifying approximation on the remaining term:
\begin{align}
    \sum_{\alpha \beta \neq \mu \nu } I_{\alpha \beta} I_{\mu \nu} \approx \left( nm \text{E}[I] \right)^2.
\end{align}
Substituting the result expressed in Eq.~(\ref{eq: one I am after}) into the variance of the ghost projection, we have:
\begin{align} \label{eq: GP Variance Approximation}
\text{Var}[P_{ij}] \approx \frac{ \left( nm \ \text{E}[R] \text{E}[I] \right)^2}{N \ \text{Var}\left[ R \right] } J_{ij}.
\end{align}

\subsection{Pseudo-correlation Coefficient}  \label{subsec:Pseudo-correlation Coefficient}

In obtaining a ghost-projection protocol by following a direct analogue of ghost imaging, a weighting coefficient for each basis member naturally arises that is proportional to its correlation with the desired image to be projected (i.e.~spatial product of $R_{ijk}I^{ij}$). However, this correlation contribution is not the correctly normalized correlation coefficient, which would take the form:
\begin{equation}
\tilde{C}_k = \frac{R_{ijk} I^{ij}}{\sqrt{(R_{stk}R^{st}_{\ \ k}) (I_{i'j'}I^{i'j'})}}.
\end{equation}
Note that the presence of the random matrix in the denominator of the correlation coefficient significantly complicates our calculations and is beyond what naturally arises (cf.~Eq.~(\ref{eq:t_k})). Instead, we will normalize by the expected denominator, keep the numerator unchanged, and call this the pseudo-correlation value $C_k$:
\begin{align}
C_k = \frac{R_{ijk} I^{ij}}{nm \sqrt{\text{E}[R^2] \text{E}[I^2]}}.
\label{eq:DefinitionForPseudoCorrelation}
\end{align} 
The pseudo-correlation coefficient still captures much of the desired characteristics of the correlation coefficient without the complication of having to normalize by each particular random matrix. A drawback of the pseudo-correlation coefficient is that it overly weights or prioritizes those random-matrix basis members that have a high average value, as opposed to a higher correlation with the desired image. However, this is a mild drawback for our purposes, with the reader being referred to Appendix \ref{AppA} for further detail. It is also worth repeating, in ghost imaging---which inspired the approach to ghost projection developed in the present section---the pseudo-correlation coefficient is directly analogous to the ghost-imaging concept of a bucket measurement, which is also an un-normalized correlation measurement. 

We now briefly elaborate on the pseudo-correlation coefficient and state useful results for later use.  We appeal to the Central Limit Theorem and assume Gaussian statistics for the pseudo-correlation distribution: 
\begin{align}
P(C_k) = \frac{1}{\sqrt{2 \pi \text{Var}[C]}} \exp \left( \frac{-(C_k -\text{E}[C])^2}{2 \text{Var}[C]} \right),
\end{align}
where the expected  pseudo-correlation value between a random-matrix basis member and the desired image is:
\begin{align}
\nonumber \text{E}[C] 
&= \text{E} \left[ \frac{R_{ijk} I^{ij}}{nm \sqrt{\text{E}[R^2] \text{E}[I^2]}} \right]\\ \nonumber 
&= \frac{\text{E}[R] J_{ij}I^{ij}}{nm \sqrt{\text{E}[R^2] \text{E}[I^2]}} \\
&= \frac{\text{E}[R] \text{E}[I]}{\sqrt{\text{E}[R^2] \text{E}[I^2]}},
\end{align}
and the variance in the pseudo-correlation value is:
\begin{align}
\nonumber 
\text{Var}[C] 
&= \text{Var}\left[ \frac{R_{ijk} I^{ij}}{nm \sqrt{\text{E}[R^2] \text{E}[I^2]}} \right]\\ \nonumber 
&= \frac{\text{Var} [R] J^{ij} I^{2}_{ \ ij}}{(nm)^2 \text{E}[R^2] \text{E}[I^2]} \\
&= \frac{ \text{Var}[R]}{nm \ \text{E}[R^2]}.
\end{align} 

\subsection{Pseudo-correlation Weighted Ghost Projection Signal-to-Noise Ratio (SNR)} \label{subsec:Naive Ghost Projection Signal-to-Noise Ratio (SNR)}

We define a pixel-wise signal-to-noise Ratio (SNR) for the truncated, pseudo-correlation weighted ghost projection case that is free from external noise contributions, such as Poisson noise, to be:
\begin{align} \nonumber
\text{SNR}_{ij} &\equiv \frac{\text{E}[P_{ij}] - nm \frac{\text{E}[R]^2 \text{E}[I]}{\text{Var}\left[ R \right]} J_{ij}}{\sqrt{\text{Var}[P_{ij}]}} \\
& \approx \frac{I_{ij} \sqrt{N \ \text{Var}\left[ R \right]}}{ nm \ \text{E}[R] \text{E}[I]}.
\end{align}
 This forms an upper bound for any experimental set-up, in which further sources of noise will be present. To obtain a global SNR expression, we can take the root-mean-square (RMS) value of the pixel-wise SNR:
\begin{align} \nonumber
\text{SNR} 
& \equiv \sqrt{\frac{1}{nm} J^{ij} \text{SNR}^{2}_{\ ij}} \\
&\approx \frac{ \sqrt{ \text{E}[I^2] N \ \text{Var}\left[ R \right]}}{ nm \ \text{E}[R] \text{E}[I]}.
\label{eq:PseudoCorrelationWeighted-GP-SNR}
\end{align}
Rearranging, we find an approximate noise-resolution uncertainty principle \cite{ceddia2018random,GureyevNRU,GureyevNRU2020} for ghost projection:
\begin{align} \label{eq: Noise-Resolution Uncertainty Principle}
(\text{SNR})^2 \times (nm)^2 \approx N \ \frac{ \text{Var}\left[ R \right] \text{E}[I^2] }{ \text{E}[R]^2 \text{E}[I]^2}.
\end{align}
 This might be taken to suggest that we would want to use as coarse a resolution as possible (or narrow a projection window as possible), with the maximum number of basis members. Recall though that this approximation is in the high-to-moderate resolution window, and therefore breaks down in the very-low-resolution limit. Furthermore, supposing we have a desired SNR in mind, we can rearrange Eq.~(\ref{eq: Noise-Resolution Uncertainty Principle}) to make $N$ the subject and determine an approximate value for how large a basis set is required to achieve our desired SNR. Hence 
\begin{align}
N \approx \text{SNR}^2 \ (nm)^2 \frac{\text{E}[R]^2 \text{E}[I]^2}{\text{Var}\left[ R \right] \text{E}[I^2]}. \label{eq:Number of masks 1}
\end{align}
 
\subsection{A Revised Pseudo-correlation Weighted Ghost Projection Scheme} \label{subsec:A Revised Naive Ghost Projection}

Upon reviewing the expression for the variance of the na\"{i}ve ghost-projection protocol in Eq.~(\ref{eq: GP Variance Approximation}) above, we see that it is proportional to the expected value of the desired image that we wish to project. To minimize this, we can seek a shifted version of our image that preserves non-negative exposure times $t_k$ (see~Eq.~(\ref{eq:t_k})). Supposing we introduce a shift to the image, we can define a new image:
\begin{equation}
I'_{ij} = I_{ij} - \delta J_{ij},  
\end{equation}
that enforces:
\begin{align}
\nonumber \text{min}[t_k] &= 0 \\ \nonumber &= \frac{1}{N \ \text{Var}\left[ R \right] } \text{min} [R_{i'j'k} I'^{i'j'}] \\ &= \frac{1}{N \ \text{Var}\left[ R \right] } \text{min} [ R_{i'j'k} (I^{i'j'} - \delta J^{i'j'})]. 
\end{align}
From this, we can determine that the optimal offset is:
\begin{equation}
\delta =\text{min} [R_{i'j'k}I^{i'j'}]/(R_{\mu \nu k}J^{\mu \nu}),
\label{eq:delta}
\end{equation}
and we can define the shifted scheme:
\begin{align} 
\nonumber P_{ij} &\equiv \frac{1}{N \ \text{Var}\left[ R \right] } R_{i'j'k} (I^{i'j'} - \delta J^{i'j'}) R_{ij}^{\ \ k} \\ &\approx I'_{ij} + nm \frac{\text{E}[R]^2 \text{E} [I']}{\text{Var}\left[ R \right]} J_{ij}.
\label{eq:shifted naive Ghost Projection} \end{align}
This shifted scheme preserves the results of Secs \ref{subsec:Naive ghost projection} and \ref{subsec:Naive Ghost Projection Signal-to-Noise Ratio (SNR)} regarding convergence and SNR, whilst performing better than the previously defined pseudo-correlation weighted ghost projection scheme. Note also that the shifted image may no longer be strictly positive. This is acceptable insofar as any negative component of the image simply falls below the additive constant and the overall projection remains physical.

\subsection{Pseudo-correlation Weighted Ghost Projection Simulation} \label{subsec:Naive Ghost Projection Simulation}

\begin{figure*}[ht!]
     \centering
     \begin{subfigure}{0.329\textwidth}
         \centering
         \includegraphics[width=\textwidth]{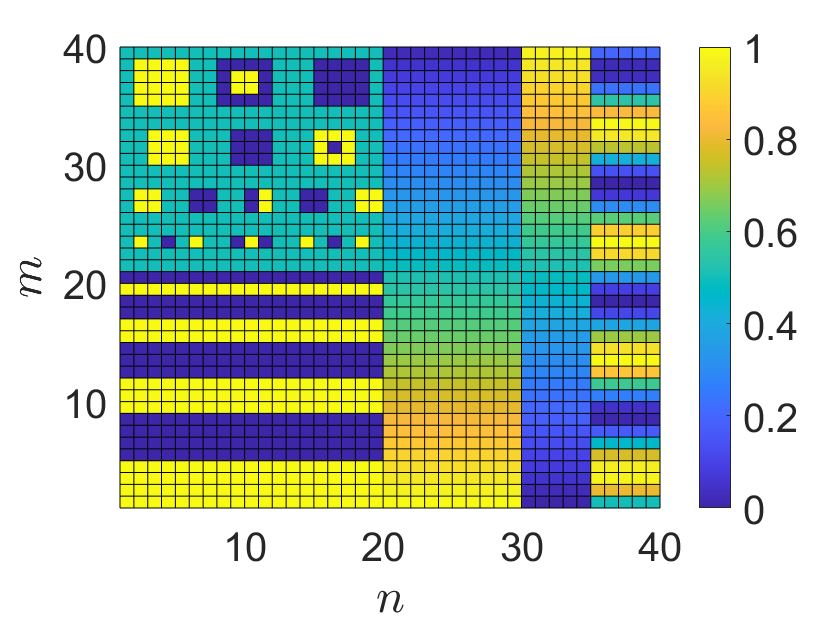}
         \caption{ }
         \label{subfig: GP1}
     \end{subfigure}
     \begin{subfigure}{0.329\textwidth}
         \centering
         \includegraphics[width=\textwidth]{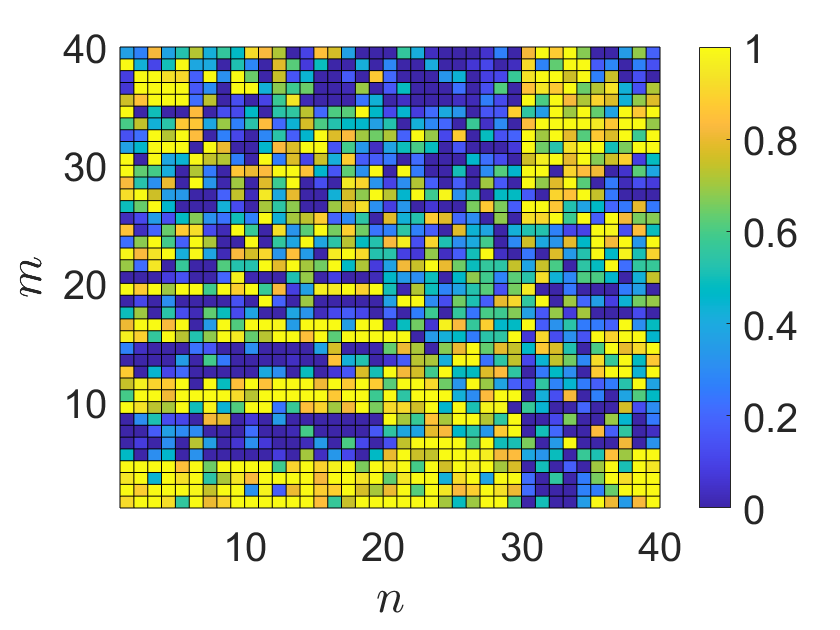}
         \caption{ }
         \label{subfig: GP2}
     \end{subfigure}
     \begin{subfigure}{0.329\textwidth}
         \centering
         \includegraphics[width=\textwidth]{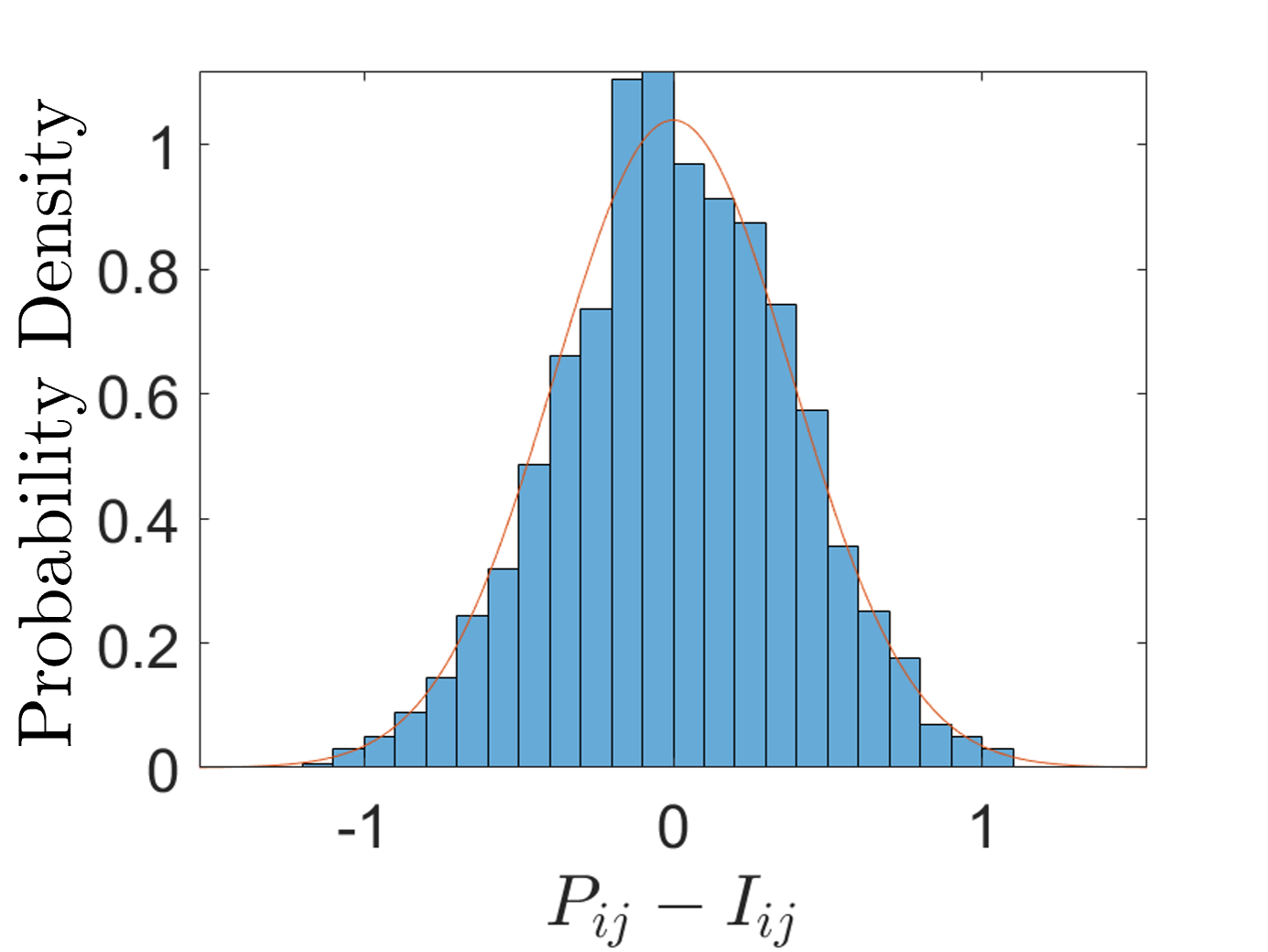}
         \caption{ }
         \label{subfig: GP3}
     \end{subfigure}
     \begin{subfigure}{0.329\textwidth}
         \centering
         \includegraphics[width=\textwidth]{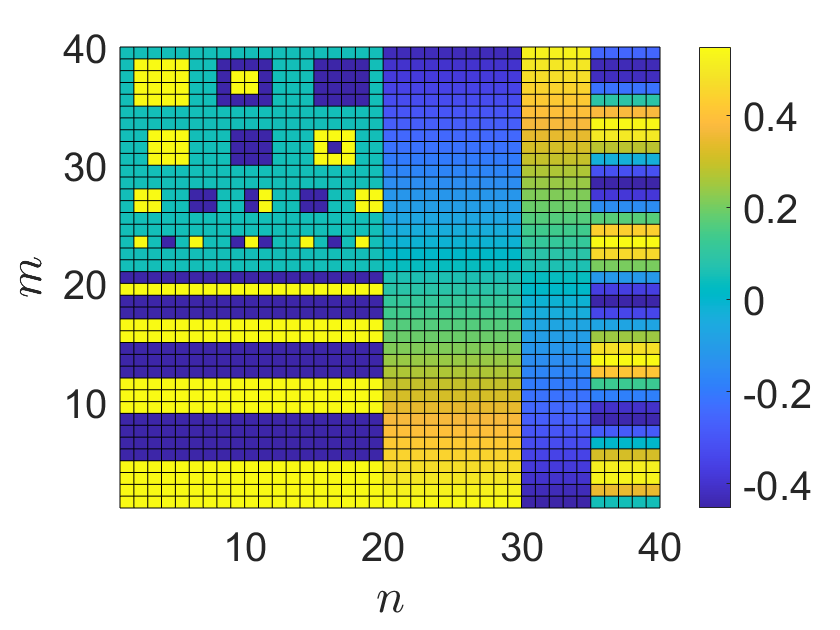}
         \caption{ }
         \label{subfig: GP1_R }
     \end{subfigure}
     \begin{subfigure}{0.329\textwidth}
         \centering
         \includegraphics[width=\textwidth]{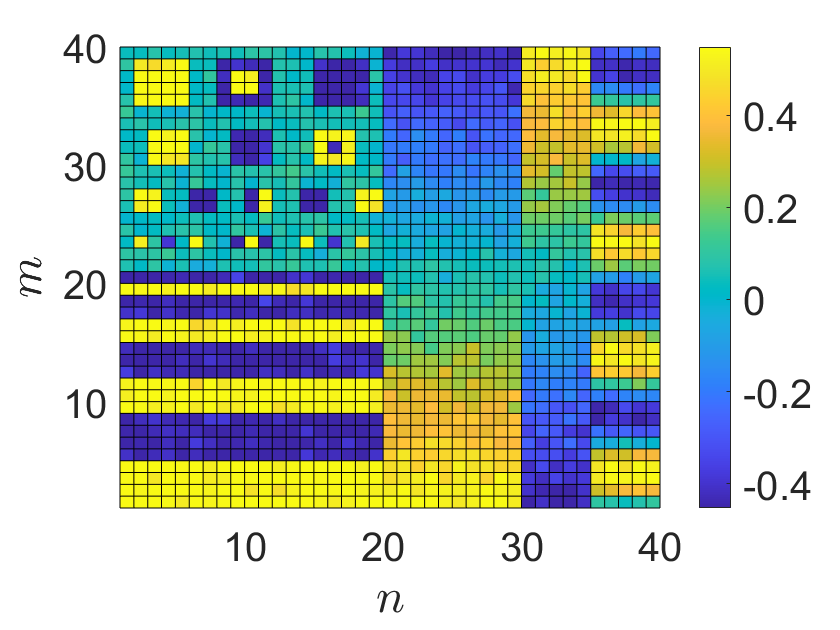}
         \caption{ }
         \label{subfig: GP2_R}
     \end{subfigure}
     \begin{subfigure}{0.329\textwidth}
         \centering
         \includegraphics[width=\textwidth]{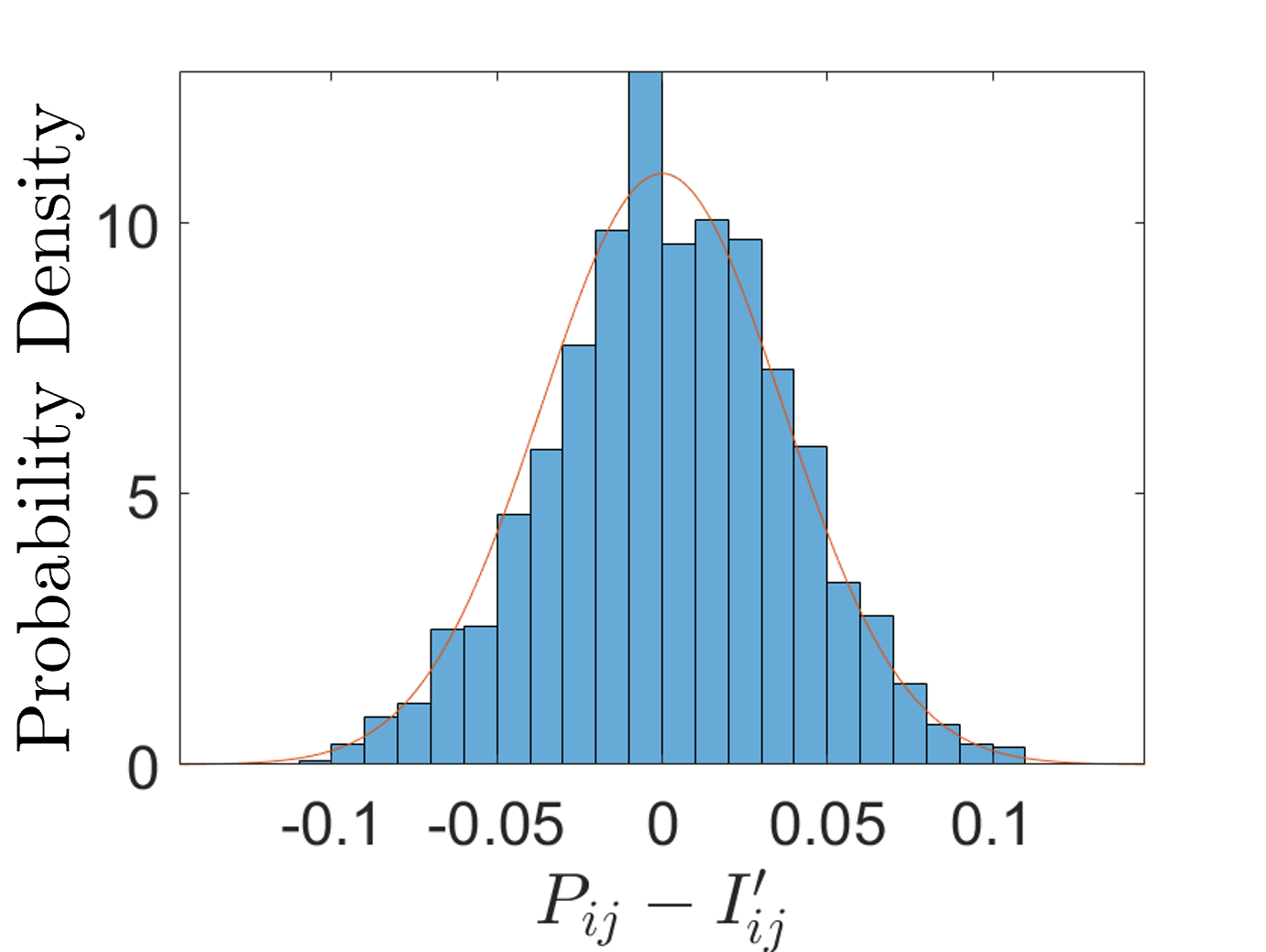}
         \caption{ }
         \label{subfig: GP3_R}
     \end{subfigure}
        \caption{(a) The desired image we wish to project, $I_{ij}$. (b) The ghost projection obtained via %the reconstruction outlined in
        Eq.~(\ref{eq: Standard Ghost Projection}), minus the expected offset of 2,400 using $N=13,046,012$. An SNR of 1.60 was expected, with an SNR of 1.64 obtained in simulation. (c) The noise obtained in the ghost projection, overlaid with the predicted noise distribution. (d) The revised desired image we wish to project, $I'_{ij}$. (e) The ghost projection obtained via Eq.~(\ref{eq:shifted naive Ghost Projection}), minus the expected offset of 228.6, also using $N=13,046,012$. An SNR of 9.83 was predicted for which an SNR of 9.99 was obtained in simulation. (f) The noise obtained in the ghost projection, overlaid with the predicted noise distribution.}
        \label{fig:six graphs}
\end{figure*}

To illustrate the analytical work performed thus far, we perform some simulations. Beginning with the target image in Fig.~\ref{subfig: GP1}, we numerically investigate how ghost projection performs in producing varying sized dots (both peaks and troughs), increasingly refined high-contrast bands, two sections of linear gradient and a sinusoidal section. The random-matrix basis used has each matrix element uniformly distributed in the real-number interval $[0,1]$. To obtain a reasonable SNR in the pseudo-correlation weighted ghost projection scheme (Sec.~\ref{subsec:Naive ghost projection}), we need to use a significant number of random-basis members, e.g.~about 13 million to obtain an SNR of 1.6 for a $40 \times 40$ image. Computationally storing such a large number of random matrices is impractical. We can instead use a statistical estimate of $\delta$ and continually overwrite each random mask after adding its contribution to the projection. To statistically estimate $\delta$, we can approximate the denominator in Eq.~(\ref{eq:delta}) with its expected value $R_{ijk}J^{ij} \rightarrow nm\text{E}[R]$, and treat the numerator as being Gaussian distributed. From there, we can say that the minimum pseudo-correlation or exposure time would be a `$1$ in $N$' event, or: 
\begin{equation}
s =  \sqrt{2} \, \text{erf}^{-1} \left( \frac{N-1}{N} \right) 
\end{equation}
standard deviations below the expected value of the numerator. Overall then, from a statistical perspective: 
\begin{equation}
\delta = \text{E}[I] - s \sqrt{\frac{\text{E}[I^2]\text{Var}[R]}{nm\text{E}[R]^2}}. 
\end{equation}
The number of random masks employed was $N=13,046,012$, which comes from Eq.~(\ref{eq:Number of masks 1}) with an SNR of 1.60. In simulation, the ghost projection achieved an SNR of 1.64, as shown in Fig.~\ref{subfig: GP2}. If we were to use the same basis set and the revised scheme for performing the ghost projection (see Fig.~\ref{subfig: GP1_R }), we can improve the situation to obtain a simulated SNR of 9.99 (see Fig.~\ref{subfig: GP2_R}), for which an SNR of 9.83 was predicted using Eq.~(\ref{eq:PseudoCorrelationWeighted-GP-SNR}). Further, we can confirm that the ghost projection converged to the desired image, pedestal and variance as expected, as evidenced by Figs.~\ref{subfig: GP3} and \ref{subfig: GP3_R}.

% These figures were generated with the MATLAB code GP_Corr.m
% The number of random masks used is 13046012 
% SNR 1 Predicted 1.600 
% SNR 1 simulated 1.636 
% SNR 2 Predicted 9.832 
% SNR 2 simulated 9.991 
% Offset 1 2400.0 
% Offset 2 228.6 

\section{Pseudo-correlation Filtered Ghost Projection with a Random-Matrix Basis} \label{sec: Pseudo-correlation Filtered Ghost Projection}

\subsection{Pseudo-correlation Filtered Ghost Projection} \label{subsec: Pseudo-correlation Filtered Random Matrix Ghost Projection}

The previous section developed a ghost-projection scheme via the somewhat na\"{i}ve protocol of weighting each random matrix by its pseudo-correlation coefficient. To more efficiently achieve a ghost projection of our desired image, we might try and filter the basis set to leave only those members that have a positive, or high degree of correlation with the desired image \cite{paganin2019writing}. Supposing we were to define a minimum pseudo-correlation value $C_{\text{min}}$ and remove all basis members with a pseudo-correlation value below this threshold, we might then ask how this filtering changes the distribution of $R'_{ijk}$, as compared to the unfiltered set $R_{ijk}$. Examining the expectation value of the filtered basis set E$[R'_{ij}]$ we can assume that if $I_{ij}$ is zero, then that pixel makes zero contribution to the  pseudo-correlation value and is, on average, unchanged by the filtering. That is, E$[R'_{ij}]=\text{E}[R]$ for $I_{ij} = 0$. From here, we suggest the ansatz:
\begin{equation}
\text{E}[R'_{ij}] = \text{E}[R] J_{ij} + I_{ij} \gamma. 
\end{equation}
That is, the expected value of the filtered random-matrix basis is linearly skewed at each pixel towards the desired value of $I_{ij}$.  To determine the scaling of this skewing, we can use the expected pseudo-correlation value of the filtered set E$[C']$ as a constraint:
\begin{align}
\nonumber \text{E}[C'] 
&= \text{E} \left[ \frac{R'_{ijk} I^{ij}}{nm \sqrt{\text{E}[R^2] \text{E}[I^2]}} \right] \\ \nonumber 
&= \frac{\text{E}[R'_{ij}]I^{ij}}{nm \sqrt{\text{E}[R^2] \text{E}[I^2]}} \\ \nonumber 
&= \frac{1}{nm \sqrt{\text{E}[R^2] \text{E}[I^2]}} (\text{E}[R]J_{ij} + I_{ij} \gamma) I^{ij} \\ 
&=  \frac{1}{\sqrt{\text{E}[R^2] \text{E}[I^2]}}  \left(\text{E}[R]\text{E}[I] + \text{E}[I^2] \gamma \right),
\end{align}
where $\text{E}[R'_{ij}]$ is the pixel-dependent expectation value of the random-matrix basis of the filtered basis set. Recalling that E$[C] = \text{E}[R]\text{E}[I]/\sqrt{\text{E}[R^2]\text{E}[I^2]}$, this implies:
\begin{equation}
\gamma = (\text{E}[C'] - \text{E}[C]) \sqrt{\text{E}[R^2]/\text{E}[I^2]}.
\end{equation}
From the form of this constant, we may retrospectively understand the effect of filtering as skewing the random-matrix basis towards the matrix $I_{ij}$ by the average correlation amount $\text{E}[C'] - \text{E}[C]$ (where we might regard $\text{E}[C]$ to be the zero-correlation value despite it not necessarily being zero owing to non-zero averages of $\text{E}[R]$ and $\text{E}[I]$). Moreover, this skewing towards $I_{ij}$ is norm-adjusted, where we divide by the image norm and multiply by the expected norm of the random-matrix basis. 

So, if we were to adopt the relatively simple ghost projection protocol of averaging over the pseudo-correlation coefficient filtered random-matrix basis set, we would obtain:
\begin{align} \nonumber
P_{ij} &\equiv  \frac{ J^{k}  R'_{ijk}}{\gamma N'} \\ \nonumber &\approx  I_{ij} + \frac{\text{E}[R]}{\gamma} J_{ij} \\  &\approx  I_{ij} + \frac{\text{E}[R]}{\text{E}[C'] - \text{E}[C]} \sqrt{\frac{\text{E}[I^2]}{\text{E}[R^2]}}  J_{ij},
\label{eq: Pseudo-correlation Ghost Projection Average}
\end{align}
where $N'$ is the number of basis members in the filtered set. The next quantity of interest is the amount of noise inherent in a ghost projection obtained via this protocol, which is quantified in the variance:
\begin{align}
\text{Var} [P_{ij}] 
= \frac{1}{(\gamma N')^2} \text{Var} [R'_{ijk}J^k] 
= \frac{1}{ \gamma^2 N'} \text{Var} [R'_{ij}].  \label{eq: Pseudo-correlation Ghost Projection Variance}
\end{align}
From here, we need to determine $\text{Var} [R'_{ij}]$, although, this quantity has a complicated behavior in general. Supposing the desired image to be projected were a single-pixel pin-hole, or Dirac delta $I_{ij} = \delta_{ii'jj'}$, then the variance of the `switched on' pixel would be proportional to the variance of the filtered pseudo-correlation coefficient $\text{Var} [R'_{i'j'}] = \text{E}[R^2] \text{Var}[C']$ and the variance of the `non switched on' pixels would remain unchanged, i.e.~$\text{Var} [R'_{ij}] = \text{Var} [R]$. Supposing the image is more complicated and many pixels contribute to the pseudo-correlation coefficient, the variance of the filtered random-matrix basis is largely unchanged, i.e.~$\text{Var} [R'_{ij}] \approx \text{Var} [R]$ for all the pixels. How $\text{Var} [R'_{ij}]$ varies between these two cases is unknown at this time and is left as a point of future work. For the purposes of this paper, we assume the latter case owing to its applicability to many foreseeable practical applications. 

Turning to E$[C']$, we can easily numerically evaluate this when given a random-matrix basis set, desired image and minimum pseudo-correlation cut-off. For optimization purposes, however, having an analytical expression will be useful:
\begin{widetext}
\begin{align}
\nonumber
\text{E}[C'] 
&= \int_{C_{\text{min}}}^{\infty} C P(C) dC \\ \nonumber
&= \sqrt{\frac{2}{\pi \text{Var}[C]}} \left[  \text{erfc} \left( \frac{C_{\text{min}} - \text{E}[C]}{\sqrt{2 \text{Var}[C]}} \right) \right]^{-1} \int_{C_{\text{min}}}^{\infty} C \exp \left(-  \frac{(C-\text{E}[C])^2}{2 \text{Var}[C]}  \right) dC  \\
&= \text{E}[C] + \sqrt{\frac{2 \text{Var}[C]}{\pi}}  \left[  \text{erfc} \left( \frac{C_{\text{min}} - \text{E}[C]}{\sqrt{2 \text{Var}[C]}} \right) \right]^{-1} \exp \left( - \frac{(C_{\text{min}} - \text{E}[C])^2}{2 \text{Var}[C]} \right).
\end{align}
\end{widetext}

\subsection{Optimum Pseudo-correlation Cut-off} \label{subsec:Pseudo-correlation Cut-off Optimum}

For fixed $N$, we might ask what is the optimum pseudo-correlation cut-off that minimizes the noise (i.e.~variance) of the ghost projection. That is, for a fixed set of $N$ random-matrix basis members, we appeal to a statistical representation of the expected filtered number $N'$ as $fN$, where $f$ is related to the cumulative distribution function (CDF) via:
\begin{align}
f(C_{\text{min}}) = 1- \text{CDF} = \frac{1}{2} \text{erfc} \left( \frac{C_{\text{min}}-\text{E}[C]}{\sqrt{2 \text{Var}[C]}} \right).
\end{align}
Substituting this into the variance expression (Eq.~(\ref{eq: Pseudo-correlation Ghost Projection Variance})) in the case that $\text{Var}[R'_{ij}] \approx \text{Var}[R]$, we have:
\begin{align}
\nonumber \text{Var} [P_{ij}] &\approx \frac{\text{Var}[R]}{\gamma^2 f N} \\
&= \frac{\text{Var}[R] }{fN} \frac{\text{E}[I^2]}{\text{E}[R^2]} \frac{1}{(\text{E}[C'] - \text{E}[C])^2}.
\end{align}
Substituting in our analytical expression for $\text{E}[C']$ and letting:
\begin{equation}
X \equiv \frac{C_{\text{min}} - \text{E}[C]}{\sqrt{2 \text{Var}[C]}},
\end{equation}
we are left with:
\begin{align}
\text{Var} [P_{ij}] &= \frac{\text{Var}[R] \text{E}[I^2]}{\text{E}[R^2] N} \frac{2\pi }{\text{Var}[C]} f(X) \exp (2 X^2).
\end{align}
To find the optimum cut-off, we take the derivative of this expression with respect to $X$ and set it equal to zero in the usual way. This yields a non-linear algebraic equation to solve for $X$ and $C_{\text{min}}$. Taking a 1-3 Pad\'{e} approximant \cite{PadeBook} for the derivative then yields: 
\begin{align}
X \approx \frac{3\sqrt{\pi}(4\pi-7)}{4(6\pi^2-15 \pi+5)} \approx 0.433. 
\end{align}
Hence the optimum pseudo-correlation cut-off is:
\begin{equation}
C_{\text{min}} \approx \text{E}[C] + 0.612 \sqrt{\text{Var}[C]}.
\end{equation}
Therefore we would expect about 27\% of the basis set $N$ to remain after filtration. Substituting our optimized value of $X$ into $\text{Var} [P_{ij}]$, as well as our expression for $\text{Var}[C]$, we have the optimized ghost-projection variance:
\begin{align}
\nonumber
\text{Var} [P_{ij}] &= \frac{ 2\pi n m \text{E}[I^2]}{ N}  f(X) \exp (2 X^2)\\ &\approx 0.393 \frac{ 2\pi n m \text{E}[I^2]}{ N} .
\end{align}
Note that the variance of a ghost projection, obtained in this way, involves the term $\text{E}[I^2]$. This term can be minimized by zero centering our image, i.e.~setting $\text{E}[I]=0$. In doing this, we might be concerned that elements of our desired image are negative, however, similar to the revised pseudo-correlation weighted ghost projection, since this dips below the additive constant, it does not lead to non-physical negative radiant exposures.

\subsection{Pseudo-correlation Filtered Ghost Projection Signal-to-Noise Ratio (SNR)}

For comparison with the pseudo-correlation weighted ghost-projection case, we develop a noise-resolution uncertainty principle for the filtered case. We define the pixel-wise SNR$_{ij}$:
\begin{align}
\nonumber
\text{SNR}_{ij} 
&\equiv \frac{\text{E}[P_{ij}] - \frac{\text{E}[R]}{\gamma} J_{ij}}{\sqrt{\text{Var}[P_{ij}]}} \\ \nonumber 
&\approx \frac{I_{ij} \gamma \sqrt{N'} }{\sqrt{ \text{Var}[R]}} \\
&\approx I_{ij}  \frac{ \sqrt{N' \ \text{E}[R^2]}}{ \sqrt{\text{Var}\left[ R \right] \text{E}[I^2]}} (\text{E}[C'] - \text{E}[C]),
\end{align}
where we have assumed that the variance of the filtered random basis is approximately unchanged by filtering, i.e.~$\text{Var}[R'_{ij}] \approx \text{Var}[R]$, which typically occurs when a reasonable number of pixels contribute to the pseudo-correlation coefficient. To obtain a global SNR, we take the RMS value of the pixel SNR:
\begin{align}
\nonumber
\text{SNR} 
& \equiv \sqrt{\frac{1}{nm} J^{ij} \text{SNR}^{2}_{\ ij}} \\ \nonumber
&\approx \sqrt{ \frac{ \text{E}[I^2] \gamma^2 N'}{\text{Var}\left[ R \right]}} \\ \nonumber
&\approx \sqrt{ \frac{ \text{E}[I^2] N'}{\text{Var}\left[ R \right]}} \  (\text{E}[C'] - \text{E}[C]) \sqrt{\frac{\text{E}[R^2]}{\text{E}[I^2]}} \\
&\approx \sqrt{ \frac{ N' \ \text{E}[R^2] }{\text{Var}\left[ R \right]}} 
\sqrt{\frac{\text{Var}[C]}{2\pi}}  \frac{ \exp \left( - X^2 \right)}{f(X)}. \label{eq: now I want this one}
\end{align}
Defining:
\begin{align}
\xi(C_{\text{min}}) \equiv  \sqrt{\frac{1}{2\pi}}  \frac{ \exp \left( - X^2 \right) }{f(X)}
\end{align}
and substituting in $\text{Var}[C] = \frac{ \text{Var}[R]}{nm \ \text{E}[R^2]}$, we have:
\begin{align} \label{eq: Pseudo-correlation Ghost Projection SNR}
\text{SNR} 
&\approx \sqrt{\frac{ N' }{nm}} \xi(C_{\text{min}}).
\end{align}
Rearranging, we find a noise-resolution uncertainty principle for pseudo-correlation filtered ghost projection as:
\begin{align}  \label{eq: Pseudo-correlation Noise Resolution Uncertainty Principle}
(\text{SNR})^2 \times nm \approx N' \xi^2(C_{\text{min}}).
\end{align}
Substituting in $N' = f(C_{\text{min}})N$ and evaluating $\xi^2(C_{\text{min}})$ at the optimum cut-off value ($\xi(C_{\text{min}}) \approx 1.22$), we obtain the best-case, noise-resolution uncertainty principle for pseudo-correlation filtered ghost projection:
\begin{align}  \label{eq:Pseudo-correlation Noise Resolution Uncertainty Principle Optimised}
(\text{SNR})^2 \times nm \approx \frac{N}{2.47}.
\end{align}
Comparing this to the ghost projection obtained by weighting each random-basis member according to its pseudo-correlation coefficient (see Eq.~(\ref{eq: Noise-Resolution Uncertainty Principle})), observe that we have indeed obtained a more favorable noise-resolution uncertainty product by filtering and averaging.  Most notable is the penalty for resolution.  For the pseudo-correlation weighted case, SNR decays linearly with $nm$. Conversely, in the pseudo-correlation filtered case, SNR decays with the square root of $nm$.

\subsection{Pseudo-correlation Filtered Ghost Projection Basis Size Estimate} \label{subsec:Pseudo-correlation Filtered Ghost Projection Basis Size Estimate}

Having calculated an SNR relationship for pseudo-correlation filtered random-matrix ghost projection, we can determine what size basis set we require to almost surely represent any arbitrary image of a given resolution to a desired SNR. Employing Eq.~(\ref{eq:Pseudo-correlation Noise Resolution Uncertainty Principle Optimised}) in reverse, supposing we have a desired SNR and resolution in mind, we can use this equation to estimate the required size $N$ of the random-matrix basis. For example, suppose we want to perform a ghost projection to an SNR of 5 at a resolution of $40 \times 40$, we would expect to be able to obtain this from an unfiltered basis set of $N \approx 9$8,800 members.

We can improve this estimate for the required basis size considering the variance in $N'$ as estimated by the binomial distribution. That is, the probability that a random-matrix basis member being kept is $f$ and the probability that it is discarded is $(1-f)$. Given that the variance in a binomial distribution is $np(1-p)$, where $n$ is the number of trials and $p$ is the probability of success, we have a variance in $N'$ of $Nf(1-f)$. That is:
\begin{equation}
N' = fN \pm \sqrt{Nf(1-f)},
\end{equation}
to within one standard deviation. Taking a worst-case estimate of $s$ standard deviations below the mean, namely 
\begin{equation}
N' = fN - s\sqrt{Nf(1-f)}, 
\end{equation}
this means that we have realized significantly fewer correlated basis members than we would reasonably expect. Inverting this result with the quadratic formula, we can estimate a larger $N$ such that any image can almost surely (to within an $s$ sigma event) be captured by an unfiltered basis set of size:
\begin{align}
\nonumber N &= \frac{N'}{f} + \frac{s}{2f} \sqrt{4(1-f)N' + s^2(1-f)^2} + \frac{s^2(1-f)}{2f} \\
&\approx \frac{N'}{f} + s \sqrt{\frac{(1-f)N'}{f^2}}.
\end{align}
Here $N' = \text{SNR}^2 nm/\xi^2$, and the approximation in moving from the first to the second line comes from assuming $N'$ to be the dominant parameter. Making this substitution and evaluating at the optimum cut-off value $C_{\text{min}}$ while focusing on dominant terms, we have:
\begin{align}
\nonumber N &\approx \frac{\text{SNR}^2 nm}{f\xi^2} + \frac{s \ \text{SNR}}{f \xi} \sqrt{nm (1-f)} \\
&\approx 2.47 \ nm \ \text{SNR}^2 + 2.59 \ s  \sqrt{nm} \ \text{SNR}.
\label{eq:ExtraSNRfactor}
\end{align}
Appending to our previous example of a ghost projection at SNR of 5 and resolution of $40 \times 40$, we might expect this ghost projection to be achievable from a basis size of $N \approx 98,800$ members. Supposing we want to be 3-sigma (99\%) confident, then according to Eq.~(\ref{eq:ExtraSNRfactor}) we require an additional 1,554 basis members.

\subsection{Realistic Dwell Time Pseudo-correlation Filtered Ghost Projection}

Supposing there is a minimum dwell time $t_{\text{min}}$ that can experimentally be achieved, in the limit of many basis members this may exclude us from using the exposure times prescribed by the analytically optimum pseudo-correlation cut-off value. In response, we could simply increase the analytical dwell times to what is experimentally achievable and obtain an intensity-dilated version of our desired image. An alternative is to use a more restrictive condition during filtration of the random basis, such that the dwell times $t_{k} = 1/(\gamma N')$ remain above the minimum. To find the pseudo-correlation minimum, we can use the statistical representations of $N'=fN$ and $\gamma$ to set up an expression which can be solved for $C_{\text{min}}$:
\begin{widetext}
\begin{align}
\frac{1}{t_{\text{min}} N} = f \gamma &=  \sqrt{\frac{\text{E}[R^2]}{\text{E}[I^2]}} \sqrt{\frac{ \text{Var}[C]}{2\pi}}  \exp \left( - \frac{(C_{\text{min}} - \text{E}[C])^2}{2 \text{Var}[C]} \right) \\
\Rightarrow  C_{\text{min}} &= \text{E}[C] +  \left[ 2 \text{Var}[C] \ln \left( t_{\text{min}} N  \sqrt{\frac{\text{E}[R^2]}{\text{E}[I^2]}}\sqrt{\frac{ \text{Var}[C]}{2\pi}} \right) \right]^{1/2}.
\end{align}
\end{widetext}
Now, for a given $N$, $nm$, $t_{\text{min}}$, E$[R]$ and Var$[R]$ we would we expect an SNR of:
\begin{align}
\text{SNR} \approx \sqrt{ \frac{ \text{E}[I^2] \gamma}{ t_{\text{min}} \text{Var}\left[ R \right]}},
\end{align}
where the approximate symbol appears owing to the use of the assumption $\text{Var}[R'_{ij}] \approx \text{Var}[R]$. 

To give an example of these results, suppose the minimum exposure time that could be experimentally achieved is $t_{\text{min}} = 1/200$. If we had a basis set of size $N = 10$0,000 with the properties $\text{E}[R] = 1/2$ and $\text{Var}[R] = 1/12$, and a $40 \times 40$ resolution image with the property $\text{E}[I^2] = 0.5$, then we would expect to filter out 11.7\% of basis members and achieve an SNR of 4.54. For comparison, the optimum exposure time would be $1/333$ with 27.0\% of basis members being filtered out and would achieve an SNR of 5.03.

\subsection{Pseudo-correlation Filtered Ghost Projection Simulation} \label{subsec:Pseudo-correlation Filtered Random Matrix Ghost Projection Simulation}

\begin{figure*}[ht!]
     \centering
     \begin{subfigure}{0.329\textwidth}
         \centering
         \includegraphics[width=\textwidth]{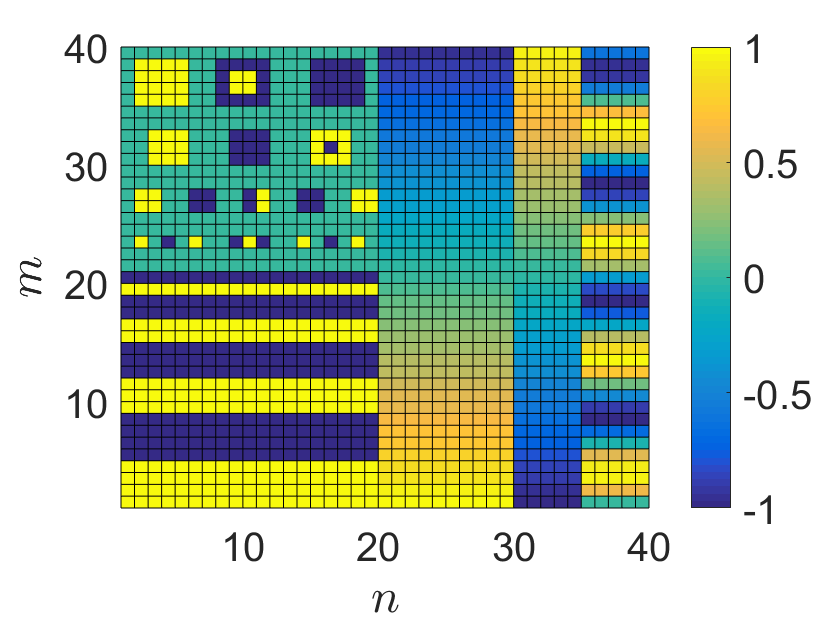}
         \caption{ }
         \label{subfig: GP1 Pseudo}
     \end{subfigure}
     \begin{subfigure}{0.329\textwidth}
         \centering
         \includegraphics[width=\textwidth]{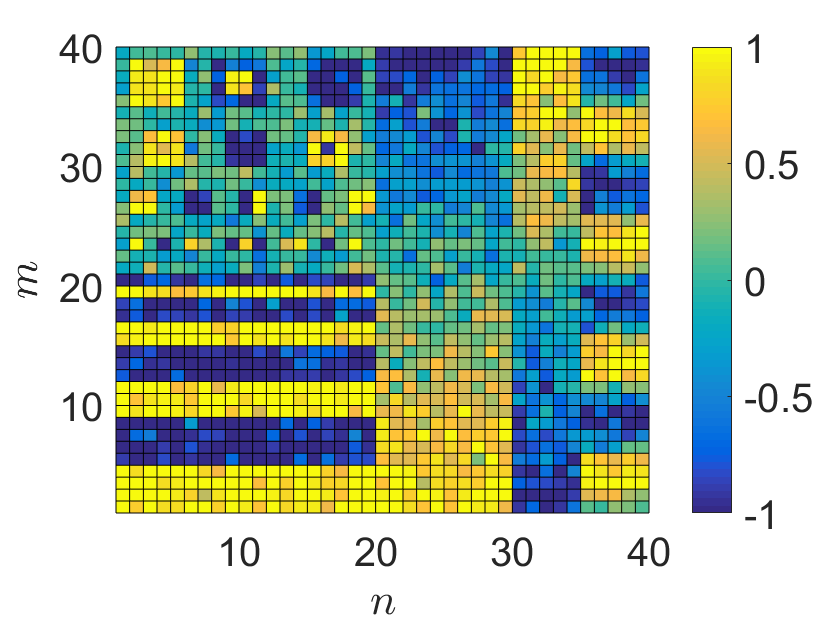}
         \caption{ }
         \label{subfig: GP2 Pseudo}
     \end{subfigure}
     \begin{subfigure}{0.329\textwidth}
         \centering
         \includegraphics[width=\textwidth]{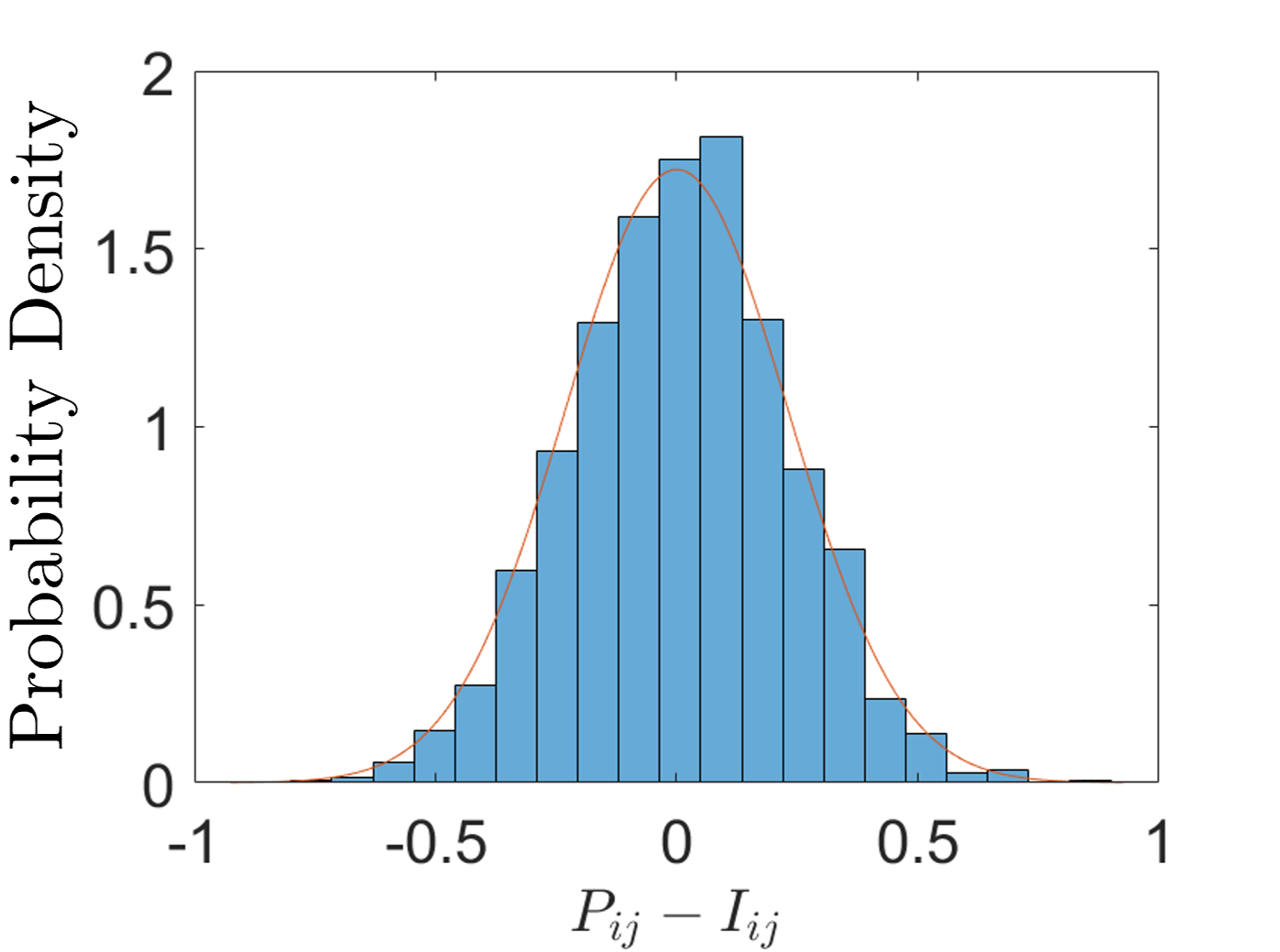}
         \caption{ }
         \label{subfig: GP3 Pseudo}
     \end{subfigure}
        \caption{(a) The desired image we wish to project expressed in terms of a transmission coefficient. (b) The pseudo-correlation filtered ghost projection obtained via the reconstruction outlined in Eq.~(\ref{eq: Pseudo-correlation Ghost Projection Average}), minus the expected offset of 34.54 for $N'=10,000$ (which was filtered from a basis $N =  37,247$) and $C_{\text{min}} = \text{E}[C] + 0.612 \sqrt{\text{Var}[C]}$. (c) The noise obtained in the pseudo-correlation filtered ghost projection, overlaid with the predicted noise distribution.}
        \label{fig:pseudo-correlation result graphs}
\end{figure*}

Using the test image in Fig.~\ref{subfig: GP1 Pseudo}, which is the same test image as in the pseudo-correlation weighted case (see~Fig.~\ref{fig:six graphs}) with altered contrast, we test the capabilities of the pseudo-correlation filtered scheme. The random basis was generated using deviates from a uniform distribution between $[0,1]$. 

From Fig.~\ref{fig:pseudo-correlation result graphs}, we can observe that our analytic prediction for the random-matrix scaling $\gamma N'$ required to obtain the image $I$ and offset $\text{E}[R]/\gamma$ of the ghost projection (see Eq.~(\ref{eq: Pseudo-correlation Ghost Projection Average})) is correct, as is the expected noise accurate to within reasonable bounds, where the noise is characterized by the variance of each pixel from the expected value $\text{Var}[P_{ij}] \approx \text{Var}[R]/(N'\gamma^2)$.

Using a filtered basis set size $N' = 10,000$ at the optimal pseudo-correlation cut-off condition $C_{\text{min}} = \text{E}[C] + 0.612 \sqrt{\text{Var}[C]}$, we would expect this filtered basis set to come from an unfiltered basis set of size $N = 37,000 \pm s \ 316$, where $s$ is the number of standard deviations we desire for confidence in our estimate. In the simulation, we needed to generate $N = 37,247$ basis members to filter out our desired $N'$ above the cut-off value (within 1 standard deviation of the expected value). Moreover, for these parameters, we expect an offset of 34.54 (so, `eating into' this by -1 still renders our projection physical) and we expect a global SNR of 3.06 (see Eq.~(\ref{eq:Pseudo-correlation Noise Resolution Uncertainty Principle Optimised})) for which an SNR of 3.16 was obtained in simulation. 

For the same simulation conditions of $N = 37,247$, changing the cut-off value for the pseudo-correlation filter to (i) the mean value $C_{\text{min}} = \text{E}[C]$ or (ii)  the mean plus one standard deviation $C_{\text{min}} = \text{E}[C] + \sqrt{\text{Var}[C]}$, respectively produced (i) an SNR of 2.72 with $N' = 18,703$ (for which an SNR of 2.73 was predicted by Eq.~(\ref{eq: Pseudo-correlation Ghost Projection SNR})) and (ii) an SNR of 2.88 with $N' = 5,844$ (for which an SNR of 2.94 was predicted). This shows that indeed, the `middle ground' of $C_{\text{min}} = \text{E}[C] + 0.612 \sqrt{\text{Var}[C]}$ strikes a balance between removing those random-basis members that are weakly correlated or uncorrelated with our desired image, whilst not being too `heavy handed' so as to remove too many basis members such that desired image, which lays hidden in the average of the basis members, may still yet emerge. 

% These figures were generated using the MATLAB code GP_filteredBasis

We have chosen a filtered basis set size of $N'=10,000$ members, which came from the unfiltered basis set having $N = 37,247$ members, with a resolution of $n=m=40$ speckles, in the hope that these are illustrative for conditions that are realistically achievable in a laboratory. Qualitatively examining the resolution results of our pseudo-correlation filtered ghost projection scheme, Fig.~\ref{subfig: GP2 Pseudo}, we can see a relatively good result for the high contrast bands in the lower left quadrant. There is a reasonable result for (i) the linear gradients and sinusoidal pattern in the right half of the ghost projection, as well as (ii) the 2-, 3- and 4-pixel dots in the upper left quadrant. However, for the single-dot features, their contrast in the ghost projection is somewhat susceptible to noise in their neighboring pixels, especially if combined with a case of particularly adverse noise in the pixel itself. As a safer approach, single pixels can be isolated if enclosed in a high contrast band. Alternatively, one could use more basis members to boost SNR and thus reduce the detrimental effect of noise on such fine-scale objects.

\section{Pseudo-correlation Filtered Ghost Projection with Poisson Noise}

\subsection{Pseudo-correlation Filtered Ghost Projection with Poisson Noise} \label{subsec:Pseudo-correlation Poisson Noise Ghost Projection Scheme}

Up to this point, we have considered our desired image in terms of a transmission coefficient between $[-1,1]$, which can be multiplied by the appropriate constant to obtain the units of interest. In this section, we shift perspective to photon counts, and model the associated shot noise with a Poisson distribution \cite{mandel1995optical,blanter2000shot}. For every $\lambda$ photons that illuminate the entrance surface of each pixel prior to attenuation by the masks, $I_{ij}\lambda$ photons are expected to contribute to the projection contrast. With these definitions in place, we  write a pseudo-correlation filtered ghost projection scheme with Poisson noise as:
\begin{align}
P_{ij} = J^{k} \hat{P} \left( \frac{\lambda R'_{ijk}}{\gamma N'} \right), \label{eq:Poisson-noise Pseudo-correlation GP}
\end{align}
where $\hat{P}(X)$ indicates that $X$ is Poisson distributed, $\lambda$ is the number of photons incident on each pixel, $R'$ denotes the pseudo-correlation filtered random-matrix basis, $\gamma$ is the scaling factor that ensures we converge to $I_{ij}$ and $N'$ is the number of filtered basis members. Taking the expected value of this scheme, we have:
\begin{align} \nonumber
\text{E}[P_{ij}] &= \text{E} \left[ J^{k} \hat{P} \left( \frac{\lambda R'_{ijk}}{\gamma N' }  \right) \right] \\ \nonumber
&= N' \text{E} \left[ \hat{P} \left( \frac{\lambda R'_{ijk}}{\gamma N' }  \right) \right] \\ \nonumber
&=  \frac{\lambda \text{E}[ R'_{ij}]}{\gamma}   \\
&= \lambda I_{ij}  +  \frac{\lambda  \text{E}[R]}{\gamma} J_{ij}. \label{eq: Expected Value of pseudo-correlation poisson noise GP}
\end{align}
This projection scheme obtains a uniform offset of $\lambda  \text{E}[R]/\gamma$ photons, and a spatial distribution of $ \lambda I_{ij}$ photons which we will call our desired image expressed in units of photon counts. Moving onto the variance that we would expect in such a scheme, we calculate this via:
\begin{align}
\text{Var}&[P_{ij}] = \text{Var} \left[ J^{k}  \hat{P} \left( \frac{\lambda R'_{ijk}}{\gamma N'} \right) \right] \\ \nonumber
&= N' \text{Var} \left[ \hat{P} \left( \frac{\lambda R'_{ijk}}{\gamma N'} \right) \right] \\ \nonumber
&= N' \text{E}\left[ \hat{P} \left( \frac{\lambda R'_{ijk}}{\gamma N'} \right)^2 \right] - N' \text{E}\left[ \hat{P} \left( \frac{\lambda R'_{ijk}}{\gamma N'} \right) \right]^2.
\end{align}
Focusing on the left term of the final line of the above equation, we can expand this out explicitly. Letting 
\begin{equation}
X(R') = \hat{P} \left( \lambda R'_{ijk}/( \gamma N')  \right)
\end{equation}
be the random variable of interest, this will have the probability distribution $\hat{P}(X(R')) P(R'_{ijk})$:
\begin{align}
\nonumber \text{E}\left[ \hat{P} \left( R'_{ij} \rho \right)^2 \right]  &= \int \sum_{X=0}^{\infty} X^2 \hat{P}(X(R')) P(R'_{ij}) dR'_{ij} \\ \nonumber
&= \int \left( \left( \frac{\lambda R'_{ijk}}{\gamma N'} \right)^2 + \frac{\lambda R'_{ijk}}{\gamma N'} \right) P(R'_{ij}) dR'_{ij} \\
&=  \left( \frac{\lambda }{\gamma N'} \right)^2 \text{E}[{R'}^{2}_{ij}] +  \left( \frac{\lambda }{\gamma N'} \right) \text{E}[{R'}_{ij}].
\end{align}
Substituting this result into our variance calculation and using the fact that $\text{E}[\hat{P}(Y)] = Y$, we can write:
\begin{align} \nonumber
\text{Var}[P_{ij}] &= \frac{\lambda^2}{\gamma^2 N'} \text{E}[{R'}^{2}_{ij}] +  \frac{\lambda }{\gamma}\text{E}[{R'}_{ij}] - \frac{\lambda^2}{\gamma^2 N'} \text{E}[ R'_{ij}]^2 \\ \nonumber
&= \frac{\lambda^2}{\gamma^2 N'} \text{Var}[{R'}_{ij}] + \frac{\lambda }{\gamma}\text{E}[{R'}_{ij}]  \\
&\approx  \frac{\lambda }{\gamma} \left( \frac{\lambda}{\gamma N'} \text{Var}[R] J_{ij} + \text{E}[{R'}_{ij}]  \right).
\end{align}
Above, we have used the approximation that $\text{Var}[R'_{ij}] \approx \text{Var}[R] J_{ij}$ in most cases of filtration. Further, in the case that $\gamma$ is small relative to $\text{E}[R]$ (i.e.~the average transmission value of the random-matrix basis is much larger than the average norm-adjusted correlation value, $\text{E}[R] \gg \gamma $, which implies $\text{E}[{R'}_{ij}] = \text{E}[R] J_{ij} + \gamma I_{ij} \approx \text{E}[R] J_{ij}$), the variance of a pseudo-correlation filtered ghost projection scheme with Poisson noise will be dominated by the constant:
\begin{align}
\text{Var}[P_{ij}] &\approx  \frac{\lambda }{\gamma} \left( \frac{\lambda}{\gamma N'} \text{Var}[R] + \text{E}[R]  \right) J_{ij}. \label{eq: Variance of pseudo-correlation poisson noise GP}
\end{align}

\subsection{Pseudo-correlation Filtered Ghost Projection with Poisson Noise SNR}

Again adapting our previously defined pixel-wise SNR$_{ij}$, we can examine this metric in the case of pseudo-correlation coefficient filtered ghost projection in the presence of Poisson noise:
\begin{align} \nonumber
\text{SNR}_{ij} 
&\equiv \frac{\text{E}[P_{ij}] - \frac{ \lambda \text{E}[R]}{\gamma} J_{ij}}{\sqrt{\text{Var}[P_{ij}]}} \\
&\approx \frac{ \sqrt{ \lambda \gamma }  I_{ij} }{\sqrt{ \frac{\lambda}{\gamma N'} \text{Var}[R] + \text{E}[R]}}.
\end{align}
The global SNR is the RMS of the pixel-wise SNR:
\begin{align}
\nonumber
\text{SNR} 
& \equiv \sqrt{\frac{1}{nm} J^{ij} \text{SNR}^{2}_{\ ij}} \\
&\approx \sqrt{ \frac{ N' \lambda \gamma^2  \text{E}[I^2] }{ \lambda \text{Var}[R] + \gamma N' \text{E}[R]}}. \label{eq:SNR for pseudo-correlation w/ Poisson}
\end{align}
Observe, in the denominator, the relative contributions due to (i) the number of photons allocated to each pixel and (ii) the number of filtered basis members. For a fixed number of photons allocated to each pixel, there is an asymptotic limit for the SNR, even if we had infinitely many basis members. That is, for $N' \rightarrow \infty$, we would still only expect an SNR of:
\begin{align}
\text{SNR} & \sim \sqrt{ \frac{ \lambda \gamma \text{E}[I^2] }{ \text{E}[R]}}.
\end{align}
Similarly, in the limit of an arbitrarily large number of photon counts, for a fixed number of basis members, we asymptote to the SNR result of the Poisson-noise-free case (cf.~the second line of Eq.~(\ref{eq: now I want this one})). 

\subsection{Optimum Pseudo-correlation Cut-off with Poisson Noise}

We extend our previous optimization of the pseudo-correlation cut-off value to achieve the maximum SNR as a function of (i) the number of photons per pixel and (ii) the number of basis members available to realize the ghost projection. In determining the optimum pseudo-correlation cut-off value that maximizes SNR in the presence of Poisson noise, we can prescribe a statistical representation to the number of filtered random-matrix basis members, i.e.~$N' =fN$, and the norm-adjusted expected pseudo-correlation of the filtered basis:
\begin{equation}
\gamma = \sqrt{\text{Var}[R]/(2\pi nm \text{E}[I^2])}\exp (-X^2)/f(X), 
\end{equation}
where 
\begin{equation*}
X = \frac{C_{\text{min}} - \text{E}[C]}{\sqrt{2 \text{Var}[C]}} \quad \textrm{and} \quad 
f(X) = \tfrac{1}{2}\text{erfc}(X). 
\end{equation*}
Further, we can seek to maximize the square of the SNR to remove the square-root from consideration. Making these substitutions and changes, we have:
\begin{align}
\nonumber
\text{SNR}^2 &\approx  \frac{ N' \lambda \gamma^2  \text{E}[I^2] }{ \lambda \text{Var}[R] + \gamma N' \text{E}[R]} \\
&\approx \frac{ N }{ 2\pi nm \left(  \exp (2X^2) + a \exp (X^2)  \right) f(X)}.  \label{eq:SNR for pseudo-correlation w/ Poisson and stat. perspective}
\end{align}
Above, we have defined a new parameter
\begin{equation}
a \equiv \frac{N}{\lambda}  \frac{\text{E}[R]}{ \sqrt{ 2\pi nm \text{Var}[R] \text{E}[I^2]}},
\end{equation}
to denote the scaled ratio of (i) the number of random-matrix basis members $N$, to (ii) the number of photons per pixel $\lambda$. Differentiating Eq.~(\ref{eq:SNR for pseudo-correlation w/ Poisson and stat. perspective}) with respect to $X$ and setting it to zero, we obtain:
\begin{align}
 -\sqrt{\pi }  X \left(a+2 e^{X^2}\right) \text{erfc}(X)+a e^{-X^2} + 1 = 0. \label{eq:Pikacu, I choose you}
\end{align}

\begin{figure*}[ht!]
     \centering
     \begin{subfigure}{0.325\textwidth}
         \centering
         \includegraphics[width=\textwidth]{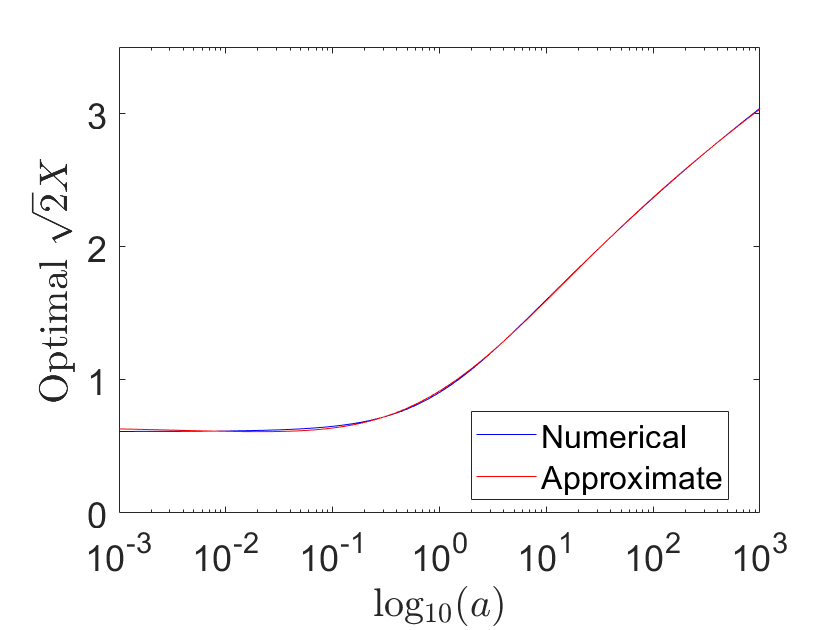}
         \caption{ }
         \label{subfig:OptX1}
     \end{subfigure}
     \begin{subfigure}{0.35\textwidth}
         \centering
         \includegraphics[width=\textwidth]{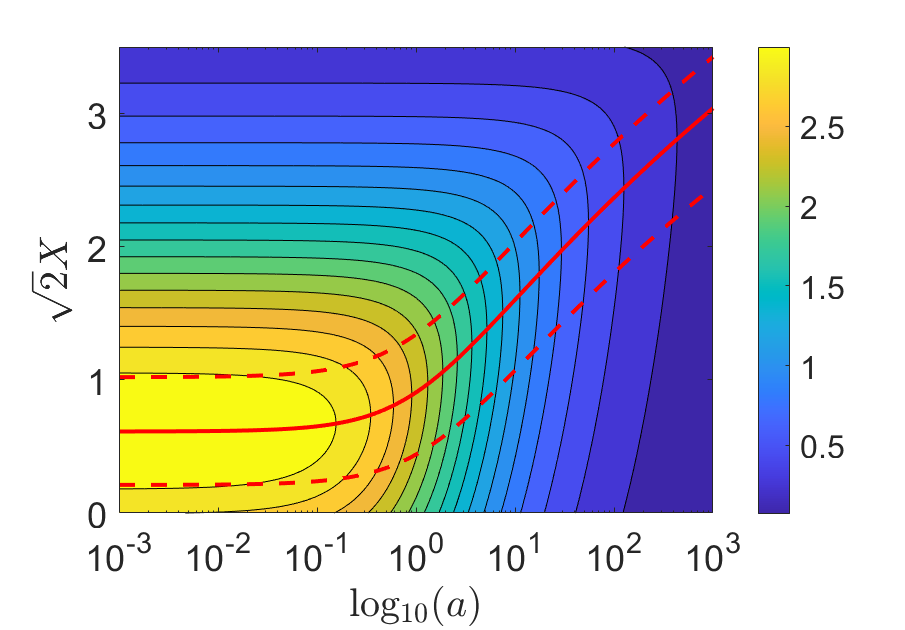}
         \caption{ }
         \label{subfig:ContourOptX2}
     \end{subfigure}
    	 \caption{(a) Plot of the number of standard deviations for the optimal pseudo-correlation cut-off $C_{\text{min}} = \text{E}[C] + \sqrt{2}X \sqrt{\text{Var}[C]}$, as a function of the scaled ratio of the number of basis members to the number of photons per pixel.  The blue curve is obtained from a numerical solution to Eq.~(\ref{eq:Pikacu, I choose you}), with the red curve given by the approximate analytic expression in Eq.~(\ref{eq:Richu, I choose you}). (b) A contour plot of SNR for a  $N=40,000$ random-matrix basis set, and a resolution of $40\times40$ projected image in the presence of Poisson noise. The red line is the $\sqrt{2}X$ that gives the optimal SNR, and the dashed red lines mark where the SNR falls to 95\% of its optimal value.}
    \label{fig: Optimum pseudo-correlation poisson cut-off}
\end{figure*}

The result of numerically solving this equation for the optimum pseudo-correlation cut-off value in the presence of Poisson noise, over several orders of magnitude of $a$, is shown in Fig.~\ref{subfig:OptX1} (blue curve). We observe two main regimes: a constant section and a linearly increasing section. Separating these two regimes is a transition point where the number of random-matrix basis members is approximately equal to the number of photons per pixel, $O(a) \approx 10^0 $. That is, in the approximately constant section to the left of the transition point, we are in a regime of more photons than basis members, and the optimal pseudo-correlation cut-off point is approximately the same as without Poisson noise. In other words, in this regime, Poisson noise is a relatively insignificant contribution to the noise and instead the dominant contribution comes from the random-matrix basis representation. In the second regime, defined by the approximately linear increase to the right of the transition point, we are limited by the number of photons. For this photon-dilute regime, we can afford to be more selective with the photons we have and seek a higher pseudo-correlation cut-off value. For example, if we were to have approximately 100 more random-matrix basis members than we have photons per pixel ($a \approx 100$), then we can afford to discard many of the basis members and instead allocate those photons to basis members that are at least approximately 2.4 standard deviations above the average pseudo-correlation value. 

We can fit a simple analytical approximation to the numerical solution for Eq.~(\ref{eq:Pikacu, I choose you}). This approximate solution is taken to be a constant $c_1$ that transitions to $c_2 \log_{10}(a) + c_3$  via a sigmoid function $S(\log_{10}(a),c_4,c_5) = 1/[1 + c_4 \exp(c_5 \log_{10}(a) )]$, giving:
\begin{align} \nonumber
    \sqrt{2} & X(a) \approx c_1 S(\log_{10}(a),c_4,c_5) \\
    &+ (c_2 \log_{10}(a) + c_3) S(-\log_{10}(a),c_4,c_5). \label{eq:Richu, I choose you}
\end{align}
Here, the parameters $c_1 = 0.6510$, $c_2 = 0.5310$, $c_3 = 1.5188$, $c_4 = 1.3682$, $c_5 = 1.2847$ were found via a gradient-descent approach. The result of this analytical approximation compared to the numerical solution is given as red and blue curves, respectively, in Fig.~\ref{subfig:OptX1}. For early experimental implementations of ghost projection, we are likely to be in the regime with more photons than basis members (left of the transition point).

For an illustrative example of how these results may be used and interpreted, suppose we have a basis set of 40,000 random matrices with which we wish to project a known $40 \times 40$ image. Deciding upon the number of photons, per pixel, that we wish to allocate to create the contrast $\lambda I_{ij}$, we can then calculate $a$. From this $a$, we can then determine the number of standard deviations above the mean that the optimum pseudo-correlation cut-off value is, via (i) numerically solving Eq.~(\ref{eq:Pikacu, I choose you}), or (ii) using the analytical approximation, Eq.~(\ref{eq:Richu, I choose you}). The SNR we expect from applying that filtering and allocating uniform exposures to those basis members that remain according to Eq.~(\ref{eq:Poisson-noise Pseudo-correlation GP}) is given by Eq.~(\ref{eq:SNR for pseudo-correlation w/ Poisson}), or from a more statistical perspective, Eq.~(\ref{eq:SNR for pseudo-correlation w/ Poisson and stat. perspective}). From the contour plot of the latter equation in Fig.~\ref{subfig:ContourOptX2}, we observe (i) the decline in SNR with increasing $a$, and (ii) the increase in the number of standard deviations the optimal pseudo-correlation cut-off is above the mean pseudo-correlation coefficient.

\subsection{Pseudo-correlation Filtered Ghost Projection with Poisson Noise Simulation}

We conclude this section with a numerical simulation to illustrate some of its key analytical results. In particular, we show that for our simulation the integrated result of uniform exposures of the pseudo-correlation filtered basis set $R_{ijk}'$, subject to Poisson noise, converges as given by Eq.~(\ref{eq: Expected Value of pseudo-correlation poisson noise GP}). Further, we will numerically illustrate that the noise present in this ghost projection, which has contributions from both the random-basis reconstruction and from Poisson noise, is accurately captured by Eq.~(\ref{eq: Variance of pseudo-correlation poisson noise GP}). Finally, we would like to see that the analytically calculated optimum minimum pseudo-correlation coefficient indeed achieves an optimum SNR in simulation, for a given number of photons per pixel $\lambda$ and unfiltered basis members $N$. 

\begin{figure*}[ht!]
     \centering
     \begin{subfigure}{0.329\textwidth}
         \centering
         \includegraphics[width=\textwidth]{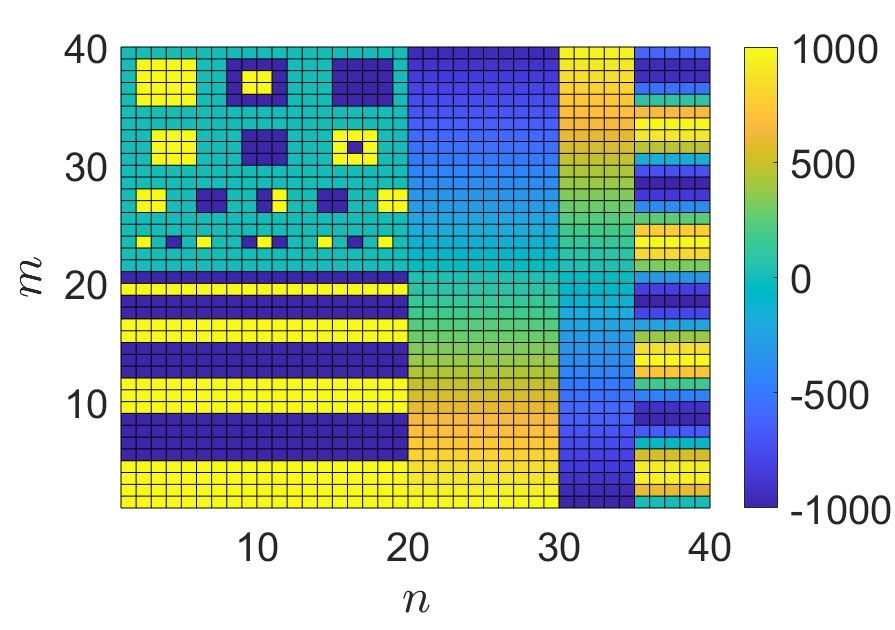}
         \caption{ }
         \label{subfig: GP1 Poisson}
     \end{subfigure}
     \begin{subfigure}{0.329\textwidth}
         \centering
         \includegraphics[width=\textwidth]{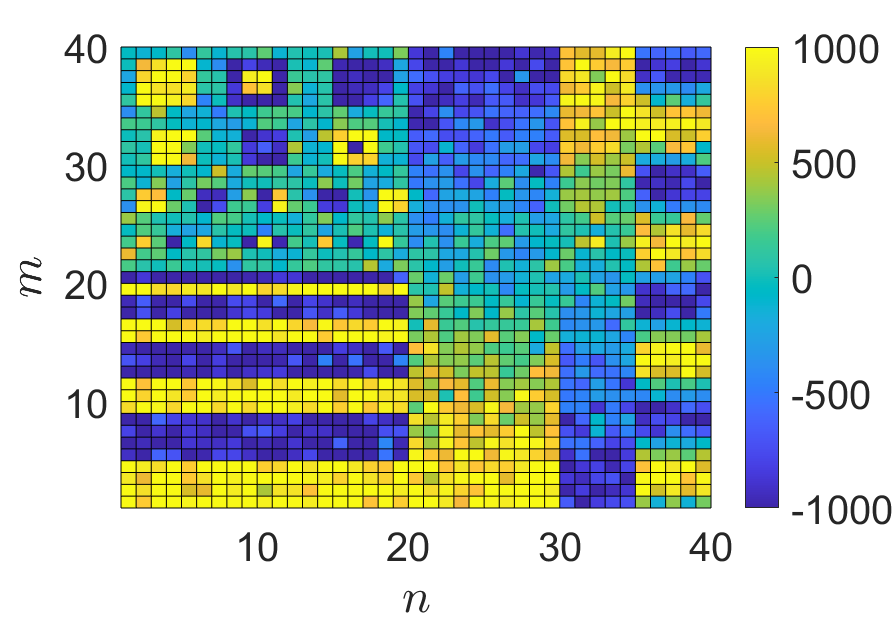}
         \caption{ }
         \label{subfig: GP2 Poisson}
     \end{subfigure}
     \begin{subfigure}{0.329\textwidth}
         \centering
         \includegraphics[width=\textwidth]{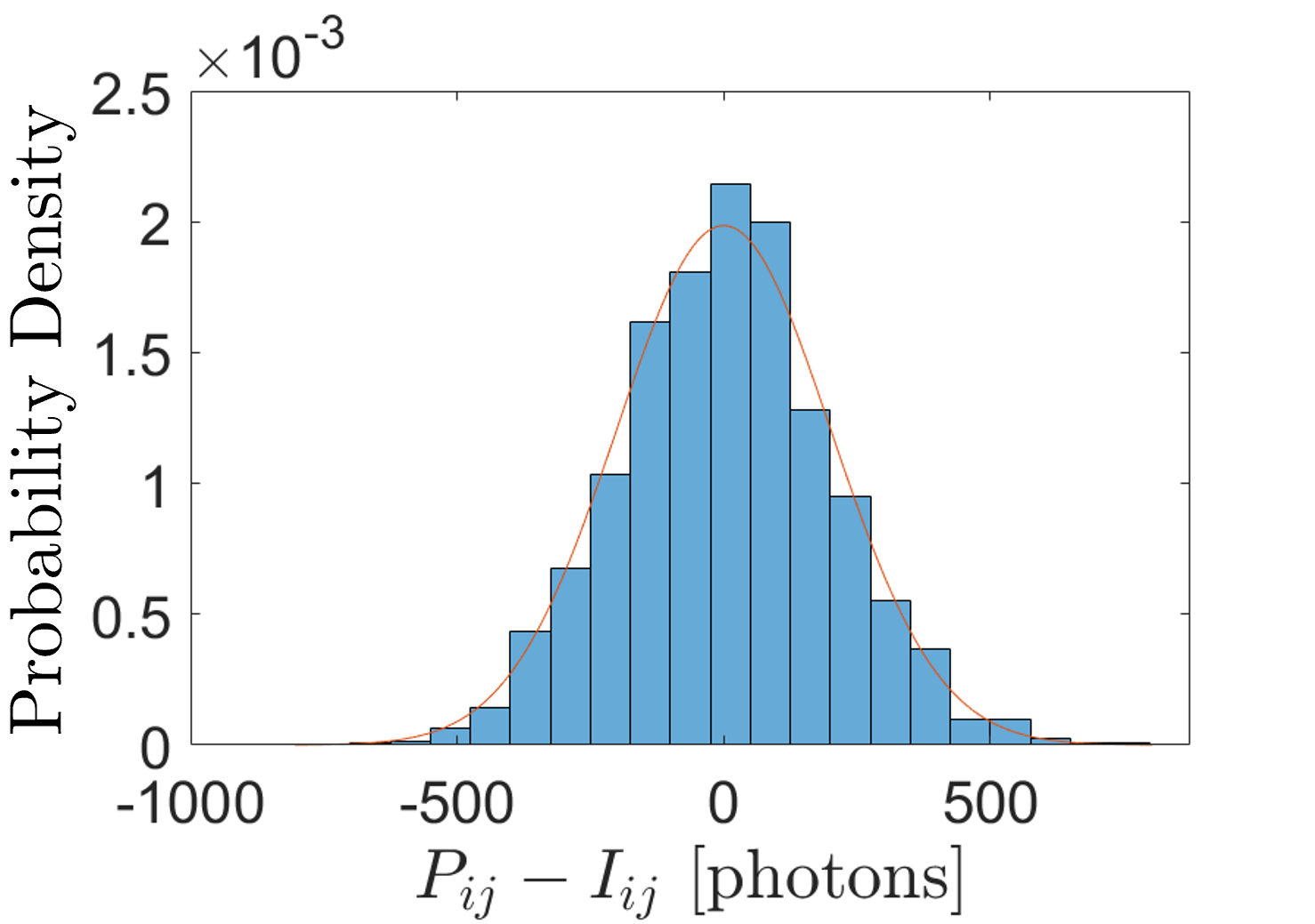}
         \caption{ }
         \label{subfig: GP3 Poisson}
     \end{subfigure}
        \caption{(a) The desired image we wish to project expressed in photon counts, minus the constant off-set of $\lambda \text{E}[R] / \gamma = 18,774$ photons. (b) The optimized pseudo-correlation filtered ghost projection subject to Poisson noise, obtained via Eq.~(\ref{eq:Poisson-noise Pseudo-correlation GP}), minus the expected offset, using $N' = 16,317$ (which was filtered from $N=100,000$ basis members) and achieved an $\text{SNR} = 3.57$ ($\text{SNR} = 3.53$ was predicted). (c) The noise obtained in the pseudo-correlation filtered ghost projection subject to Poisson noise, overlaid with the predicted noise distribution given by Eq.~(\ref{eq:SNR for pseudo-correlation w/ Poisson and stat. perspective}).}
        \label{fig:pseudo-correlation Poisson noise result graphs}
\end{figure*}

Our simulation employs the same test image that was previously used in the pseudo-correlation filtered ghost projection case, namely Fig.~\ref{subfig: GP1 Pseudo}.  When multiplied by $\lambda = 1,000$, we obtain our desired image expressed in terms of photon counts, as seen in Fig.~\ref{subfig: GP1 Poisson}. The random basis is constructed of binary random matrices, $R_{ijk} \in \{0,1\}$, in which the values of zero and one are equally likely. 

Starting with $N=100,000$ basis members, as a baseline, let us in the first instance neglect  Poisson noise considerations and adopt the perspective of ghost projection via pseudo-correlation filtering (Sec.~\ref{sec: Pseudo-correlation Filtered Ghost Projection}). That is, use $\sqrt{2}X = 0.612$ (which is the optimum pseudo-correlation cut-off in the plentiful-photon limit). With this, we filtered out 27,175 basis members. Including Poisson noise in the simulation, we obtained an SNR of 3.44 which can be compared to a predicted SNR of 3.42. Changing the cut-off value to reflect that there are a finite number of photons which display Poisson noise ($a=1.40$), we can update $\sqrt{2}X$ to be 0.985. With these conditions, starting with $N=100,000$ basis members yielded $N' = 16,317$ above the minimum pseudo-correlation cut-off value. Further, the simulation yielded an SNR of 3.57 for which an SNR of 3.53 was predicted, as shown in Fig.~\ref{fig:pseudo-correlation Poisson noise result graphs}. This amounts to a 3.2\% improvement in SNR having accounted for the Poisson noise in this case -- although, we are still in a regime with an $a$ value on the order of unity, near the transition point. Should we reduce the number of photons and keep the same number of basis members, or increase the number of basis members for a fixed number of photons (i.e.~increase $a$), then we would expect to see a greater level of relative improvement in SNR, having accounted for using a finite number of photons exhibiting Poisson noise.

% These figures were generated with the MATLAB code called GP_PseudoPoisson.m

\section{Color Ghost Projection} \label{sec:Colour Ghost Projection}

Motivated by color ghost imaging \cite{colourGI}, suppose we have a random-matrix basis that is comprised of $c$ color channels.  This could be achieved, for example, by illuminating a thin or thick spatially-random screen using white light emanating from a sufficiently-small source in order to create correlated or uncorrelated superimposed speckle fields, respectively, for a variety of different energy bands \cite{FinkSpeckles2012}. Employing this means for generating speckle fields across independent energy channels, one could aspire to project a color image. That is, for the sake of argument, we could construct the visible-light three-channel color ghost projection scheme:
\begin{align}
P_{ijc} = t_{k} R_{ij \ c}^{\ \ k} = t_{k}  \left( R_{ij \ r}^{\ \ k} , R_{ij \ g}^{\ \ k},  R_{ij\ b}^{\ \ k} \right),
\label{eq:ThreeColourChannelsA}
\end{align}
where $c$ indexes the color channel and $r,g,b$ denote the red, green and blue (RGB) random-matrix color channels that make up the random-matrix color mask. Here, we would like to construct the color ghost-projection image:
\begin{equation}
I_{ijc} = (I_{ijr},I_{ijg},I_{ijb}). 
\label{eq:ThreeColourChannelsB}
\end{equation}

If we could isolate each illumination color channel, we could simply perform three independent cases of monochromatic ghost projection, for which all of the results of the preceding sections are applicable. Supposing, for the sake of time or due to inherent constraints of the illumination source, we illuminate the color masks all at once, then how do we choose a global $t_k$ such that we achieve our desired color image without erroneous dilation of the color palette? Furthermore, there is the question of how we filter our color random-matrix basis. Should we pick only those members that are positively correlated in all color channels?  Assuming the correlation of each of three color channels is independent, we would only expect $(\tfrac{1}{2})^3=\tfrac{1}{8}$ of the overall RGB color basis to be positively correlated with the desired color ghost projection. Perhaps, for the sake of including more basis members, it would be worth including a basis member that is only slightly negatively-correlated with one channel of the required color image, if it is highly correlated in another channel. Such considerations would be captured by a global color pseudo-correlation coefficient. Indeed, even for the case of independently-filtered color channels, when we sum those basis members, they will converge to the color image with a scaling determined by the global color pseudo-correlation value too. 

Borrowing the ghost projection scheme from the pseudo-correlation filtered case, which uses uniform dwell times, we can construct the color ghost projection:
\begin{align} \label{eq:ColourProjection}
P_{ijc} = \frac{J_k R_{ij \ c}^{\ \ k} }{N' \gamma}. 
\end{align}
Here, $\gamma$ is the appropriately normalized, expected correlation of the color basis and $N'$ is the number of basis members that make it through the filtering process. This has advanced us towards answering the question of what global $t_k$ is appropriate to achieve an un-dilated color palette, although we still need to determine the functional form of $\gamma$ and decide between an independently or globally filtered color basis for $N'$.

\subsection{Color Pseudo-correlation Coefficient}

Consider the global color pseudo-correlation coefficient:
\begin{align}
C_{k} = \frac{R_{ijkc}I^{ijc}}{ 3 n m \sqrt{ \text{E}[R^2]\text{E}[I^2]}},
\end{align}
where we have again used three color channels (although this is an arbitrary choice and may just as easily be any number of color channels) and, in this case, $\text{E}[R^2] = R_{ijkc} R^{ijkc} /(nmN3)$ and $\text{E}[I^2] = I_{ijc}I^{ijc}/(nm3)$. This coefficient will have the expected value:
\begin{align} \nonumber
\text{E}[C] &= \text{E} \left[  \frac{R_{ijkc}I^{ijc}}{ 3 n m \sqrt{ \text{E}[R^2]\text{E}[I^2]}} \right] \\ \nonumber
&= \frac{ \text{E}[R] J_{ijc} I^{ijc}}{ 3 n m \sqrt{ \text{E}[R^2]\text{E}[I^2]}} \\
&= \frac{ \text{E}[R] \text{E}[I]}{\sqrt{ \text{E}[R^2]\text{E}[I^2]}}.
\end{align}
Further, it will have the variance: 
\begin{align} \nonumber
\text{Var}[C] 
&= \text{Var}\left[ \frac{R_{ijkc} I^{ijc}}{nm3 \sqrt{\text{E}[R^2] \text{E}[I^2]}} \right]\\ \nonumber
&= \frac{\text{Var} [R] J^{ijc} I^{2}_{ \ ijc}}{(3nm)^2 \text{E}[R^2] \text{E}[I^2]} \\
&= \frac{\text{Var}[R]}{ 3 nm \ \text{E}[R^2]}.
\end{align} 
These results are very similar to the pseudo-correlation coefficient results for the non-color case in Sec.~\ref{subsec:Pseudo-correlation Coefficient}, with the only difference being to update the definitions of the expectation results (i.e.~$\text{E}[I]$, $\text{E}[I^2]$, $\text{E}[R]$, $\text{E}[R^2]$) and reduce the variance of the pseudo-correlation value by dividing by the number of color channels. 

From here we can substitute these updated definitions into our definition of the appropriately normalized, expected correlation of the color basis $\gamma$, i.e. 
\begin{equation}
\gamma = (\text{E}[C'] - \text{E}[C]) \sqrt{\frac{ \text{E}[R^2]}{ \text{E}[I^2] }}, 
\end{equation}
where $\text{E}[C']$ is the average global color pseudo-correlation coefficient of the filtered basis, $\text{E}[C]$ is the zero reference point of the global color pseudo-correlation coefficient and $\sqrt{ \text{E}[R^2] / \text{E}[I^2]}$ is a norm-adjustment factor along the direction $I_{ijc}$.

\subsection{Independent versus Global Filtration of the Color Channels}

Consider the two options of (i) independently filtering the color basis based upon the pseudo-correlation coefficient of each color channel, or (ii) globally filtering the color basis based upon the global color pseudo-correlation coefficient. For each case, we will have a different number of basis elements $N'$ that remain, post filtration. Moreover, the average global color pseudo-correlation coefficient post filtration will be different in each case. To decide between these two options, we can compare their predicted SNR results.  

If we were to employ the independently-filtered method and only include those basis members that have a greater-than-average pseudo-correlation coefficient in all of their respective channels, then we have already pointed out that we can expect $\tfrac{1}{8}=12.5\%$ of the original basis set to remain post-filtration, for an RGB color arrangement. Allowing for a global filtration, we can borrow from Sec.~\ref{subsec:Pseudo-correlation Cut-off Optimum} and expect 27\% of the original basis set to remain post-filtration. 

We now move onto concerns regarding the expected global color pseudo-correlation of the filtered basis, namely $\text{E}[C'] - \text{E}[C]$, in each case. For the independently-filtered case, we can infer an effective correlation based upon the result that $f = 0.125 = (1/2)\text{erfc}(X)$, where $X = (C_{\text{min}} - \text{E}[C])/\sqrt{2\text{Var}[C]}$, which implies $X\approx 0.8134$. For the globally-filtered case, we have the previously obtained result $X \approx 0.433$. Substituting in these estimates of $X$ into the expression for the expected global color pseudo-correlation value of the filtered basis, we obtain:
\begin{align} \nonumber
\text{E}[C']-  \text{E}[C] &= \sqrt{\frac{2 \text{Var}[C]}{\pi}}  \left[  \text{erfc} \left( X \right) \right]^{-1} \exp \left( - X^2 \right) \\
&= \sqrt{ \frac{2 \text{Var}[R]}{ 3 nm \pi \text{E}[R^2]} }  \left[  \text{erfc} \left( X \right) \right]^{-1} \exp \left( - X^2 \right),
\end{align}
which, for the sake of comparison, if we define $\text{Var}[R] = 1/12$ and $\text{E}[R^2] = 1/3$, gives the expected global color pseudo-correlation for the independently-filtered case as 0.823/$\sqrt{3nm}$ and the globally-filtered case as 0.612/$\sqrt{3nm}$. Note, for the independently filtered case, this assumes those members that are selected belong to the upper 12.5\% of the global members in terms of color pseudo-correlation coefficient, which is not necessarily the case. That is, just because we expect 12.5\% of basis members to be positively correlated in each channel, it does not mean that they necessarily belong to the 12.5\% most-globally-correlated members---this estimate would form a best-case scenario for the independently filtered case and we may achieve less than this in practice. 

Defining a color SNR as the RMS of the pixels and channels:
\begin{align}
\text{SNR} = \sqrt{\frac{1}{3nm} J^{ijc} \text{SNR}_{ijc}^{2}},
\end{align}
and making the substitution for each color channel that
\begin{equation}
\text{SNR}^2_{ijc} \approx \frac{ I^2_{ijc} \gamma^2 N'}{\text{Var}\left[ R \right]}
\end{equation}
where the approximation arises owing to $\text{Var}[R'_{ij}] \approx \text{Var}[R] $, we obtain:
\begin{align} \nonumber
\text{SNR} &\approx \sqrt{ \frac{ \text{E}[I^2] \gamma^2 N'}{\text{Var}[R]}} \\
&\approx \sqrt{ \frac{ \text{E}[R^2] N'}{\text{Var}[R]}}  (\text{E}[C'] - \text{E}[C]).\label{eq:ColourGhostProjectionSNR}
\end{align}
Substituting in our results for the number of filtered basis members and expected color pseudo-correlation coefficient for each case yields an SNR-uncertainty relationship prediction of $\text{SNR}^2 \times 3nm \approx 0.34N $ and  $\text{SNR}^2 \times 3nm \approx 0.40 N $ for the independently and globally filtered basis set, respectively, where we have again used $\text{Var}[R] = \tfrac{1}{12}$ and $\text{E}[R^2] = \tfrac{1}{3}$ for the sake of comparison. Based upon these results, we would expect the SNR of the globally filtered basis set to be about 9\% better than the independently filtered set. 

\subsection{Color Ghost Projection Simulation}

\begin{figure*}[ht!]
    \centering
    \begin{subfigure}{0.35\textwidth}
         \centering
         \includegraphics[width=0.8\textwidth, clip,trim={41cm 26.5cm 42cm 10cm}]{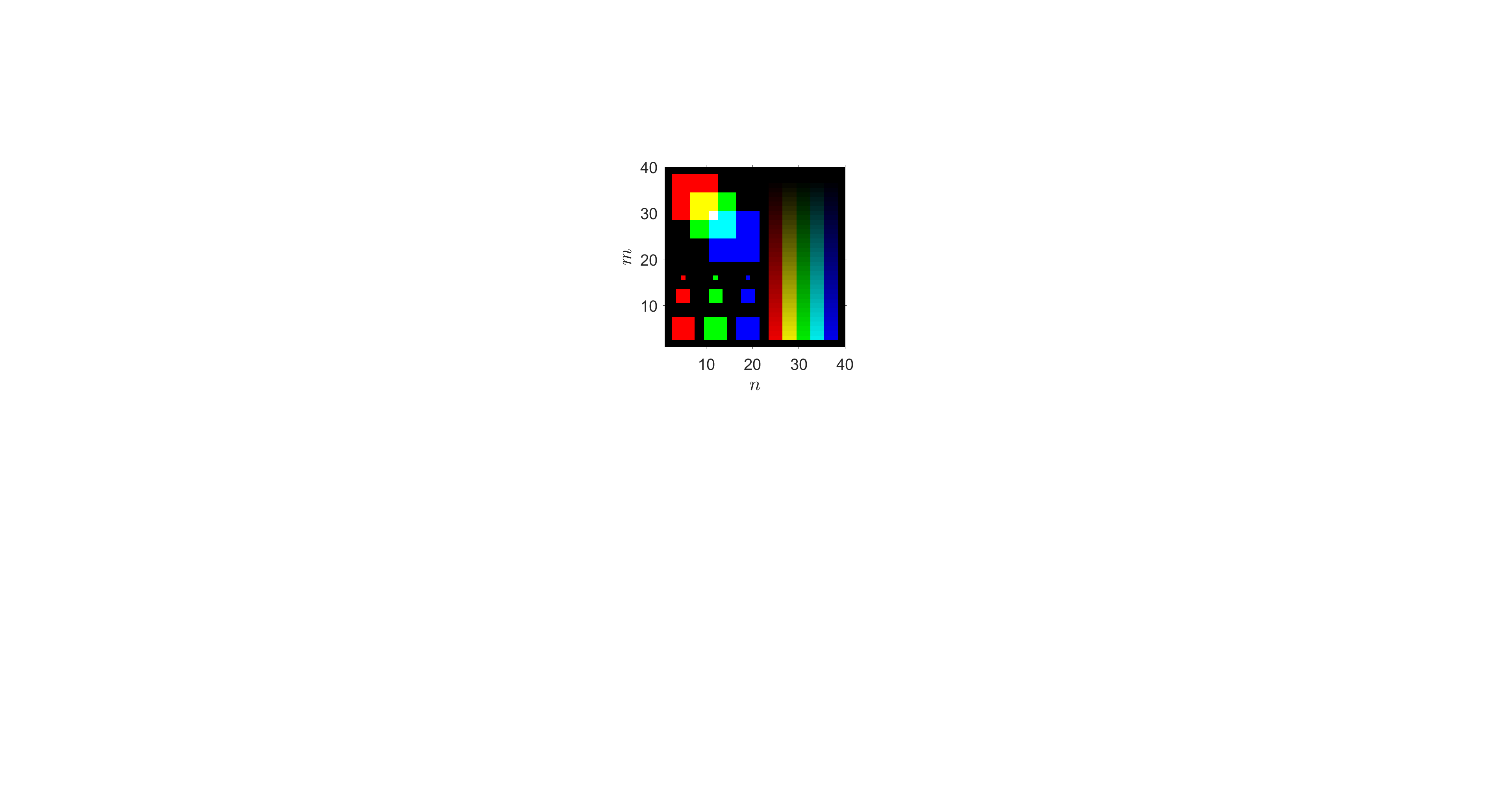}
         \caption{ }
         \label{subfig:CGP0}
     \end{subfigure}
     \begin{subfigure}{0.35\textwidth}
         \centering
         \includegraphics[width=0.8\textwidth, clip,trim={41cm 26.5cm 42cm 10cm}]{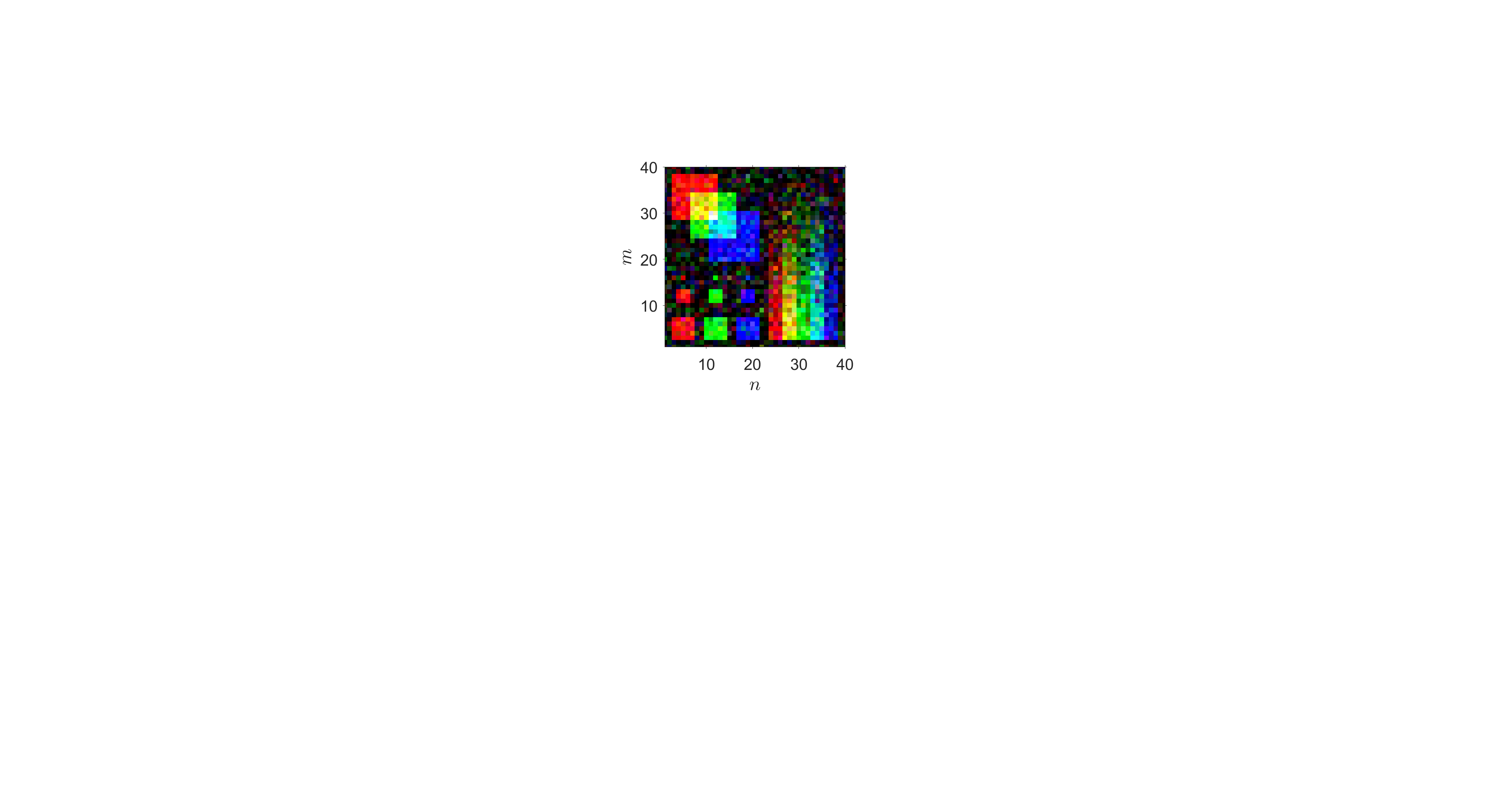}
         \caption{ }
         \label{subfig:CGP1}
     \end{subfigure}
     \begin{subfigure}{0.35\textwidth}
         \centering
         \includegraphics[width=0.8\textwidth, clip,trim={41cm 26.5cm 42cm 10cm}]{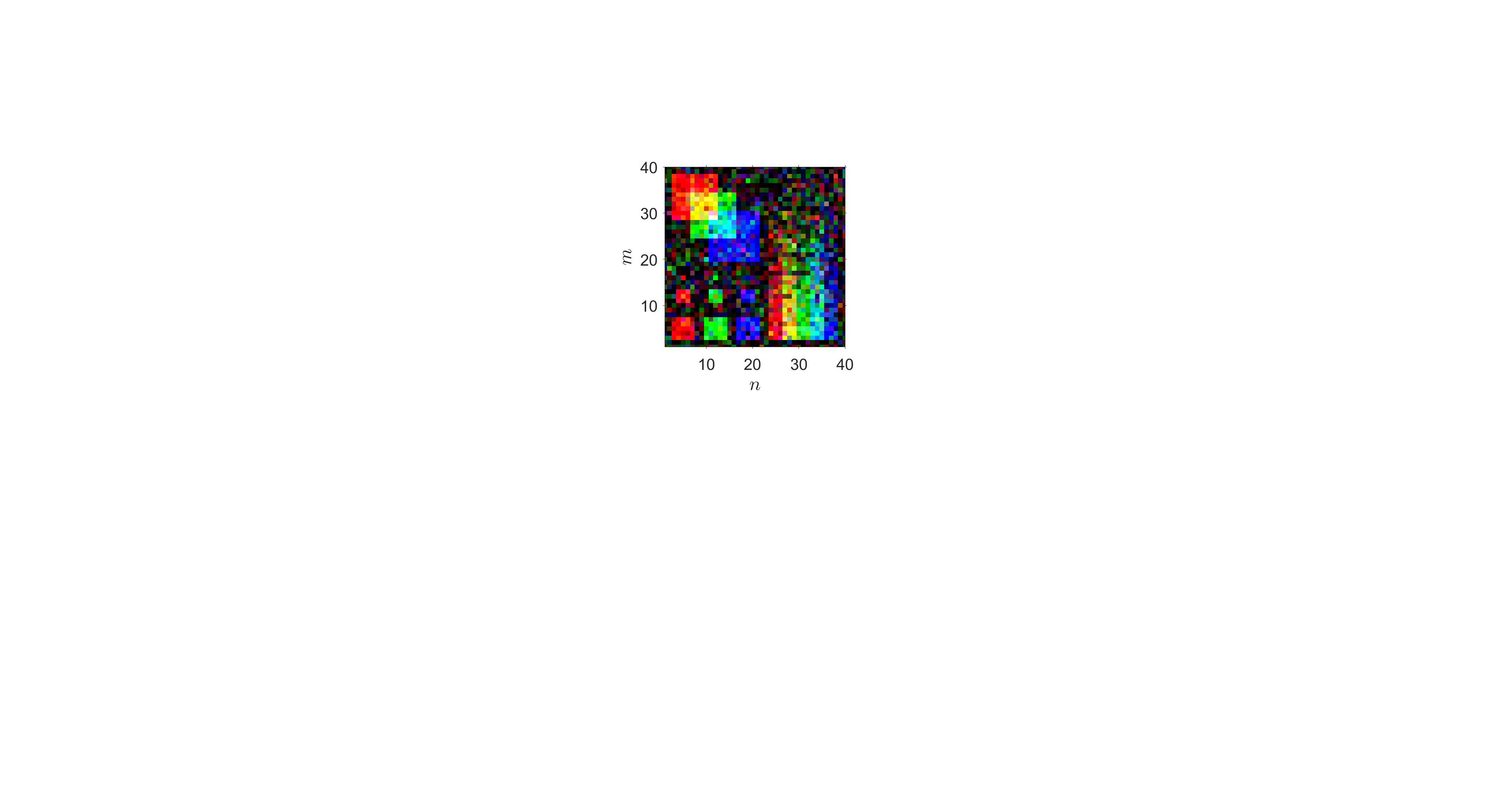}
         \caption{ }
         \label{subfig:CGP2}
     \end{subfigure}
     \begin{subfigure}{0.35\textwidth}
         \centering
         \includegraphics[width=0.9\textwidth]{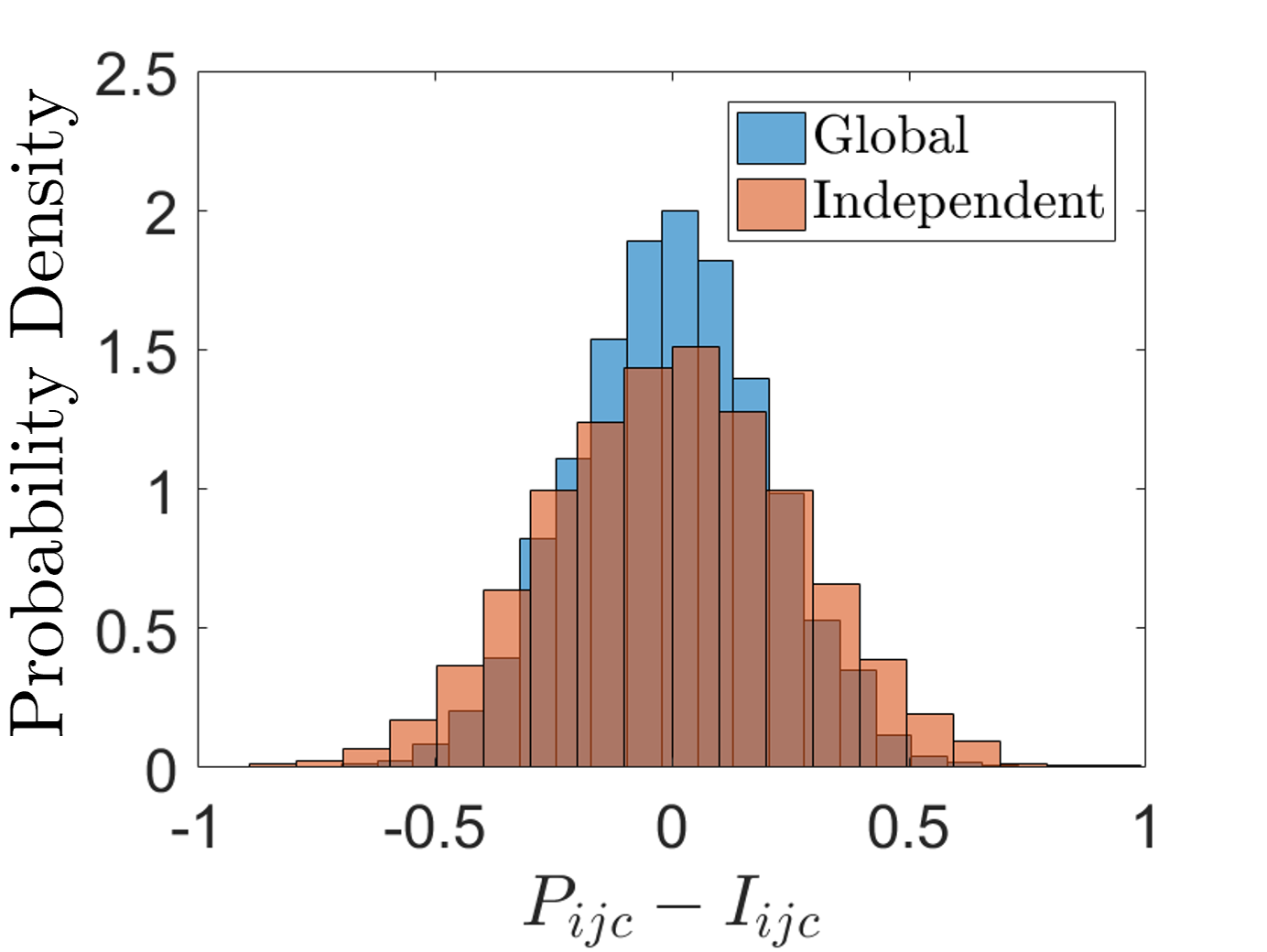}
         \caption{ }
         \label{subfig:CGP3}
     \end{subfigure}
        \caption{ (a) Plot of the desired color image we wish to project, expressed as an attenuation coefficient. There are three independent color channels $c$: red $r$, green $g$ and blue $b$ (see Eqs.~(\ref{eq:ThreeColourChannelsA}) and (\ref{eq:ThreeColourChannelsB})). (b) Globally-filtered color ghost projection with the constant offset of 35.77 subtracted, using $N' = 10$,680 filtered from $N=4$0,000 which achieved an SNR of 1.82. (c) Independently-filtered color ghost projection with the constant offset of 31.77 subtracted, using $N' = 4$,905 filtered from $N=4$0,000 which achieved an SNR of 1.38. (d) The random-basis reconstruction noise obtained in the globally-filtered and independently-filtered color ghost projections, with $P_{ijc}$ given by Eq.~(\ref{eq:ThreeColourChannelsA}) and $I_{ijc}$ given by Eq.~(\ref{eq:ThreeColourChannelsB}). }
        \label{fig:Colour Ghost Projection}
\end{figure*}

Consider the $40 \times 40$ resolution, test color image in Fig.~\ref{subfig:CGP0}. We wish to color ghost project this using an independently and globally filtered color basis of size $N=40,000$. The color basis consists of three channels that are each populated with uniformly random values between 0 and 1. The results are given in Fig.~\ref{fig:Colour Ghost Projection}b,c,d.

% These plots were made with the MATLAB code GP_Colour.m

In simulation, the global filtration selected 10,680 basis members of the 40,000, where 10,800 was expected (i.e.~27\%). For the independently-filtered basis set, 4,905 members were selected where 5,000 was expected (i.e.~12.5\%). The predicted SNR of the globally filtered color ghost projection was 1.82 (see Eq.~(\ref{eq:ColourGhostProjectionSNR})) and the simulated SNR was also 1.82. For the independently filtered color ghost projection, if we use Eq.~(\ref{eq:ColourGhostProjectionSNR}) with the numerically calculated $(\text{E}[C'] - \text{E}[C])$, then the SNR was predicted to be 1.39 (which is lower than that predicted as our best-case scenario of 1.68, where the independently filtered basis members were assumed to belong to the top 12.5\% pseudo-correlated values). The independently filtered color ghost projection obtained an SNR of 1.38 in simulation (which is to suggest Eq.~(\ref{eq:ColourGhostProjectionSNR}) is reasonably accurate, although our best-case assumption was askew). In these simulation results, the globally-filtered color ghost projection scheme did 32\% better than the independently-filtered scheme.

%Predicted Globally Filtered SNR 1.821536 
%Simulated Globally Filtered SNR 1.821598 
%Predicted Independently Filtered SNR 1.389823 
%Simulated Independently Filtered SNR 1.378372 
%Global is 32.155777 percent better than independent 

\section{Numerically Optimized Ghost Projection}

In the analytical approaches adopted thus far, we have considered weighting the basis, and filtering the basis. A natural progression is to consider weighting and filtering the basis at the same time. This is discussed in Appendix \ref{AppB:Pseudo-correlation Filtered Ghost Projection with Linear Weights}, which suggests only modest improvements that are offset by a significant increase in analytical complexity. Instead of taking this approach, we here investigate numerical optimization as a more profitable line of inquiry. 

\subsection{Ghost Projection with Numerically Derived Weights}

Given that we know {\em a priori} both the realized random-basis set and the particular desired image to be ghost projected, we now explore numerical optimization of the random-matrix exposures.  This illustrates the significant improvement in ghost-projection SNR that is possible by using numerically-optimized weights, rather than the analytical weights of the preceding sections.  The analytical schemes of the preceding sections are more physically and conceptually transparent but less efficient.  Conversely, the numerical-optimization schemes of the present section are less conceptually transparent but more efficient. 

For simplicity, to explore ghost projection with numerically derived weights, we restrict ourselves to the zero-averaged case of the desired image, where  $\text{E}[I] = 0$. We define the ghost projection scheme:
\begin{align} 
P_{ij} \equiv  R_{ij}^{\ \ k} w_k \rightarrow I_{ij} + \bar{P}, 
\end{align}
where $R_{ij}^{\ \ k}$ is our particular random-matrix basis, $w_{k}$ are weights to be numerically optimized such that the above sum approaches the image, plus an offset that is equal to the average $\bar{P} = J^{ij} P_{ij} /(nm) = N' \text{E}[w] \text{E}[R]$. We can render this in a common form by vectorizing our random-matrix basis and setting each member equal to the columns of the matrix $M$. Further, we absorb $\bar{P}$ into the left-hand side by subtracting the average of each column:
\begin{equation}
M = [R_{ij1} - \overline{R_{ij1}};R_{ij2}- \overline{R_{ij2}};\cdots ;R_{ijN}- \overline{R_{ijN}}]. 
\end{equation}
With this, we can now express our ghost projection as:
\begin{align}
M \vec{w} \rightarrow \vec{I},\label{eq:Numerical Ghost Projection}
\end{align}
where $\vec{I}$ is the vectorized version of our desired image. With this definition of ghost projection via numerically derived weights, the function to be optimized is:
\begin{align}
\text{SNR} = \sqrt{\frac{\text{E}[I^2]}{\text{Var}[M\vec{w} - \vec{I}]}}, \label{eq:Numerical Ghost Projection SNR}
\end{align}
namely the RMS of a pixel-wise SNR$_{ij}$.    

\subsection{Ghost Projection with Numerically Derived Weights Simulation}

\begin{figure*}[ht!]
     \centering
     \begin{subfigure}{0.329\textwidth}
         \centering
         \includegraphics[width=\textwidth]{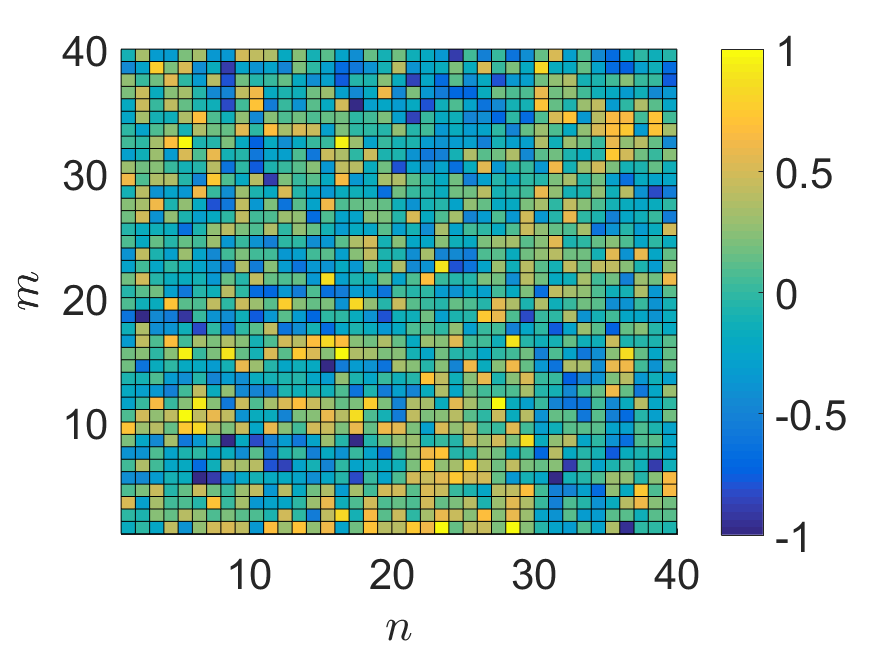}
         \caption{ }
         \label{subfig:GP1 Numerical NNLS}
     \end{subfigure}
     \begin{subfigure}{0.329\textwidth}
         \centering
         \includegraphics[width=\textwidth]{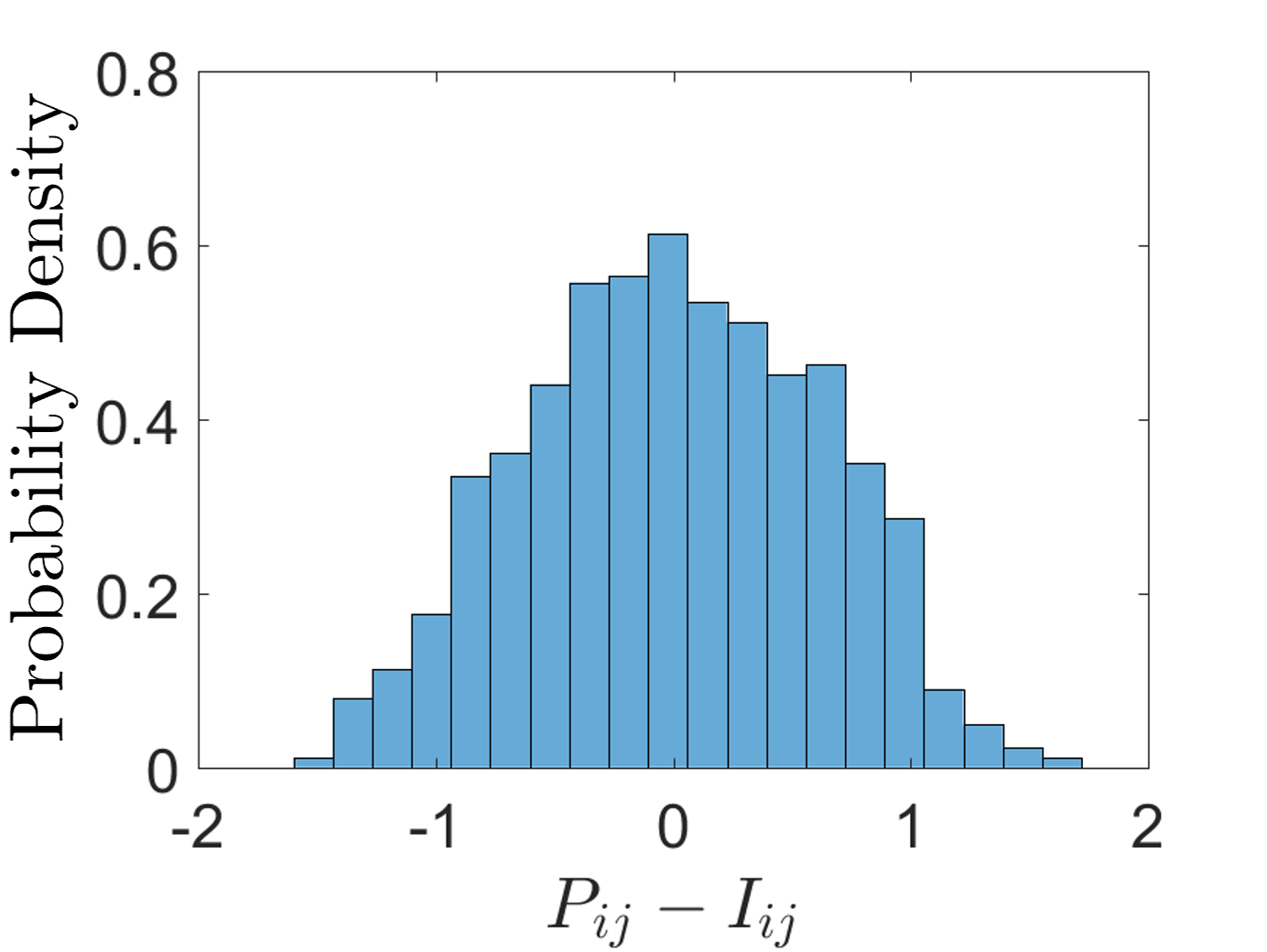}
         \caption{ }
         \label{subfig:GP2 Numerical NNLS}
     \end{subfigure}
     \begin{subfigure}{0.329\textwidth}
         \centering
         \includegraphics[width=\textwidth]{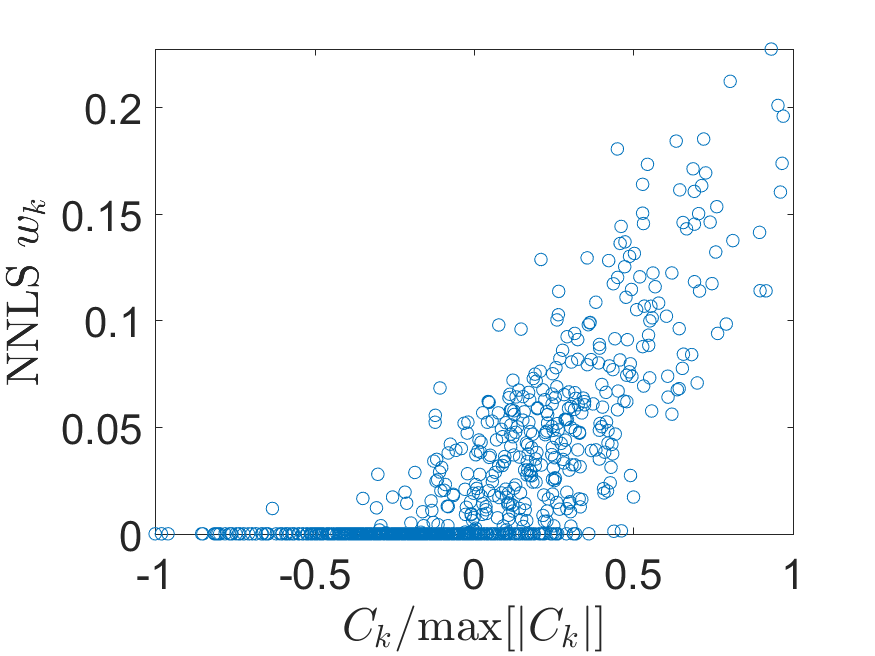}
         \caption{ }
         \label{subfig:GP3 Numerical NNLS}
     \end{subfigure}
     \begin{subfigure}{0.329\textwidth}
         \centering
         \includegraphics[width=\textwidth]{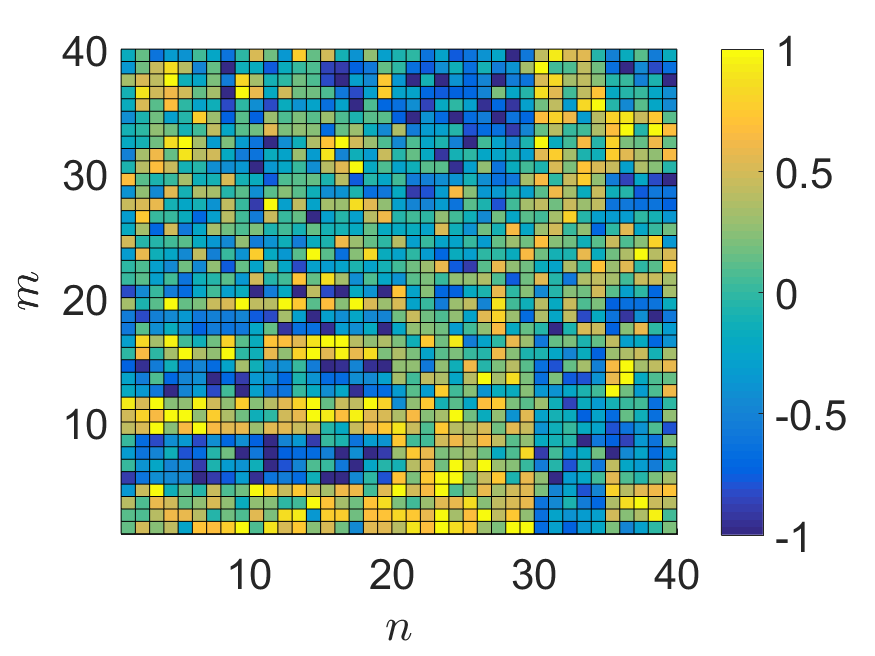}
         \caption{ }
         \label{subfig:GP4 Numerical NNLS}
     \end{subfigure}
     \begin{subfigure}{0.329\textwidth}
         \centering
         \includegraphics[width=\textwidth]{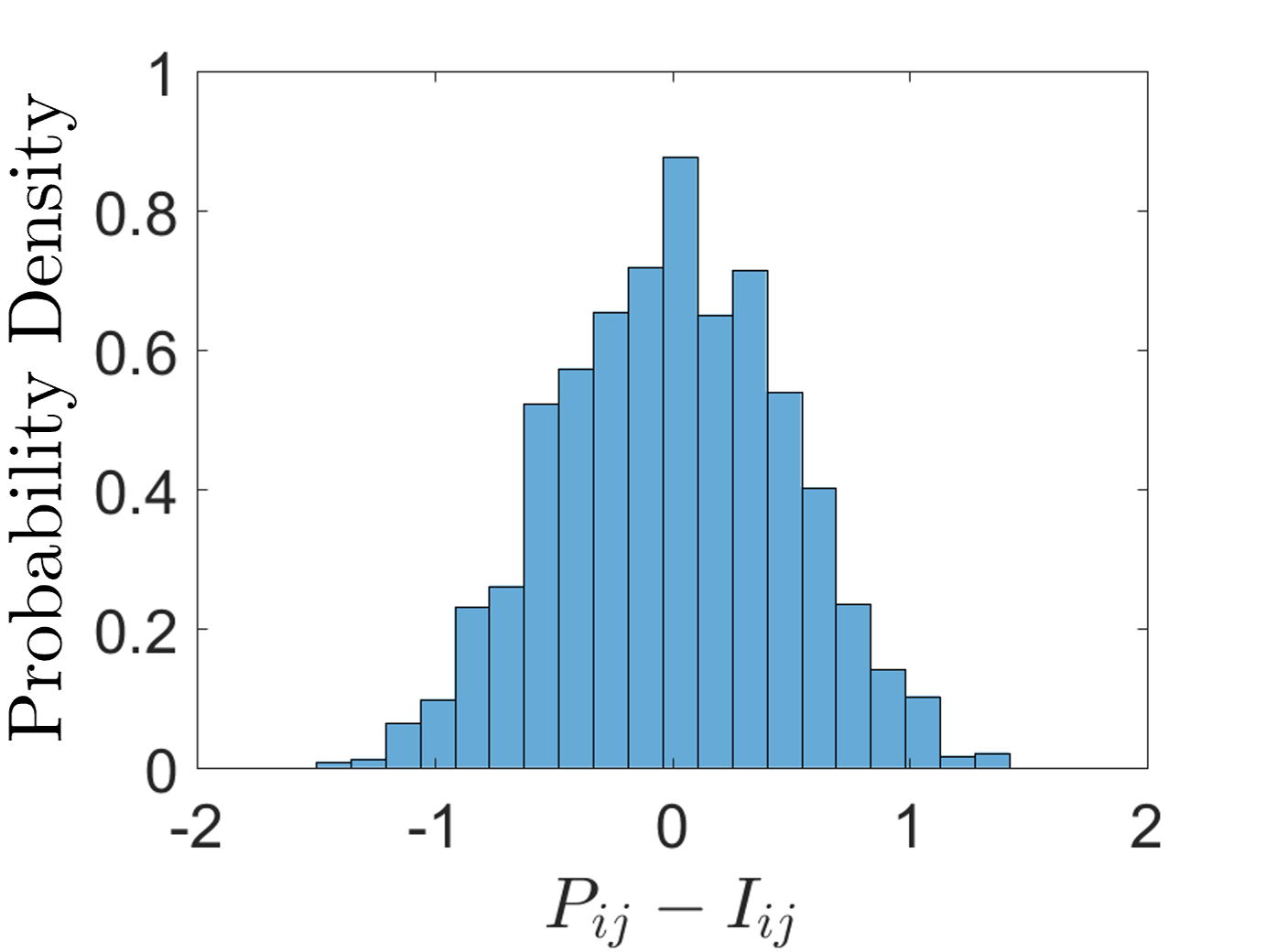}
         \caption{ }
         \label{subfig:GP5 Numerical NNLS}
     \end{subfigure}
     \begin{subfigure}{0.329\textwidth}
         \centering
         \includegraphics[width=\textwidth]{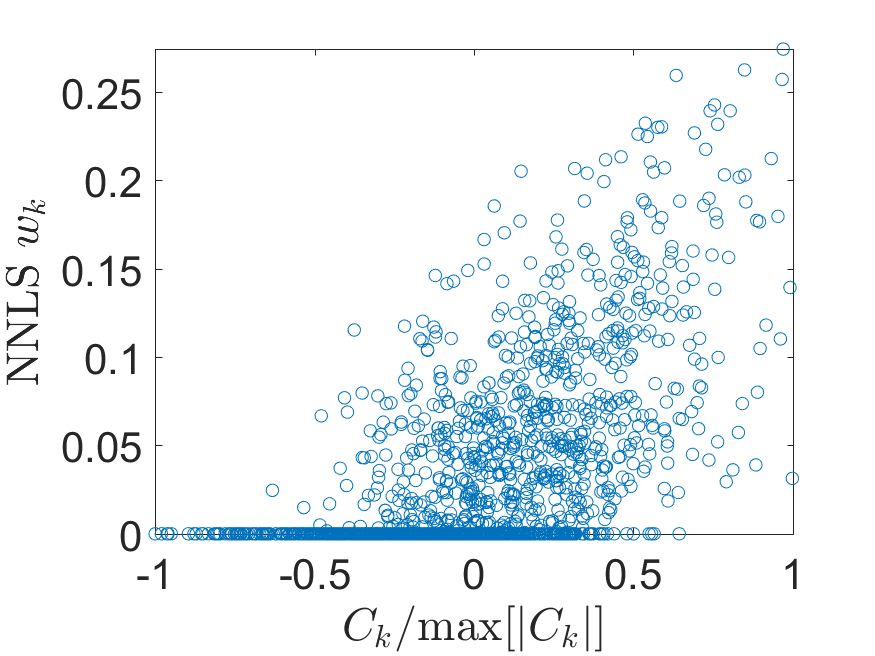}
         \caption{ }
         \label{subfig:GP6 Numerical NNLS}
     \end{subfigure}
     \begin{subfigure}{0.329\textwidth}
         \centering
         \includegraphics[width=\textwidth]{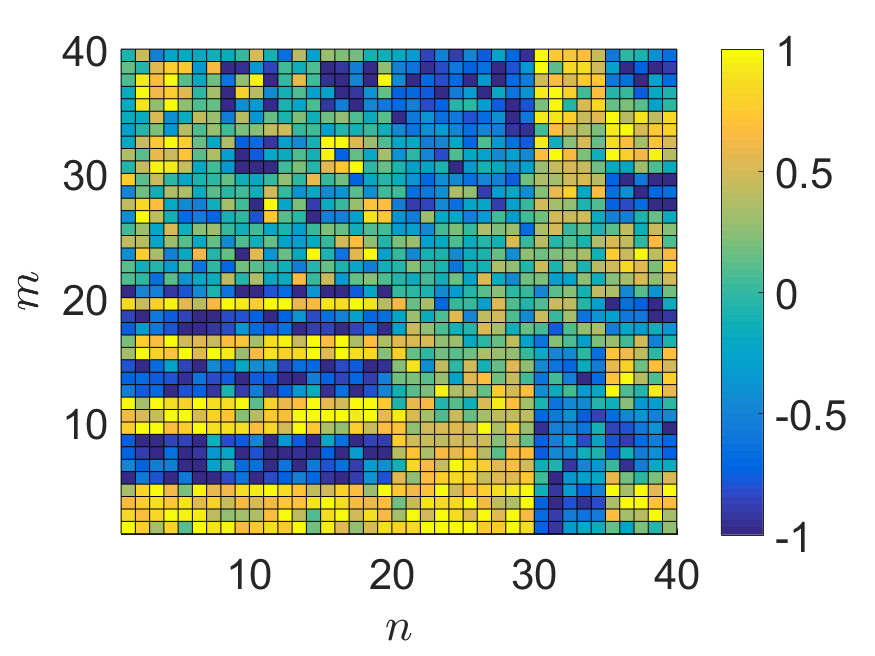}
         \caption{ }
         \label{subfig:GP7 Numerical NNLS}
     \end{subfigure}
     \begin{subfigure}{0.329\textwidth}
         \centering
         \includegraphics[width=\textwidth]{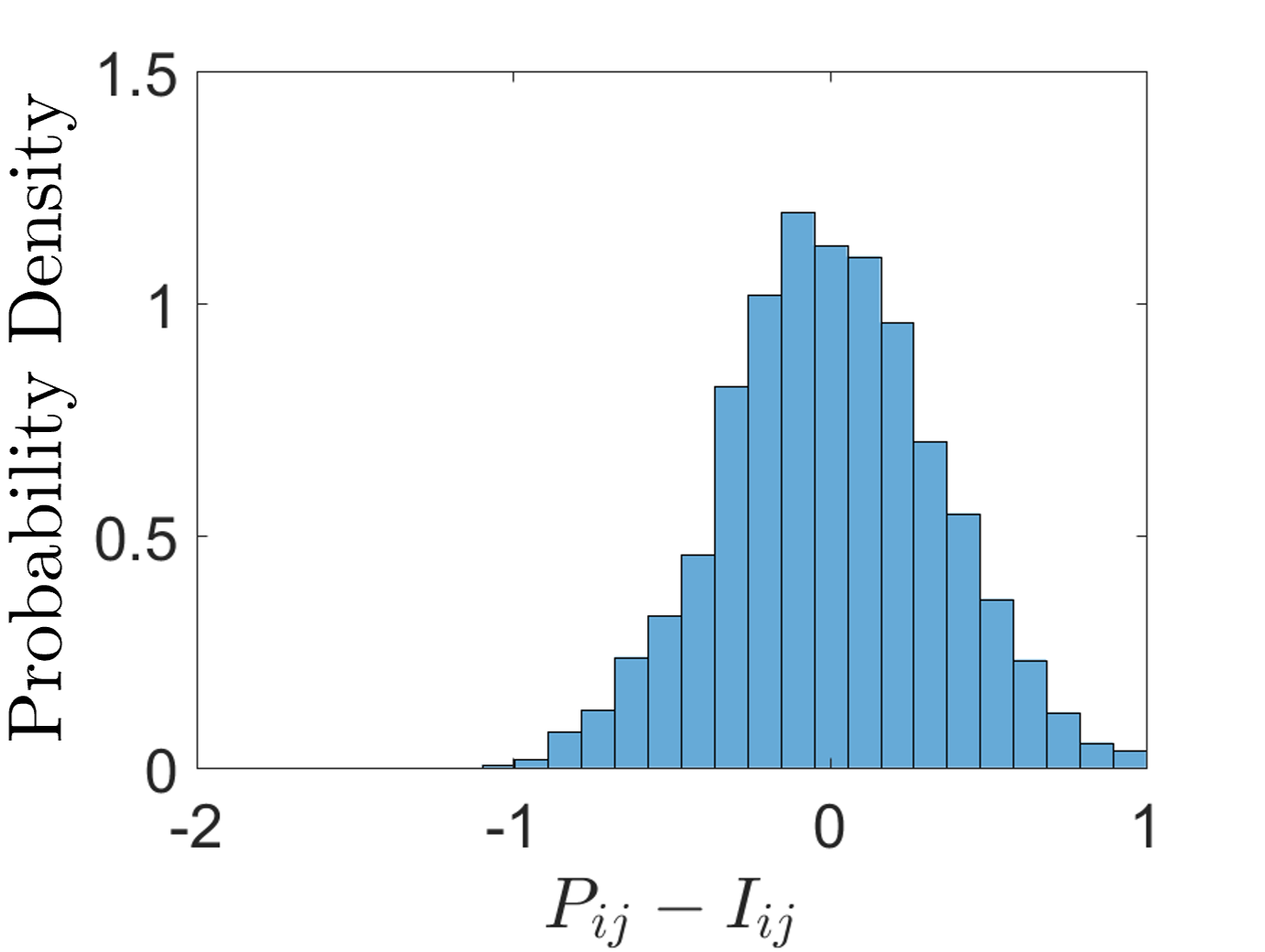}
         \caption{ }
         \label{subfig:GP8 Numerical NNLS}
     \end{subfigure}
     \begin{subfigure}{0.329\textwidth}
         \centering
         \includegraphics[width=\textwidth]{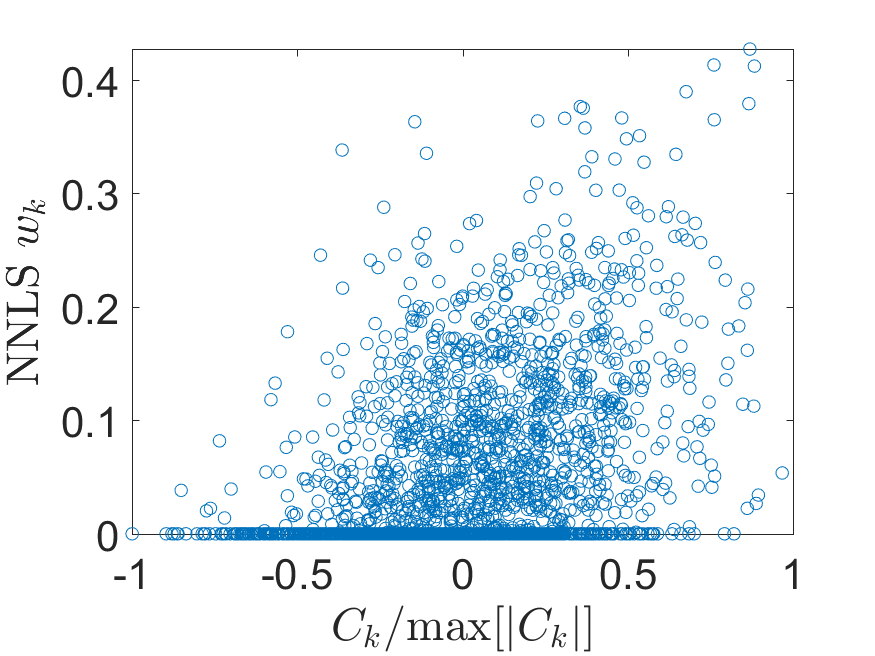}
         \caption{ }
         \label{subfig:GP9 Numerical NNLS}
     \end{subfigure}
     \begin{subfigure}{0.329\textwidth}
         \centering
         \includegraphics[width=\textwidth]{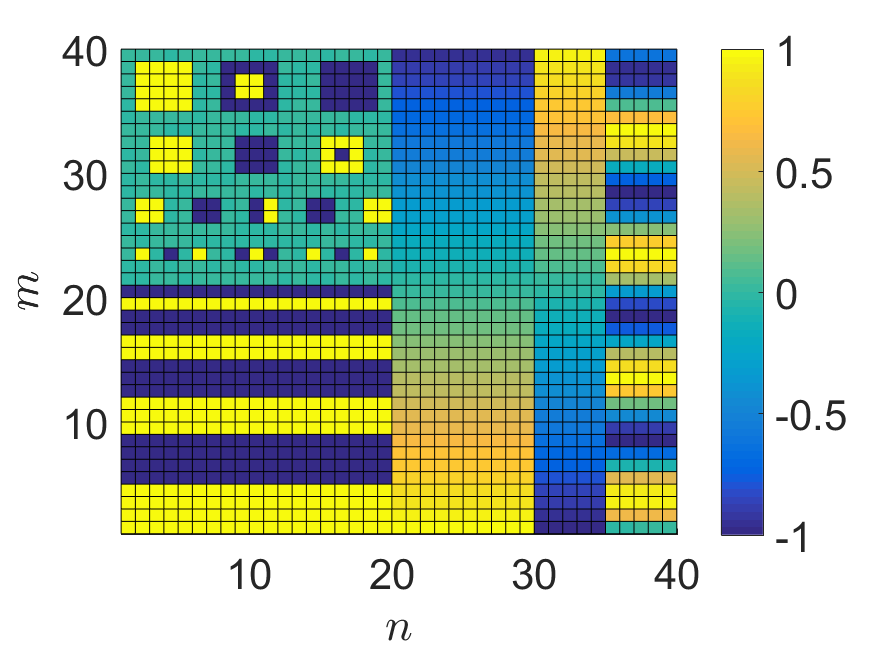}
         \caption{ }
         \label{subfig:GP10 Numerical NNLS}
     \end{subfigure}
     \begin{subfigure}{0.329\textwidth}
         \centering
         \includegraphics[width=\textwidth]{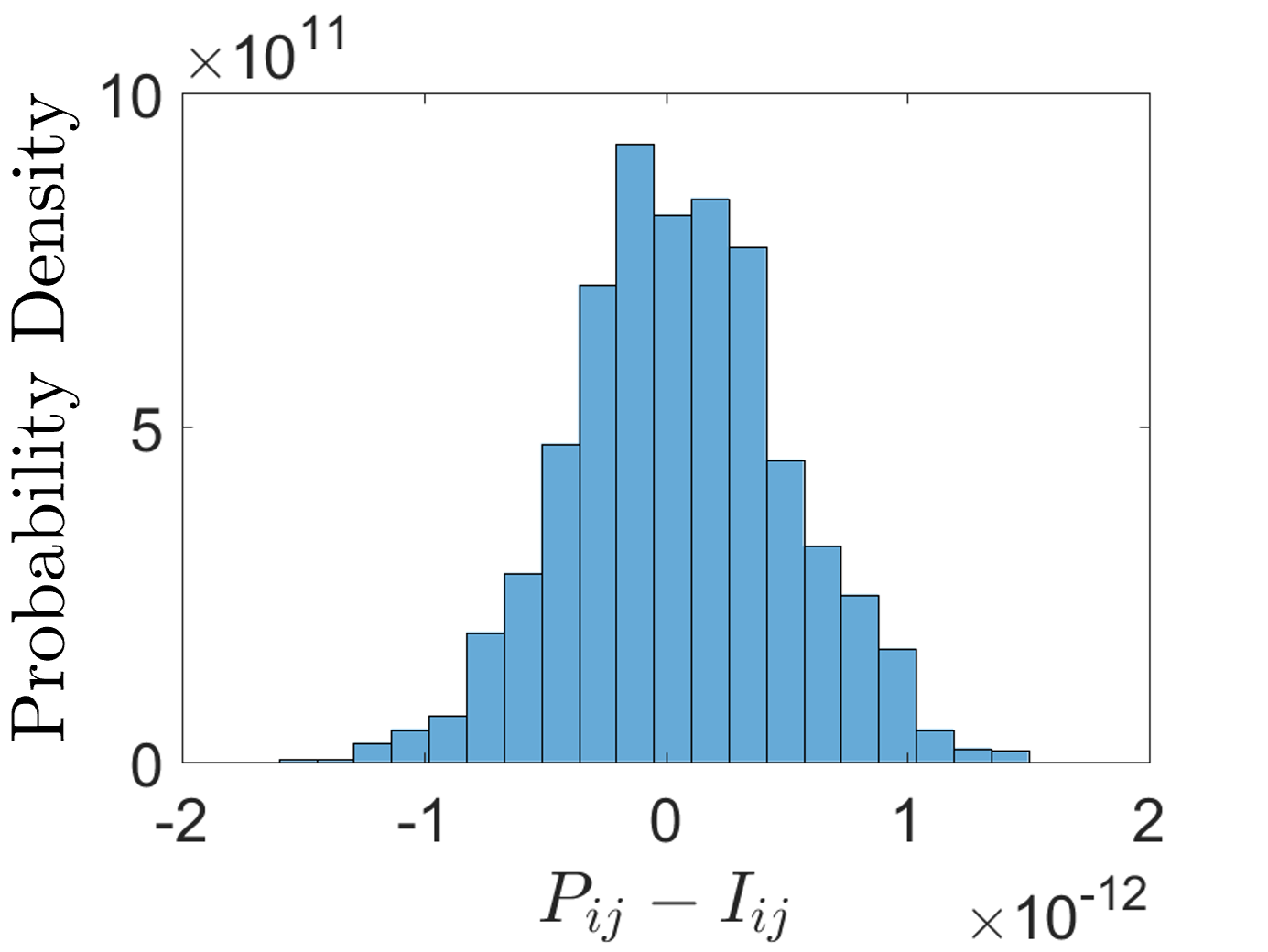}
         \caption{ }
         \label{subfig:GP11 Numerical NNLS}
     \end{subfigure}
     \begin{subfigure}{0.329\textwidth}
         \centering
         \includegraphics[width=\textwidth]{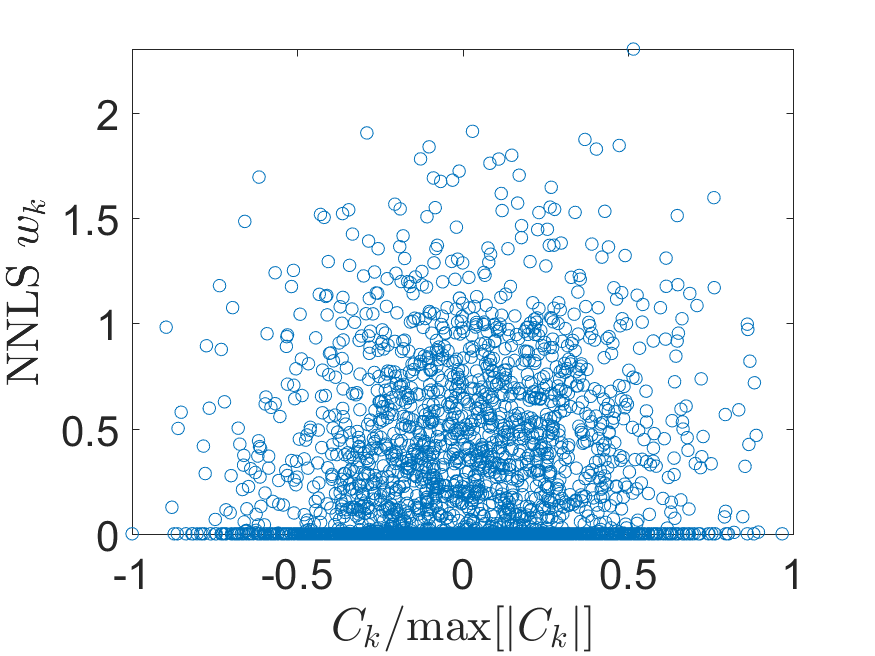}
         \caption{ }
         \label{subfig:GP12 Numerical NNLS}
     \end{subfigure}
        \caption{(a,d,g,j) NNLS ghost projections with $N =\{0.5,1,1.5,2\}nm$, respectively. (b,e,h,k) Random-matrix reconstruction noise present in their respective ghost projections, which achieved an SNR of 1.16, 1.47, 2.10 and $1.57 \times 10^{12}$, respectively. (c,f,i,l) Scatter plots of numerical weights against the normalized pseudo-correlation coefficient of that basis member. Note, each row of the figure represents one realization of a stochastic process and these results will vary from run to run. Typically, roughly $nm$ basis members were given non-zero weighting coefficients in the NNLS optimization. The constant offset in the ghost projection increased with each of the four cases $N =\{0.5,1,1.5,2\}nm$, and had values of $\{10.6,27.8,60.2,395.9\}$, respectively.}
        \label{fig:NNLS Refined Ghost Projection}
\end{figure*}

We use Non-Negative Least Squares (NNLS) \cite{NumericalRecipes} to obtain the weights that optimize SNR. That is, we seek the vector of weights $\vec{w}$ corresponding to:
\begin{align}
\text{arg~min} \|M\vec{w} - \vec{I}\|,
\end{align}
subject to the constraint that $w_k \geq 0$. Using the $\mathtt{MATLAB}$ programming language NNLS routine $\mathtt{lsqnonneg}$  to perform numerically optimized ghost projection of our desired image in Fig.~\ref{subfig: GP1 Pseudo}, with a basis set of four different sizes $N = $\{0.5,1,1.5,2\}$nm$, we see modest improvements for the first three cases and a drastic improvement for the final case, as shown in Fig.~\ref{fig:NNLS Refined Ghost Projection}. For the first case, $N = 0.5nm$, Fig.~\ref{subfig:GP3 Numerical NNLS} shows (i) a somewhat linear relationship between the numerically optimized non-zero weights and the pseudo-correlation coefficient, as well as (ii) some filtering out of certain basis members, many of which have $C_k<0$, as indicated by the row of zero weights along the horizontal axis.  Conceptually, this aligns well with our previously-considered analytical approaches of pseudo-correlation weighting and filtering. We note, however, that the NNLS scheme includes some masks with negative $C_k$, namely masks that are anti-correlated with the desired image, and which may be loosely considered as `position-dependent erasers' that contribute to the ghost projection by suppressing rather than establishing correlations.   For the final case ($N=2nm$), it seems that amongst the twice-over-complete basis set, an almost exact non-negative-weights representation exists. As we progress from the first to the final row of Fig.~\ref{fig:NNLS Refined Ghost Projection}, we see from panels (c,f,i,j) that the near-linear relationship between pseudo-correlation coefficient and numerically optimized weighting coefficient dissolves with increasing basis size. For a sufficiently large number of members in our random ghost-projection basis, there exist non-negative weights that can achieve a near-exact representation or projection of our desired image, up to an additive constant. 

%These figures were produced with the MATLAB file GP_NNLS.m
%Non-negative Least Squares SNR = 1.164843 for N = 0.50 nm 
%Non-negative Least Squares SNR = 1.465618 for N = 1.00 nm 
%Non-negative Least Squares SNR = 2.098343 for N = 1.50 nm 
%Non-negative Least Squares SNR = 1574691054515.843700 for N = 2.00 nm

\subsection{Ghost Projection with Poisson Noise}

Suppose we have a ghost projection with numerically derived weights that, in a noise-free environment, obtains a near-perfect projection. In the present subsection, we study how robust this reconstruction is against Poisson noise \cite{mandel1995optical}. To model the effect of Poisson noise, consider the scheme:
\begin{align}
P_{ij} =  J^{k} \hat{P}( \lambda w_k R_{ijk}  ).
\end{align}
Here, $\hat{P}$ denotes a quantity that is Poisson distributed, $w_k$ are our numerically derived weights which correspond to random-mask exposures, $R_{ijk}$ is our random-matrix basis and $\lambda$ is the number of photons per pixel, per unit exposure (although, since our weights are normalized to give one unit exposure of the image in total, $\lambda$ is really the number of photons per pixel). Supposing we are in the limit that noise from the random-matrix basis is insignificant, we model the degradation due to the Poisson noise alone. Taking the expected value, we obtain:
\begin{align} \nonumber
\text{E}[ P_{ij} ] &= \text{E}[ J^{k} \hat{P}( \lambda w_k R_{ijk}  ) ] \\ \nonumber 
&= \lambda w_k R_{ij}^{\ \ k}  \\
&= \lambda I_{ij}   +  \lambda \bar{P} J_{ij},
\end{align}
where $\bar{P} = J^{ij} P_{ij} /(nm) = N' \text{E}[w] \text{E}[R]$ for $\text{E}[I]=0$, and we have used the Poisson expectation value property $\text{E}[\hat{P}(X)] = X$. Moving onto the variance, we have:
\begin{align} \nonumber
\text{Var} [ P_{ij} ] &= \text{Var}[ J^{k} \hat{P}(  \lambda w_k R_{ijk} ) ] \\ \nonumber
&= \lambda w_k R_{ij}^{\ \ k}  \\
&= \lambda I_{ij}   + \lambda \bar{P} J_{ij},
\end{align}
where we have used the Poisson variance property $\text{Var}[\hat{P}(X)] = X$. For almost all cases of ghost projection, we would expect $I_{ij}$ to be much less than $\bar{P}$, and we can approximate the variance by the constant result:
\begin{align}
\text{Var} [ P_{ij} ] & \approx  \lambda \bar{P} J_{ij}. \label{eq:Variance result for NW w/ Poisson Noise}
\end{align}
Combining this into a pixel-wise SNR, we have:
\begin{align} \nonumber
\text{SNR}_{ij} &\equiv  \frac{\text{E}[P_{ij}] - \lambda \bar{P} }{\sqrt{\text{Var}[P_{ij}]}} \\
&\approx  \sqrt{\lambda} \frac{I_{ij}}{\sqrt{\bar{P}}}.
\end{align}
The RMS of the pixel-wise SNR gives the global SNR:
\begin{align}
\nonumber
\text{SNR} &= \sqrt{\frac{1}{nm} J^{ij} \text{SNR}_{ij}^2} \\
&\approx \sqrt{\frac{ \lambda  \text{E} [I^2] }{ \bar{P}}} \approx \sqrt{\frac{ \lambda  \text{E} [I^2] }{ \text{E}[w] \text{E}[R] N'}}. \label{eq:SNR result for numerical weights with Poisson noise}
\end{align}
Above, we see that SNR decreases proportionally to the square-root of the pedestal $\sqrt{\bar{P}}$ or $\sqrt{\text{E}[w] \text{E}[R] N'}$, and increases with both the area-normalized image 2-norm $\text{E}[I^2]$ and the number of photons, per pixel, used. Based upon the above result, we can see how critical it is to minimize the pedestal for the sake of minimizing the Poisson noise.

\subsection{Ghost Projection with Poisson Noise Simulation}

Starting with $N=2.50nm$ uniformly distributed random matrices, we use a NNLS solver to find an optimized weighting scheme that minimizes the random-matrix reconstruction noise. In Poisson-noise-free simulation we obtained a weighting scheme that produced an SNR of $1.36 \times 10^{10}$ and a pedestal of $\bar{P} = 100.2$. Subsequently adding Poisson noise to the reconstruction drastically degraded the reconstruction SNR down to 5.06 in simulation, which compared well with the predicted an SNR of 5.03 based on  Eq.~(\ref{eq:SNR result for numerical weights with Poisson noise}).

\begin{figure*}[ht!]
     \centering
     \begin{subfigure}{0.329\textwidth}
         \centering
         \includegraphics[width=\textwidth]{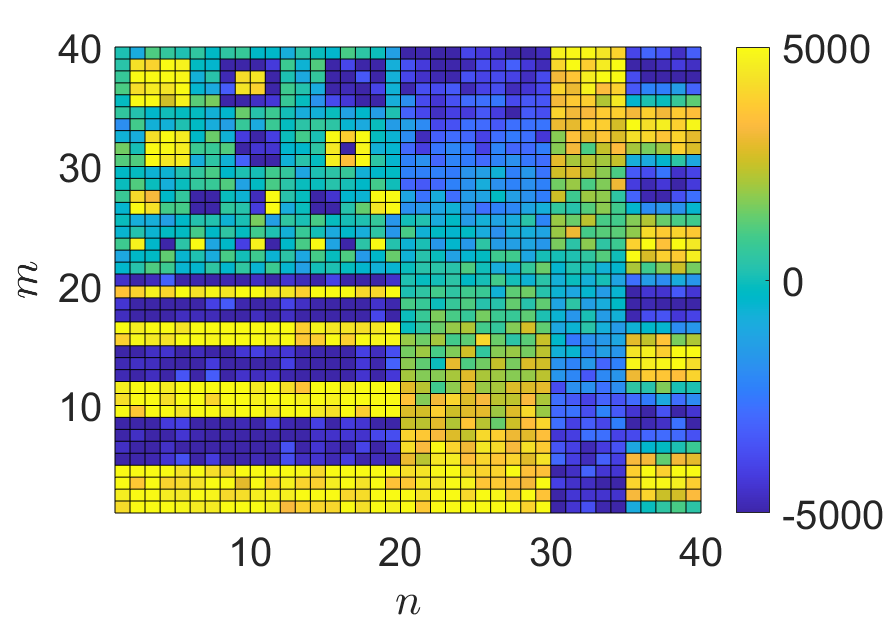}
         \caption{ }
         \label{subfig:PoisNoise1}
     \end{subfigure}
     \begin{subfigure}{0.329\textwidth}
         \centering
         \includegraphics[width=\textwidth]{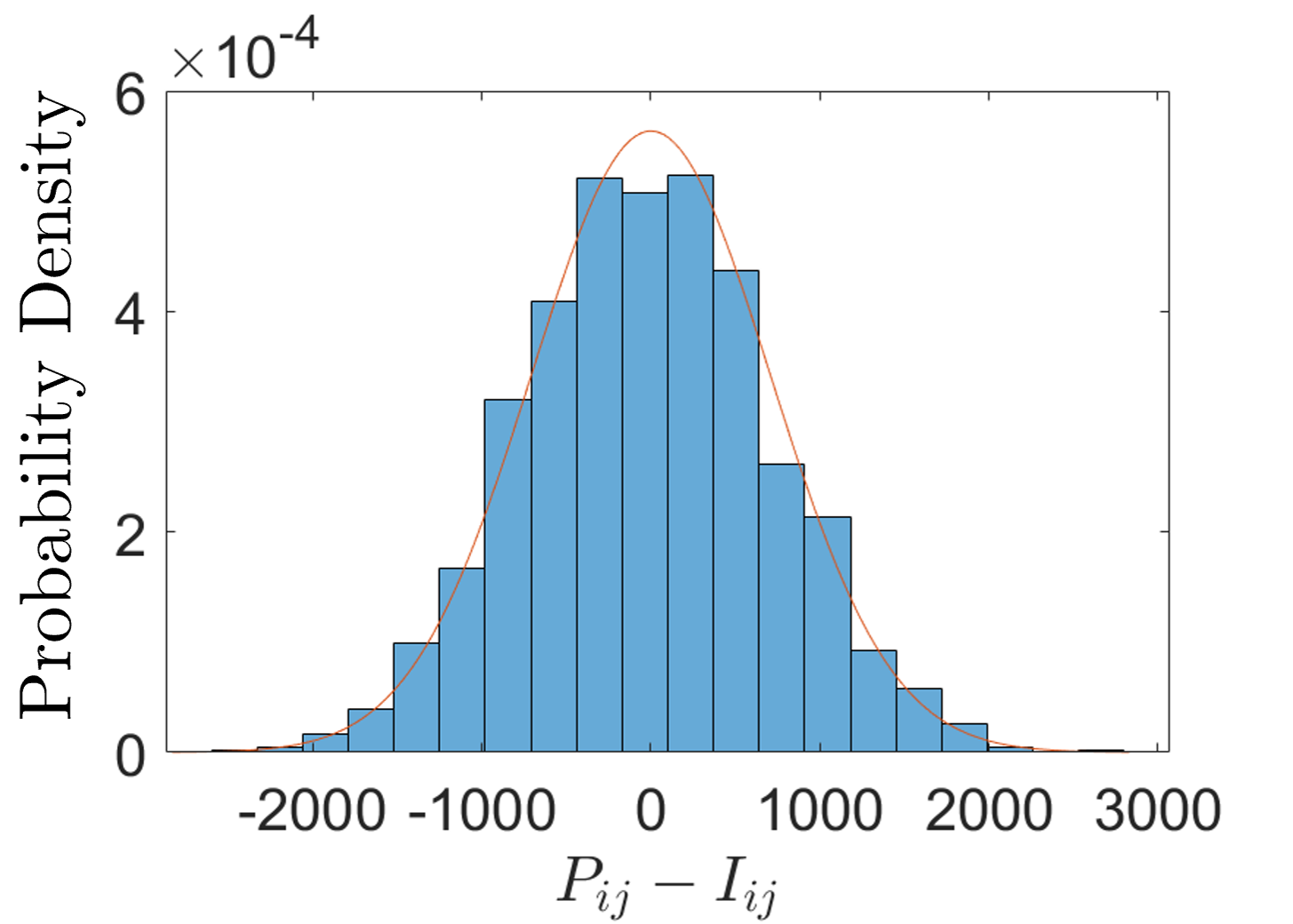}
         \caption{ }
         \label{subfig:PoiNoise2}
     \end{subfigure}
        \caption{(a) Numerically optimized ghost projection with subsequently-added Poisson noise, predicted SNR = 5.03 based on Eq.~(\ref{eq:SNR result for numerical weights with Poisson noise}),  and simulated SNR = 5.06. The NNLS reconstruction SNR, in the absence of Poisson noise, was  $\text{SNR} = 1.36 \times 10^{10}$ for $N = 2.50 nm$ and $\bar{P} = 100.2$. (b) A plot of the effect of Poisson noise, overlaid with the expected behavior based on Gaussian statistics and the variance result of Eq.~(\ref{eq:Variance result for NW w/ Poisson Noise}).}
        \label{fig:Poisson noise fig}
\end{figure*}

% These figures were made with the MATLAB  code GP_NNLS.m 
% Predicted Poisson Noise SNR = 5.029652 and simulated SNR = 5.06 
% Non-negative Least Squares SNR = 13556516301.348505 for N = 2.50 nm 
% Pbar = 100.2

\subsection{Ghost Projection with Exposure Noise}

Again consider a ghost projection with numerically derived weights that, in a noise-free environment, obtains a near-perfect projection. To model exposure noise, namely the fluctuations in exposure time that will be present in any experimental realization of ghost projection, consider the scheme:
\begin{align}
P_{ij} =  \tilde{P}( w_k ) R_{ij}^{\ \ k},
\end{align}
where $\tilde{P}$ denotes a quantity that is normally distributed, so that $\tilde{P}(w_k)$ denotes a Gaussian for each exposure which has an expectation value of the weight $w_k$ and a variance $\sigma_w^2$, and $R_{ij}^{\ \ k}$ is our random-matrix basis. Supposing we are in the limit that noise from the random-matrix-basis reconstruction and Poisson noise are insignificant, we can isolate the degradation due to exposure noise alone. Taking the expectation value of the scheme, we achieve:
\begin{align} 
\text{E} [ P_{ij} ] = \text{E} [ \tilde{P}( w_k ) R_{ij}^{\ \ k} ] 
=w_k  R_{ij}^{\ \ k}  
= I_{ij}  + \bar{P} J_{ij}.
\end{align}
The corresponding variance is:
\begin{align}
\text{Var}[P_{ij} ] = \text{Var} [ \tilde{P}( w_k ) R_{ij}^{\ \ k} ] = \sigma_{w}^2 N' \text{Var}[R]  J_{ij}.\label{eq:Exposure noise variation result}
\end{align}
Note, this is the variance in the pixels for a single realization of exposure noise in all weights. Despite having $N'$ realizations of exposure noise, these impact all of the pixels and the induced variance is dominated by that of summing random deviates of the basis, $N' \text{Var}[R]$, which are then skewed by the expected amount of $\text{E}[\tilde{P}(w_k)^2] = \sigma_w^2$. Note also that this ignores perturbations from the pedestal, which do not impact the desired contrast, hence we are assuming these perturbations to be insignificant. Should we want the more general variance of this quantity, then this would be $ \sigma_{w}^2 N' \text{E}[R^2]$ (which would correspond to the variance of many realizations of the set of weights). Combining the above results into a pixel-wise SNR, in this case we have:
\begin{align}
\text{SNR}_{ij} &= \frac{I_{ij}}{\sqrt{N' \sigma_{w}^2 \text{Var}[R]  }},
\end{align}
leading to the global SNR:
\begin{align}
\text{SNR} = \sqrt{  \frac{\text{E}[I^2]}{N' \sigma_{w}^2 \text{Var}[R] }}. \label{eq:SNR of NNLS GP w/ Exposure Noise}
\end{align}
Hence minimizing the variance from our desired exposures $w_k$ is essential for preserving a quality projection.

\subsection{Ghost Projection with Exposure Noise Simulation}

\begin{figure*}[ht!]
     \centering
     \begin{subfigure}{0.329\textwidth}
         \centering
         \includegraphics[width=\textwidth]{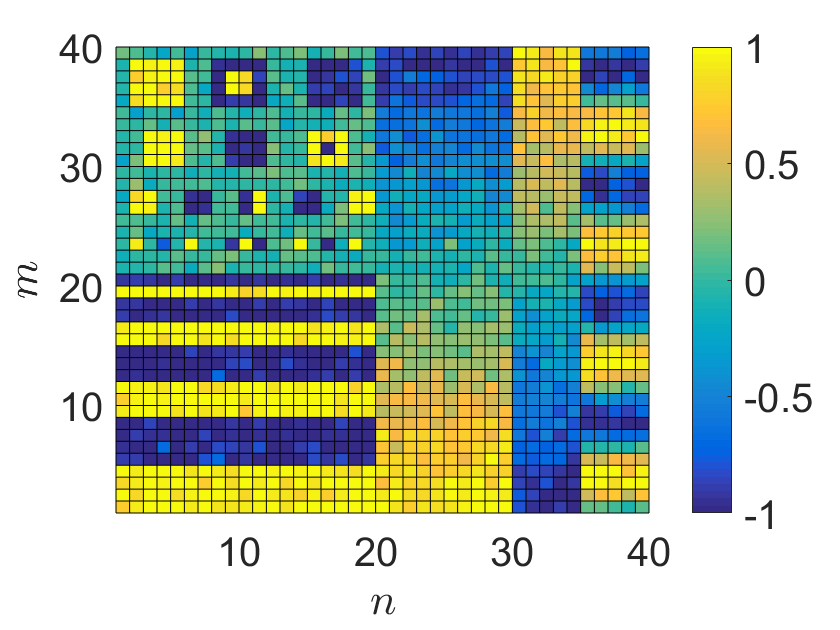}
         \caption{ }
         \label{subfig:ExpNoise1}
     \end{subfigure}
     \begin{subfigure}{0.329\textwidth}
         \centering
         \includegraphics[width=\textwidth]{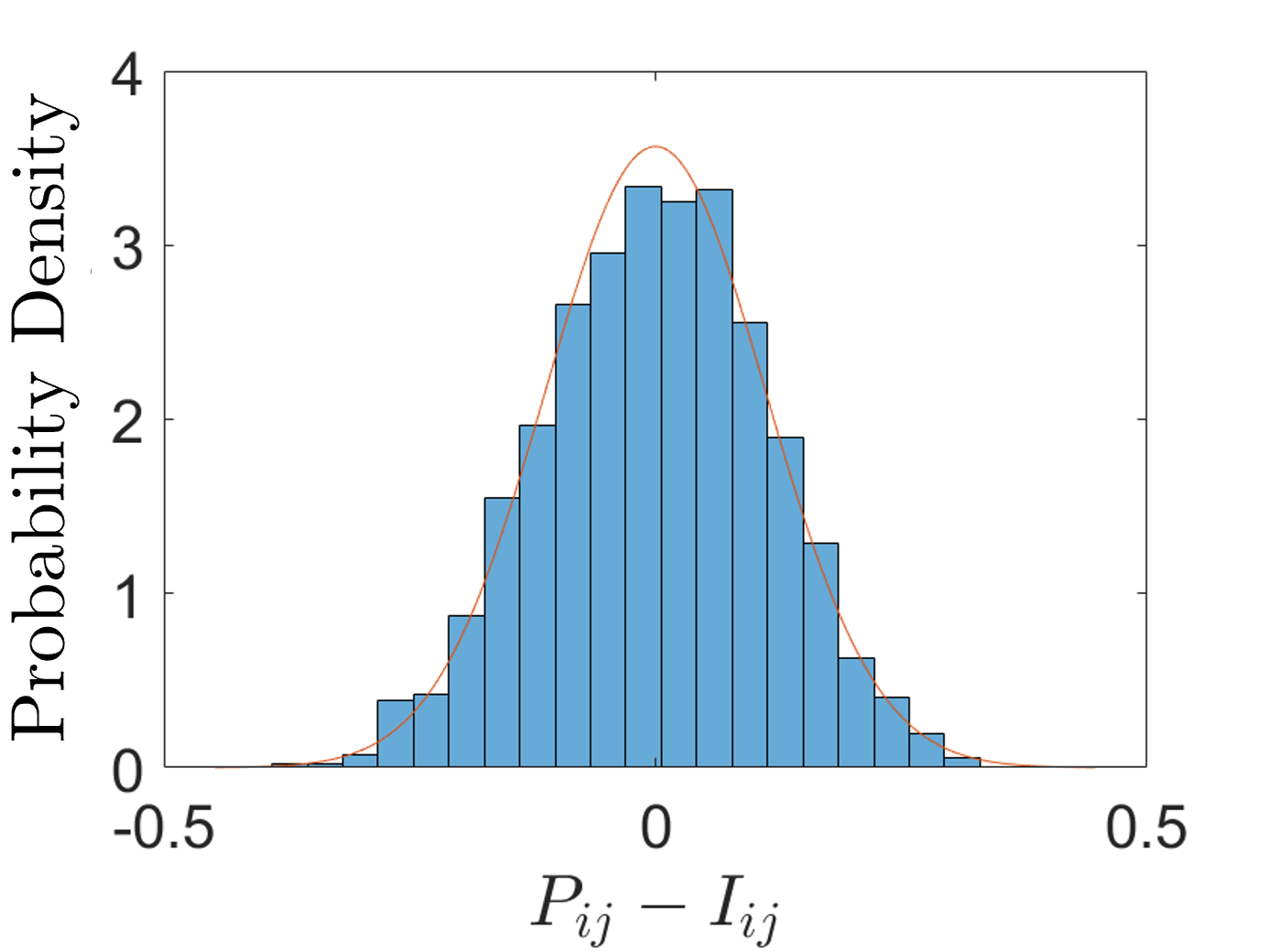}
         \caption{ }
         \label{subfig:ExpNoise2}
     \end{subfigure}
        \caption{(a) Ghost projection with a shutter speed standard deviation of $\sigma_w=1/100$ and a simulated SNR of 6.52, for which an SNR of 6.50 was predicted using Eq.~(\ref{eq:SNR of NNLS GP w/ Exposure Noise}). The NNLS reconstruction SNR, in the absence of Poisson noise, was $1.57 \times 10^{12}$ and had a pedestal of $\bar{P} = 395.9$. (b) A plot of the effect of exposure noise, overlaid with the expected behavior based on Gaussian statistics and the variance result of Eq.~(\ref{eq:Exposure noise variation result}).}
        \label{fig:Exposure noise fig}
\end{figure*}

Suppose we start with a NNLS ghost projection which has a Poisson-noise-free SNR of $1.57 \times 10^{12}$ and a pedestal of $\text{E}[w] \text{E}[R] N' = 395.9$. If we suggest that the weights realized in a realistic experiment have $\sigma_{w} = 1/1000$, then the SNR we would expect from Eq.~(\ref{eq:SNR of NNLS GP w/ Exposure Noise}) is 63.2. In simulation, we obtained an SNR of 63.1.  If we increase the exposure noise to  $\sigma_{w} = 1/100$, we obtain the results shown in Fig.~\ref{fig:Exposure noise fig}, for which the simulated SNR of 6.52 may be compared to the predicted SNR of 6.50. From this assessment, so long as we can keep the shutter speed standard deviation on the order of $\sigma_w=1/1000$, we might expect exposure noise to be a subdominant contribution when compared to the noise due to a random-matrix basis representation sought with a small pedestal and Poisson noise.

\subsection{Ghost Projection with both Poisson and Exposure Noise}

Modeling the combined influence of Poisson and exposure noise in the regime of a near-perfect random-basis representation, we have the scheme:
\begin{align}
P_{ij} =  J^{k} \hat{P} ( \lambda  \tilde{P}( w_k ) R_{ijk} ),
\end{align}
where $\hat{P}$ denotes a quantity that is Poisson distributed, $\tilde{P}$ denotes a quantity that is Gaussian distributed, $w_k$ are our numerically derived weights which correspond to exposure, $R_{ijk}$ is our random-matrix basis and $\lambda$ is the number of photons per pixel. The expectation value of this scheme is:
\begin{align}
\nonumber
\text{E}[P_{ij}] &= \text{E} [J^{k} \hat{P} ( \lambda \tilde{P}( w_k ) R_{ijk}  )] \\ \nonumber
&= J^{k} \text{E} [ \lambda  \tilde{P}( w_k ) R_{ijk}  ] \\ \nonumber
&=  \lambda w_k R_{ij}^{\ \ k}  \\
&= \lambda I_{ij}  + \lambda \bar{P} ,
\end{align}
where $\bar{P} =  J^{ij} P_{ij} /(nm) = N' \text{E}[w] \text{E}[R] $ is our pedestal for the case $\text{E}[I] = 0$. Moving onto the variance, we can utilize our previous working from the pseudo-correlation Poisson noise case, in Sec.~\ref{subsec:Pseudo-correlation Poisson Noise Ghost Projection Scheme}, as well as our working from the pure exposure noise case, to yield:
\begin{align}
\nonumber
\text{Var}[P_{ij}] &= \text{Var} [J^{k} \hat{P} ( \lambda \tilde{P}( w_k ) R_{ijk}  )] \\
&= N' \sigma_w^2 \lambda^2 \text{Var}[R]  + N'  \lambda \text{E}[w] \text{E}[R]. \label{eq:GPwPoisson and Exposure Noise Variance Result}
\end{align}
To the right of the equals sign, we can recognize that the first term is the exposure noise contribution and the second term is the Poisson noise contribution. Combining the above two results into a pixel-wise SNR relationship, we have:
\begin{align}
\text{SNR}_{ij} &= \frac{\lambda I_{ij} }{\sqrt{N' \sigma_w^2 \lambda^2 \text{Var}[R]  + N' \lambda \text{E}[w] \text{E}[R] }}.
\end{align}
The RMS of the above expression gives the global SNR:
\begin{align}
\text{SNR} = \sqrt{\frac{ \lambda \text{E}[I^2] }{N' \sigma_w^2 \lambda \text{Var}[R]  + N' \text{E}[w] \text{E}[R]}}. \label{eq:SNR result for NNLS w/ both Poisson and Exposure noise}
\end{align}
In this expression, we see precisely the behavior we would expect based on the analytical results obtained in the above subsections. That is, for increasingly large $\lambda$ (i.e. in the case of many photons per pixel) we see the SNR asymptotically approach that of the pure-exposure-noise case. Conversely, if we are in the regime where the shutter-speed standard deviation $\sigma_w$ becomes increasingly small, we asymptotically approach the SNR behavior of the pure-Poisson-noise case.

\subsection{Ghost Projection with both Poisson and Exposure Noise Simulation}

\begin{figure*}[ht!]
     \centering
     \begin{subfigure}{0.329\textwidth}
         \centering
         \includegraphics[width=\textwidth]{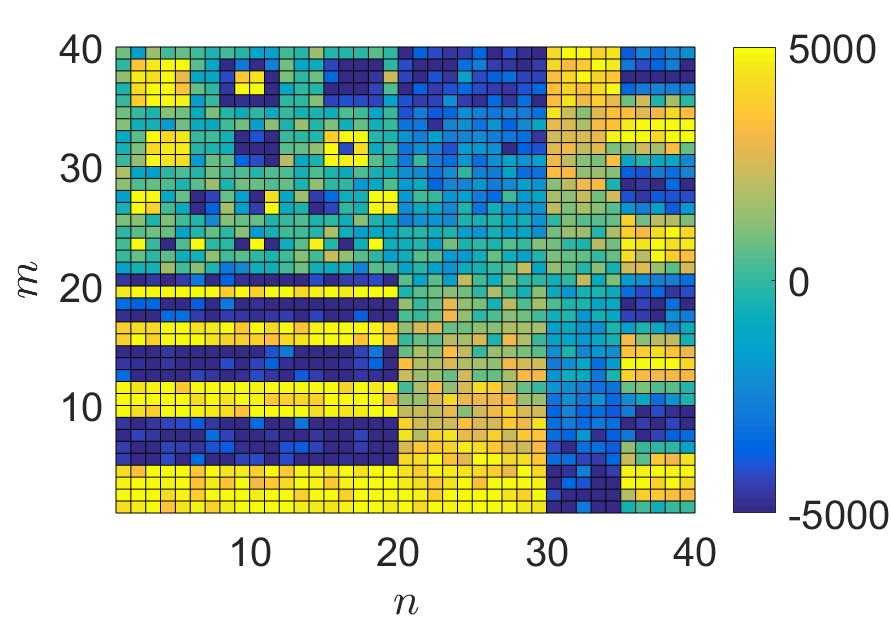}
         \caption{ }
         \label{subfig:PoiExpNoise1}
     \end{subfigure}
     \begin{subfigure}{0.329\textwidth}
         \centering
         \includegraphics[width=\textwidth]{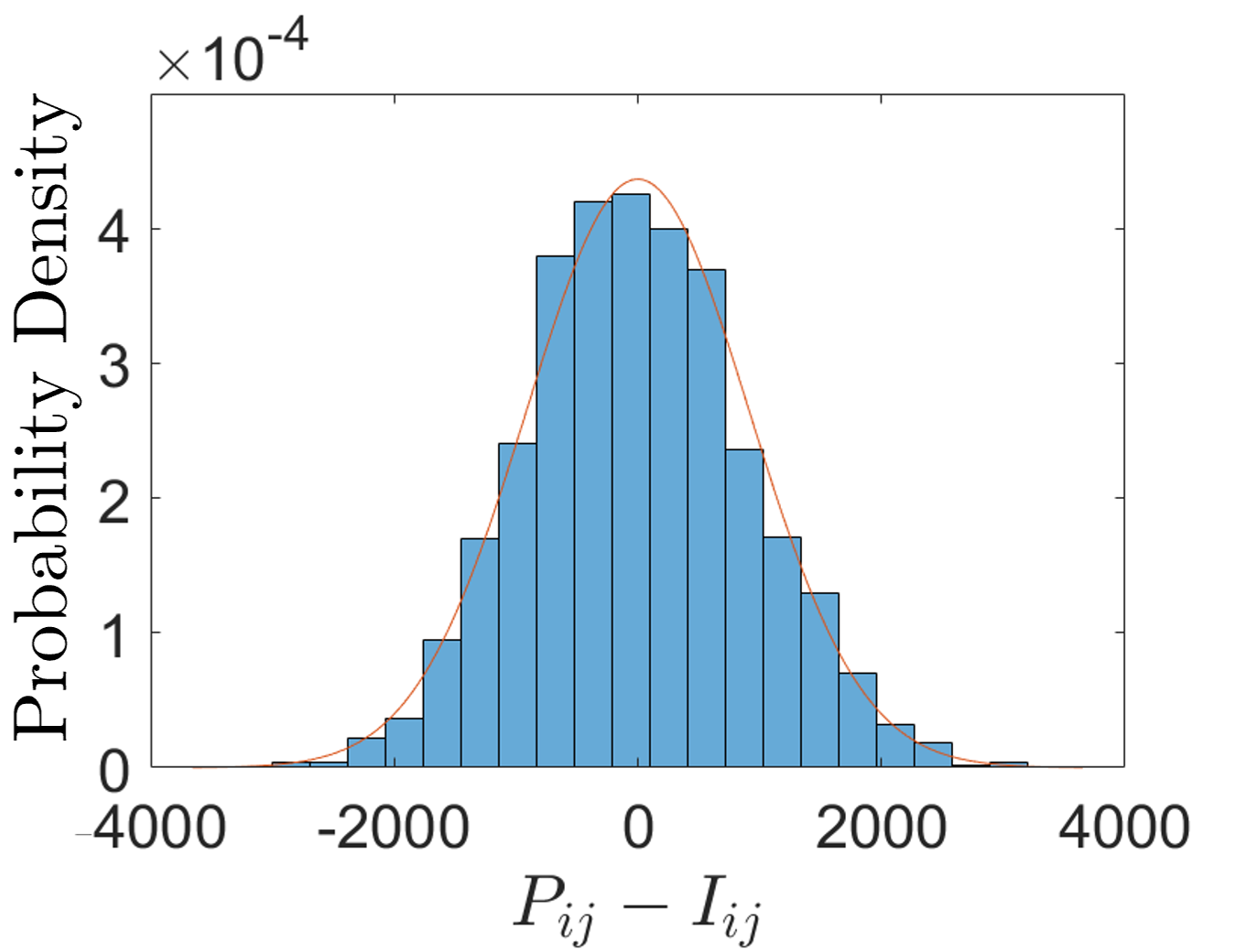}
         \caption{ }
         \label{subfig:PoiExpNoise2}
     \end{subfigure}
        \caption{(a) Ghost projection with Poisson noise in the photon count for $\lambda = 5,000$, and a shutter speed standard deviation of $\sigma_w=1/100$ which produced an SNR of 3.92 in simulation, for which an SNR of 3.90 was predicted by Eq.~(\ref{eq:SNR result for NNLS w/ both Poisson and Exposure noise}). For these conditions, if it were Poisson noise alone contaminating the projection, we would expect an SNR of 4.93. If it were exposure noise alone contaminating the projection, we would expect an SNR of 6.50. This projection had a pedestal of $\bar{P} = 104.9$, which can be multiplied by $\lambda$ to obtain the pedestal in units of photon counts. (b) A plot of the combined effect of Poisson and exposure noise, overlaid with the expected behavior based on Gaussian statistics and the variance result of Eq.~(\ref{eq:GPwPoisson and Exposure Noise Variance Result}).}
        \label{fig:Poisson/Exposure noise fig}
\end{figure*}

Consider a $N=2.5nm$ uniformly-random basis set and corresponding numerically derived weights that obtains a near-perfect random-basis projection, with a noise-free SNR of $9.22 \times 10^{9}$ and pedestal of $\bar{P} = 104.9$. Choosing the number of photons per pixel to be $\lambda = 5,000$ and the standard deviation in the shutter speed to be $\sigma_w = 1/100$, we can perform a simulation to confirm the working of the previous subsection. Once both contributions of experimentally motivated noise were added, an SNR of 3.92 was obtained in simulation, for which an SNR of 3.90 was predicted using Eq.~(\ref{eq:SNR result for NNLS w/ both Poisson and Exposure noise}). For these conditions, if it were Poisson noise alone contaminating the projection, we would expect an SNR of 4.93. If it were exposure noise alone contaminating the projection, we would expect an SNR of 6.50. The combined result can be observed in Fig.~\ref{fig:Poisson/Exposure noise fig}.

\subsection{Ghost Projection with Poisson Noise and Numerically Derived Weights}

Within this section, we have first seen that by using numerically optimized weights, an image can be projected using random matrices to a near perfect reconstruction (supposing the initial basis set is large enough, e.g.~$N > 2nm$). We then went on to examine the case of ghost projection with a near perfect reconstruction, under the influence of Poisson noise. In that, we saw the dominant cause of Poisson noise is the pedestal, which deposits photons without contributing to the desired contrast. Arising from these two observations is that there exists a reconstruction which balances the random-basis reconstruction noise with minimizing the pedestal and Poisson noise contribution. To find this balance, we can derive the SNR for an imperfect random-basis reconstruction, under the influence of Poisson noise, which can subsequently be numerically optimized. The variance of an imperfect random-basis reconstruction, under the influence of Poisson noise, is:
\begin{align} \nonumber
    \text{Var}[P_{ij}] &= \text{E} \left[ \left\{ \hat{P} \left( w_{k}R_{ij}^{\ \ k}  \lambda \right)  - \lambda (I_{ij} + \bar{P}) \right\}^2 \right] \\ \nonumber
    &=  \text{E} \left[ \hat{P} \left( w_{k}R_{ij}^{\ \ k} \lambda \right)^2 \right] - 2 \lambda (I_{ij} + \bar{P})    \\ \nonumber
    & \ \times \text{E} \left[ \hat{P} \left( w_{k}R_{ij}^{\ \ k} \lambda \right)\right] +  \lambda^2 (I_{ij} + \bar{P})^2 \\ \nonumber
    &=  \left( w_{k}R_{ij}^{\ \ k} \lambda \right)^2 +\left( w_{k}R_{ij}^{\ \ k} \lambda \right) - 2 \lambda^2 (I_{ij} + \bar{P})   \\ \nonumber
    & \ \times w_{k}R_{ij}^{\ \ k}   +  \lambda^2 (I_{ij} + \bar{P})^2 \\
    &=  \lambda^2 \left( w_{k}R_{ij}^{\ \ k} - (I_{ij} + \bar{P}) \right)^2 +  \lambda w_{k}R_{ij}^{\ \ k}, \label{eq:NumericallyOptimisedGPwPoissonNoise}
\end{align}
where $\bar{P} = J^{ij}w_{k}R_{ij}^{\ \ k}/(nm) = N' \text{E}[w] \text{E}[R] $. We recognize the left-hand term as the variance contribution of the random-basis reconstruction, scaled by the number of photons squared.  The right-hand term is the contribution due to Poisson noise -- i.e.~the more photons we allocate to create the contrast, the more the second term decays in contribution. This result still contains a pixel-wise dependence, whereas we seek an expression for the global variance of the ghost projection. To determine this, we can numerically approximate it by taking the average of the above result. We can then use this in a pixel-wise SNR expression, which can then be transformed into a global SNR by numerically taking the RMS.

\subsection{Ghost Projection with Poisson Noise and Numerically Derived Weights Simulation}

\begin{figure*}[ht!]
     \centering
     \begin{subfigure}{0.329\textwidth}
         \centering
         \includegraphics[width=\textwidth]{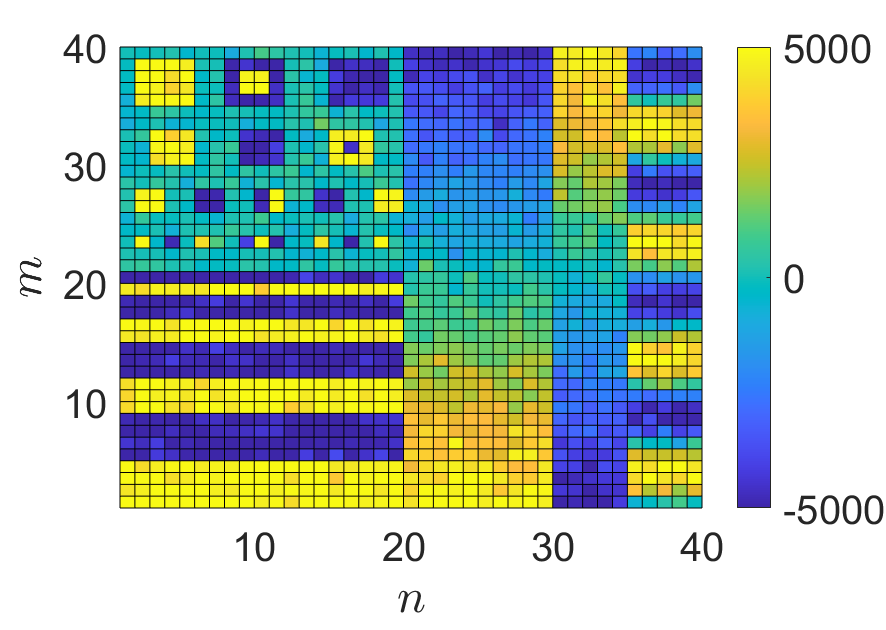}
         \caption{ }
         \label{subfig:Poisson Optimised 1}
     \end{subfigure}
     \begin{subfigure}{0.329\textwidth}
         \centering
         \includegraphics[width=\textwidth]{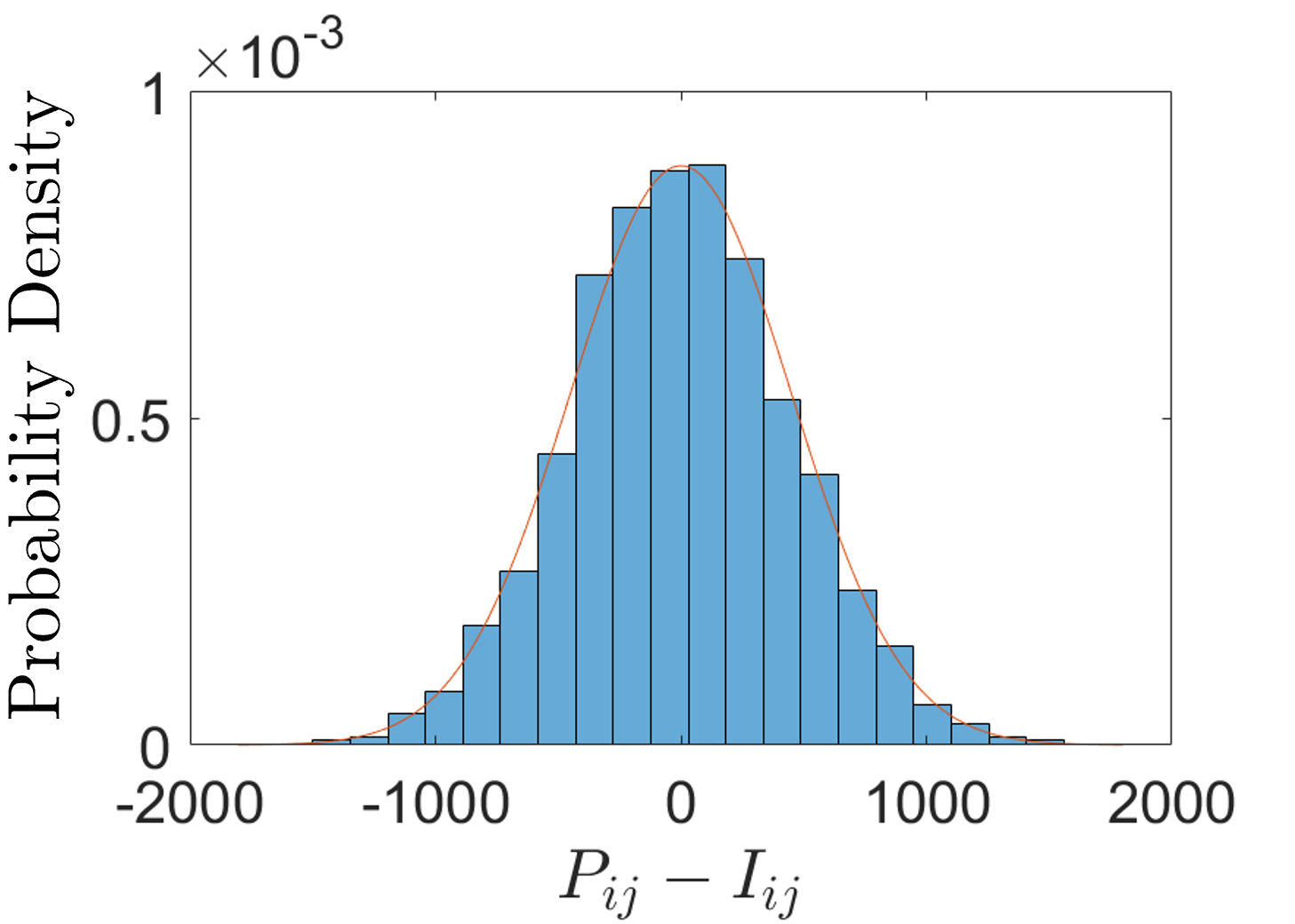}
         \caption{ }
         \label{subfig:Poisson Optimised 2}
     \end{subfigure}
     \begin{subfigure}{0.329\textwidth}
         \centering
         \includegraphics[width=\textwidth]{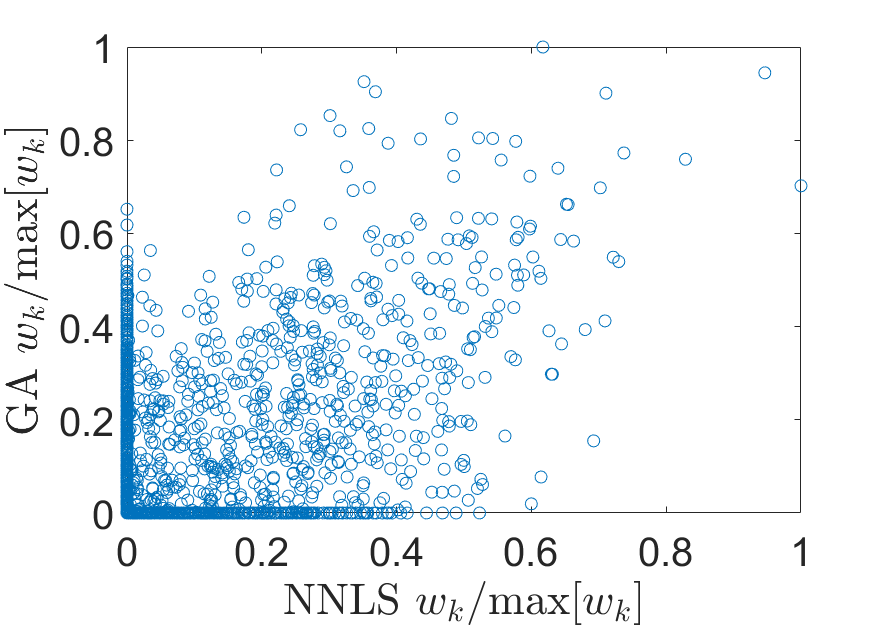}
         \caption{ }
         \label{subfig:Poisson Optimised 3}
     \end{subfigure}
        \caption{(a) Numerically optimized ghost projection with Poisson noise for $\lambda = 5,000$, which produced an SNR of 7.90 in simulation and had a pedestal of $\bar{P} = 40.25$ (which can be multiplied by $\lambda$ to obtain the pedestal in units of photon counts). (b) The noise obtained in the ghost projection subject to Poisson noise, overlaid with the predicted noise distribution, i.e. Gaussian statistics with a variance obtained by spatially averaging Eq.~(\ref{eq:NumericallyOptimisedGPwPoissonNoise}). (c) Comparison of NNLS numerical weights which optimized the random-matrix reconstruction and the GA weights which started from the NNLS weights and included Poisson noise in the numerical optimization.}
        \label{fig:Poisson Optimised}
\end{figure*}

In the context of ghost projection with Poisson noise and numerically derived weights, we have two points of comparison: the lower-bound will be the numerically optimized reconstruction which subsequently has Poisson noise added, and the upper-bound will be the SNR obtainable from a conventional, direct mask-based projection in the presence of Poisson noise. Starting, for example, with a basis $N=5nm$, we can seek the weights that optimize the random-matrix basis reconstruction of the desired image. This was done using $\mathtt{MATLAB}$'s inbuilt NNLS solver, thereby obtaining a noise free SNR of $4.58 \times 10^{9}$ and a pedestal of $\bar{P} = 58.07$. Once Poisson noise was included, however, the SNR decayed to 6.49 where $\lambda = 5$,000. The maximum attainable SNR, with this number of photons and for this image, is $\sqrt{\lambda} \approx 70.71$ (using a result derived later in Sec.~\ref{subsec:Comparison of Direct Projection and Ghost Projection}). Employing a Gradient Ascent (GA) algorithm to search for those weights that can optimally trade-off the pedestal's Poisson noise with excess SNR in the random-matrix basis reconstruction, we obtained an SNR of 7.90 and a pedestal of $\bar{P} = 40.25$ (see Fig.~\ref{subfig:Poisson Optimised 1}), which represents a 21.7\% improvement in SNR over that obtained from the weights derived from the NNLS solver, but a reduction in SNR by a factor of 8.95 compared to the maximum possible. Note, this comparison of SNR improvement---in going from (i) NNLS-derived weights that subsequently have Poisson noise added to (ii) GA-derived weights including Poisson noise---is particular to each random-matrix basis realization and desired image, and is only illustrative of what general improvements can be obtained.  Further, the numerically-obtained histogram in Fig.~\ref{subfig:Poisson Optimised 2} is consistent with the curve corresponding to the average of Eq.~(\ref{eq:NumericallyOptimisedGPwPoissonNoise}).  Finally, we can observe the difference in the numerically optimized weights obtained for the Poisson-noise-free case (NNLS solver) and those obtained in the presence of Poisson noise (GA solver), in Fig.~\ref{subfig:Poisson Optimised 3}. Interestingly, there is next to no relationship between the two sets of weights. Moreover, many basis members that were assigned a zero weight in the Poisson-noise-free case (NNLS solver) were assigned a non-negative weight when in the presence of Poisson noise (GA solver).

% These figures were generated with the MATLAB code called GP_GradAscPoisson.m
% NNLS SNR with Poisson Noise = 6.494611 
% NNLS Pbar = 58.070116 
% NNLS SNR without Poisson Noise = 4575979879.111251 
% Gradient Ascent SNR with Poisson Noise = 7.902867 
% Gradient Ascent Pbar = 40.249856 
% Gradient Ascent SNR without Poisson Noise = 86.098200 
% an SNR improvement of 21.68 percentage

\section{Discussion}

\subsection{Basis Representations} \label{subsec:Basis Representations}

Ghost projection synthesizes signals using a random basis, and may therefore be viewed from this more general perspective.  To this end, consider an $nm$-length vector $\vec{I}$, where $\vec{I}$ could just as easily be a reshaped matrix, or tensor. Further, define $M$ to be a matrix with vectorized basis members as columns, and let $\vec{w}$ be a vector of weighting coefficients.  We then have the following list of vector representations: 
\begin{enumerate}

    \item A orthonormal basis comprising $nm$ members, where $\vec{w} = M^{\text{T}} \vec{I} =M^{-1}\vec{I}$, such that $M\vec{w} = \vec{I}$. 
    
    \item A non-orthogonal basis with $nm$ linearly-independent members (which could be transformed into a orthonormal basis via e.g.~the Gram–Schmidt process), with weights found via $\vec{w} = (M^{\text{T}}M)^{-1}M^{\text{T}}\vec{I}$ such that $M\vec{w} = \vec{I}$.
    
    \item An over complete, non-orthogonal basis comprising greater than $nm$ basis members. Such a basis set will admit a family of solutions of $\vec{w} = (M^{\text{T}}M)^{-1}M^{\text{T}}\vec{I}$ such that $M\vec{w} = \vec{I}$. %An example of such is the coherent-state basis employed in quantum optics, quantum mechanics, quantum field theory, condensed matter physics and quantum aspects of general relativity \cite{mandel1995optical}.
    
    \item An over-complete, non-orthogonal basis where the weights are the correlation coefficients between the basis members and desired image, such as in the simplest forms of ghost imaging \cite{ceddia2018random}. That is, the weights in this case are $\vec{w} = \frac{1}{ N \text{Var}[R]} ( M^{\text{T}} \vec{I} - \overline{M^{\text{T}} \vec{I}} \hspace{0.05cm})$ such that $M\vec{w} \sim \vec{I}$ in the limit of large $N$. Moreover, a finite set of non-orthogonal basis members with correlation-coefficient weights can always be transformed into the non-orthogonal projection weights via $(M^{\text{T}}M)^{-1}$.
    
    \item A set of $(nm+1)$ basis members arranged according to Appendix \ref{App:C Minimum number of basis vectors}, which details a proof for the existence of non-negative weights and constitutes the minimum number of members required to guarantee a non-negative solution. Assuming the criterion for the arrangement of basis members is met, the representation of $\vec{I}$ is unique. To find the weights in this case, see the next item in this list. 
    
    \item A twice-orthonormal basis comprising two orthonormal sets, one being the negative (mirrored version) of the other, where the weights are restricted to be non-negative values $w_k \geq 0 \ \forall \  k \in [1,2nm]$. In this construction, there will exist a unique representation of the vector $\vec{I}$. Finding the weights in this case can be achieved by taking the projection of $\vec{I}$ in one orthonormal basis set and then, for each negative weight, switching that basis member for its mirrored counterpart in the other set and removing the negative sign on that weight. Interestingly, in the space of non-negative coefficients, unique solutions exist in $nm$-dimensions for $nm+1$ basis members and $2nm$ basis members. To find the weights in the $(nm+1)$ case, in Appendix \ref{App:C Minimum number of basis vectors}, we detail how to decompose the twice-orthonormal set in terms of the $(nm+1)$ basis. That is, the weights of $\vec{I}$ in the $(nm+1)$ case are: those obtained by taking an orthonormal projection, switching the negative weights for the negative basis member, and multiplying these by the decomposition coefficients.
    
    \item A twice over-complete basis comprising two sets of $nm$ non-orthogonal linearly-independent basis vectors, where one set is a mirrored version of the other and the weights are restricted to non-negative values $w_k \geq 0 \ \forall \ k \in [1,2nm]$. Since this construction can be transformed into the previous construction of this list, for this case too there will exist a unique representation for $\vec{I}$. The weights in this case can be found by taking the non-orthogonal projection in one basis, and switching those basis members which receive a negative weight with those in the mirrored set.
    
    \item A more than twice over-complete, non-orthogonal basis comprising more than $2nm$ basis members, where the weights are non-negative $w_k \geq 0 $ for all $k$. Note, each basis vector can have a constant added to it, such that the entire basis set is non-negative too. This will have the effect of reproducing the image up to an additive constant or pedestal $\bar{P}$, i.e.~$M\vec{w} = \vec{I} + \bar{P} \vec{J}$. This representation has applications for cases which require positive weights and positive basis members, and are insensitive to an additive constant.
    
    \item An over complete, non-orthogonal basis where the weights are asymmetrically skewed towards those basis members that are correlated with the desired image. This could be as in the first analytical case of this paper and linear weights, or, as in the second analytical case, a step function that filters out basis members and prescribes either zero or uniform, positive value. More generally still, any weighting scheme that displays appropriate asymmetry in the weights should asymptotically converge to the desired image, up to an additive constant, in the limit of a sufficiently large number of basis members \cite{paganin2019writing}.
    
\end{enumerate}

\subsection{Comparison of Direct Projection and Ghost Projection} \label{subsec:Comparison of Direct Projection and Ghost Projection}

Conventional image projection illuminates a specifically fabricated mask for an appropriate length of time. Such a scheme, with the inclusion of Poisson noise, has the SNR relationship:
\begin{equation*}
\text{SNR}_{ij} = \lambda I_{ij} /\sqrt{\lambda I_{ij}} = \sqrt{\lambda I_{ij}}, 
\end{equation*}
where $\lambda$ is the number of photons per pixel in the illumination beam, and $I_{ij}$ is the desired image expressed as a transmission coefficient between 0 and 1. This leads to the global SNR of $\sqrt{\lambda \text{E}[I]}$. Comparing to the ghost-projection case of:
\begin{equation*}
\text{SNR} \approx \sqrt{ \frac{\lambda  \text{E} [I^2]}{ N' \text{E}[w] \text{E}[R] }}, 
\end{equation*}
where $I$ is here considered to be zero-centered, we concede that it is unlikely that ghost projection will ever outperform direct projection techniques.  Similar conclusions have been drawn in the context of classical ghost imaging \cite{ceddia2018random,Gureyev2018,LaneRatner2020,Kingston2021}. This comparison between direct and ghost projection does presuppose, however, that a purpose built mask for direct projection already exists, or is readily constructible. Moreover, there are some regimes of electromagnetic radiation or matter-wave fields where constructing such a specific mask can be physically infeasible due to unacceptably high aspect ratios and/or precision machining required to create the correct attenuation of the transmitted field. Whilst ghost projection may be suboptimal compared to direct projection when considering the deleterious effects of Poisson noise, it has the advantage of being universally applicable for any desired image, using any illuminating field for which a spatially-random mask is available.  

Another advantage is that ghost projection is inherently free from the need for proximity correction, associated with the transverse redistribution of transmitted-beam energy upon coherent propagation from the exit surface of the mask to the projection plane \cite{Bourdillon2000,Bourdillon2001,Vladimirsky1999}. In a ghost-projection setting, propagation distance is relative to where one measures the random-matrix basis, and not the position of the mask used to generate the random-matrix basis. This means that if the substrate being projected onto is placed in the same location as the measuring plane of the random-matrix basis (speckle field), no propagation effects need be accounted for. Indeed, propagation effects may be used advantageously to sharpen speckles produced by a spatially-random mask, via propagation-based phase contrast \cite{Bremmer1952,Snigirev1995,Wilkins2014}. 

\subsection{Practical Random Masks}

Here we offer several comments regarding the practicalities of constructing random masks for ghost projection.

\begin{enumerate}

\item A single spatially-random master mask may be transversely scanned, to create a random-mask ensemble for the purposes of ghost projection \cite{paganin2019writing}.  Moreover, as pointed out by K.~S.~Morgan, two or more consecutive simultaneously-illuminated  independently-transversely-scanned known random master masks may be employed.\footnote{Private communication from Kaye S.~Morgan (Monash University, Australia) to the authors, on June 24, 2021.}  This exponentially increases the number of known random masks that may be generated using a very small number of known random master masks. That is, we may image the master masks independently and then computationally reconstruct the correspondingly much larger data-base of possible random masks -- the experimental requirements scale linearly whilst the gain in number of masks is exponential in the number of consecutive master masks.  For example, if a single random mask can be scanned to $10^3$ different transverse locations in each of two orthogonal transverse directions, then (i) $(10^3)^2=10^6$ different random masks can be generated using one transversely scanned random master mask, but (ii) $(10^6)^2=10^{12}$ different masks can be generated using two consecutive independently-translatable random master masks with the same transverse positioning capability. For ghost-projection schemes in which basis-filtration is employed, using say $10^4$ post-filtration random masks, then for the example two-master-mask scheme given above, it is feasible that $10^4/10^{12}=0.000001\%$ may be retained, out of the set of all possible masks.  Such extreme filtration may assist with optimizing quality metrics such as SNR and contrast-to-offset ratio.      

% Note, in the context of the private communication that is mentioned above: On June 25, 2021, Kaye Morgan emailed the following text: (a) ``I was thinking about how two (or more) consecutive masks might make it easier to maintain zeros, since `mask 2' could block the bright areas in `mask 1'.''; (b) ``Multiple phase masks placed at specific distances could further address this ...''.

\item  For speckle-field realizations of the spatially-random ghost-projection masks, K.~S.~Morgan has noted that there is some flexibility in choosing the form of the mask spatial power spectrum.\footnote{Ibid.} It is likely that the manner in which radial spatial power spectra decay, with high radial spatial frequency, will influence ghost-projection quality metrics such as SNR and spatial resolution.  Power-law decays, associated with random-fractal screens, may be more favorable than e.g.~Gaussian-type decay, for the purposes of ghost-projection resolution. Some parallels may be drawn, here, with recent work on the use of pink-noise speckle patterns in the context of computational ghost imaging \cite{PinkNoiseGI2021}. 

%Cf. Kaye Morgan's email from June 25, 2021: `` ... are there certain types of masks that are most helpful – e.g. ones that have every feature size/spatial frequency present? Or relatively smooth masks to minimise high-frequency noise? It is likely hard to manufacture a particular pattern, but easier to choose the length scales present in a random pattern.''

\item  D.~Pelliccia and S.~D.~Findlay have noted that it would be interesting to investigate the tolerance of ghost projection, to inaccuracies in the transverse positioning of the illuminated mask or masks.\footnote{Private communications from Daniele Pelliccia (Instruments \& Data Tools, Australia) and Scott D.~Findlay (Monash University, Australia) to the authors, on June 23 and June 24, 2021, respectively.}

\item  For color ghost projection,  a small polyenergetic source can be used to illuminate a thick spatially-random screen, namely a screen that is sufficiently thick for strong multiple scattering to be operative.  For each transverse position of such a source-plus-screen combination, an independent speckle field will be generated for each energy band \cite{FinkSpeckles2012}.  

\end{enumerate}

\subsection{Combined raster scanning and ghost projection}

Since SNR is typically inversely proportional\footnote{In the analytical cases of weighting and filtering the basis according to the pseudo-correlation coefficient, $\text{SNR} \propto 1/(nm)$ and $\text{SNR} \propto 1/\sqrt{nm}$, respectively. For the numerically optimized case, the relationship between SNR and resolution is a more complex one which depends on the number of basis members available. If $N \gg nm$, then the SNR will not necessarily be degraded, although the computational expense of optimization will significantly increase. If $N \approx O(nm)$, then the SNR will degrade with increased resolution.} to the resolution $nm$ of the projected image, a combination of raster scanning and ghost projection could be employed to improve SNR. This would give a contiguous array of ghost projections, in a manner analogous to the ghost-imaging scheme in Kingston {\em et al.}~\cite{Kingston2020}. 

\subsection{Ghost 2D and 3D Photocopier}

An application of ghost projection would be to construct a ghost photocopier where one could ghost image and ghost project an object in the same set up. Interestingly, in this process one would not necessarily need to image the random-basis set, as explained below.  

\begin{figure*}[ht!]
    \hspace{2cm}
     \begin{subfigure}{\textwidth}
         \includegraphics[width=0.7\textwidth]{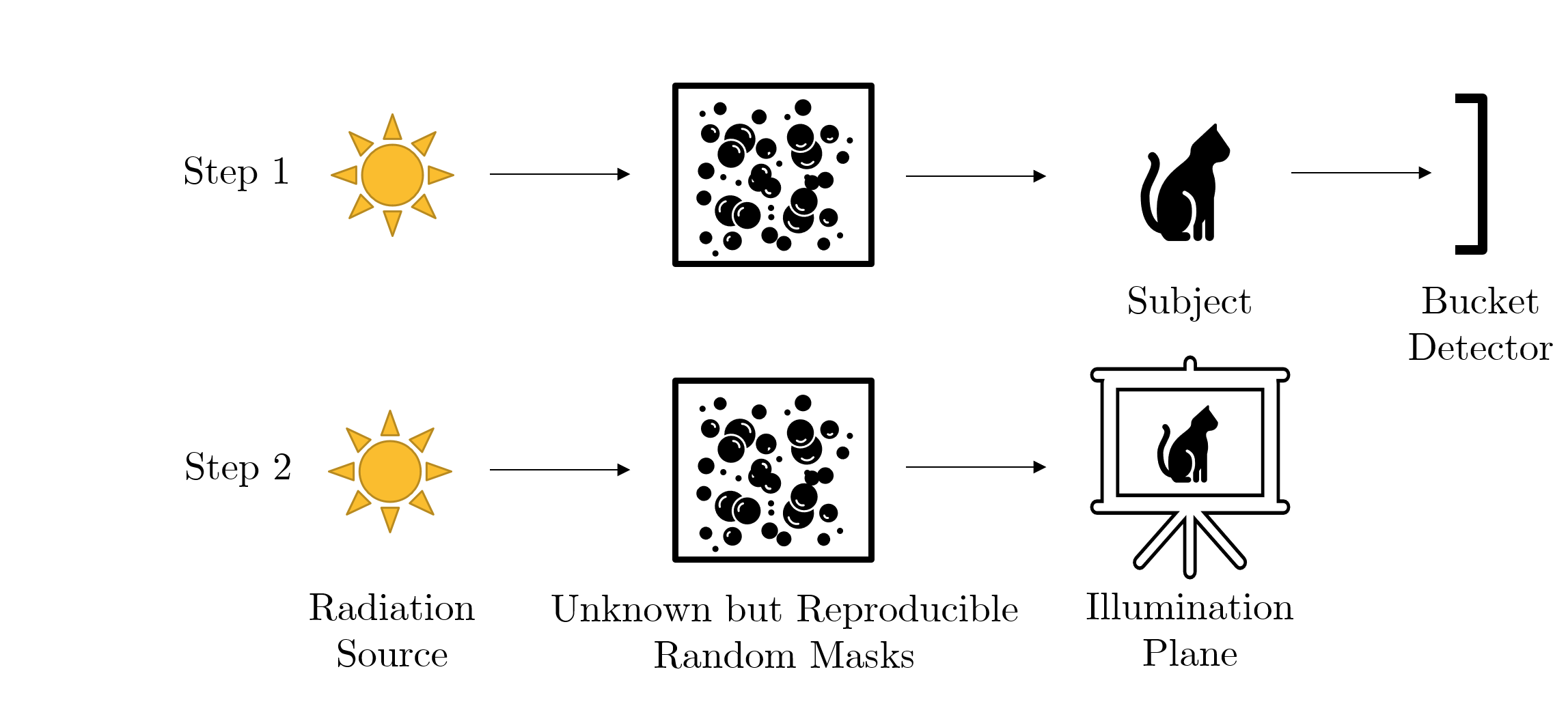}
         \caption{ }
         \label{subfig:discussion schematic 1}
     \end{subfigure} 
     \begin{subfigure}{\textwidth}
         \includegraphics[width=0.7\textwidth]{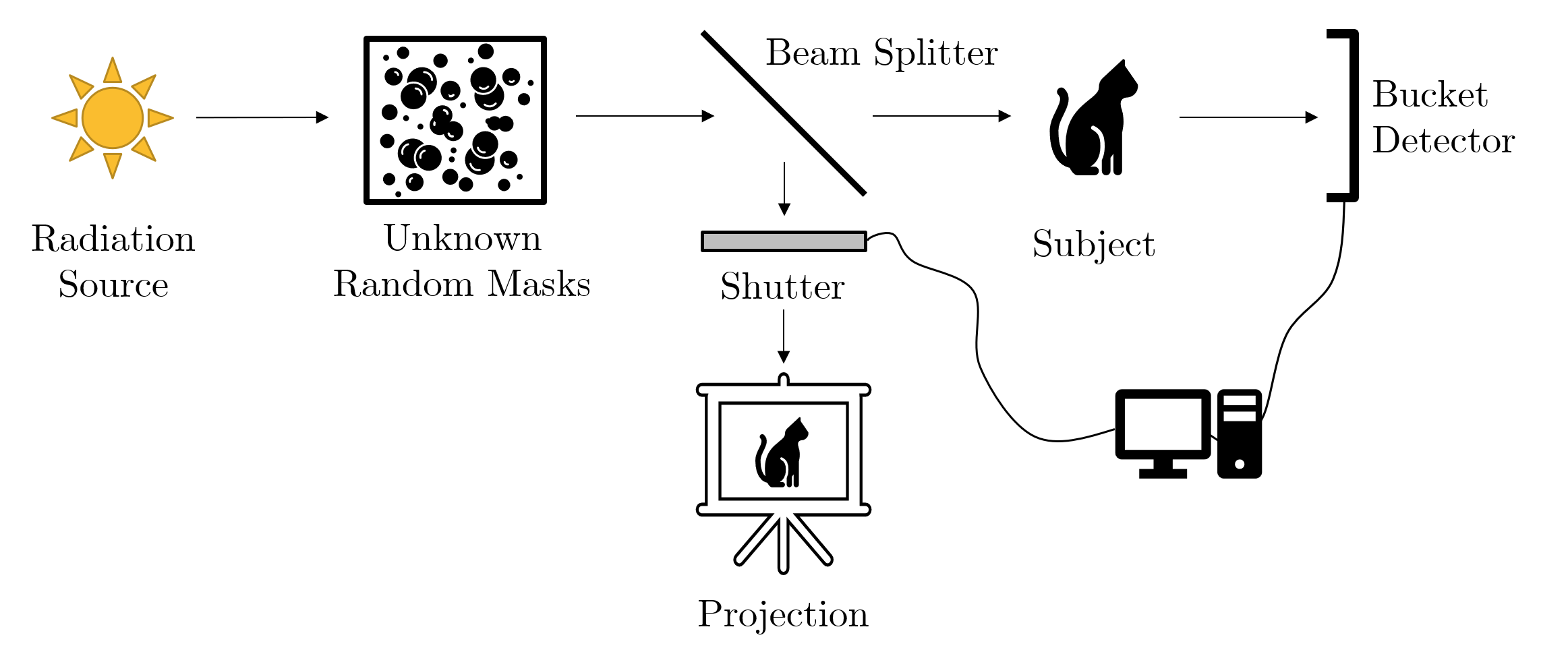}
         \caption{ }
         \label{subfig:discussion schematic 2}
     \end{subfigure}
\caption{Ghost-photocopier schemes for ghost projection. (a) Two-step scheme.  In step 1, a ghost-imaging experiment is performed, with an unknown sample illuminated by a set of reproducible but otherwise unknown random masks $\{R_{ijk}\}$, to give an associated set of bucket signals $\{C_k\}$. In step 2, the measured bucket signals are used to determine the illumination times $\{t_k\}$ for the ghost-projection step of the ghost photocopier. (b) Single-step scheme, in which the time-dependent bucket signal from the ghost-imaging arm is used to determine the shutter exposure time in the ghost-projection arm.  
\label{fig:GhostPhotocopier}}
\end{figure*}

A two-step ghost-photocopier scheme is shown in Fig.~\ref{fig:GhostPhotocopier}(a).  Step 1 corresponds to a conventional mask-based classical ghost-imaging setup using a reproducible but otherwise unknown set of  random masks $\{R_{ijk}\}$.  This enables the pseudo-correlation coefficients $\{C_k\}$, which are equivalent to suitably normalized bucket signals in the terminology of ghost imaging (cf.~Eq.~(\ref{eq:t_k})), to be measured.  Step 2 of this scheme uses the same set of masks to perform ghost projection, with mask exposure times $\{t_k\}$ chosen based on the previously measured pseudo-correlation coefficients (see e.g.~Eq.~(\ref{eq: Standard Ghost Projection})).  If the masks are not known, one of the analytical schemes in the present paper can be used for the ghost-projection step.  Conversely, if the masks are previously measured or known {\em a priori}, any of the schemes (analytical or numerical) could be used for the ghost-projection step.

A single-step ghost-photocopier scheme is shown in Fig.~\ref{fig:GhostPhotocopier}(b).  Here a radiation source illuminates a series of unknown random masks, with the beam transmitted by each mask being passed through a beamsplitter.  An object is placed in one arm of the beamsplitter, enabling a time-dependent bucket signal to be measured.  Since this bucket signal is proportional to the pseudo-correlation coefficient, it can be used to drive the exposure time of a shutter in the other, ghost projection, arm of the setup.  Moreover, if the coherence time of the source is comparable to or larger than the shutter response time, one could dispense with the masks altogether and instead employ the spatiotemporal speckles associated with partially coherent sources, to generate the required ensemble of random masks \cite{Alperin2019,PaganinSanchezDelRio2019,PaganinPelliccia2021}.  

Finally, we mention that the schemes outlined here could be performed in a 2D sense, as already indicated, or made into a 3D-printing version for e.g.~volumeteric additive manufacturing \cite{Beer2019,TomographyInReverse2019}, via an extension of the concept of ghost tomography \cite{KingstonOptica2018,KingstonIEEE2019,paganin2019writing}.

\subsection{Numerical Weights} \label{subsec:discussion numerical weights}

Given the orders-of-magnitude improvement in ghost-projection SNR using numerically derived weights in comparison to the  analytical treatment, we can trade some of this SNR improvement for additional desired attributes. Examples include:
\begin{itemize}
\item a reduction of the number of members in the required basis set, 
\item a reduction in the magnitude of the pedestal to which the projected image is added, 
\item imposing the constraint of a minimum mask-weight value, corresponding to a minimum non-zero mask exposure time, 
\item maximizing SNR for the case of uniform weights being applied to all masks. 
\end{itemize}
Supposing the number of elements in the starting basis to be sufficiently large (e.g.~on the order of 5 times the resolution of the desired ghost projection), it may be realistic to seek many of the listed attributes to be attained, while maintaining an acceptably high SNR in the ghost projection. 

For the analytical treatments in Secs.~\ref{sec:Naive Ghost Projection with a Random Matrix Basis}-\ref{sec:Colour Ghost Projection}, we derived results for the expected SNR given the basis size, projection resolution and various contaminating noise contributions. Several of these questions remain open for ghost projection with numerically-derived weights: 
\begin{itemize}
\item Given $N$, $n$, $m$ and $I$, what SNR should we expect? 
\item How do the numerical-weights and ghost-projection SNR depend on the statistical characteristics of the random-matrix basis, e.g.~on the choice of statistical distribution (uniform, binomial, Poissonian, Gaussian, etc.), variable overlap between matrices, the spatial power spectrum of the random basis, or the expected value $\text{E}[R]$ and variance $\text{Var}[R]$ of the basis?
\item How do the numerical-weights and ghost-projection SNR depend on aspects of the desired image such as its norm $\text{E}[I^2]$ and entropy \cite{Yu2017}? 
\item What noise-resolution uncertainty principle applies to numerical-weights ghost projection?  
\item What size basis set will, to within a specified confidence interval, ghost-project  any image to a desired SNR?
\item How might the numerical-optimization ghost-projection scheme be extended to the case of color-image ghost projection?
\end{itemize}

\section{Conclusion}

Ghost projection is an indirect projection technique that creates a desired distribution of radiant exposure via a suitably-weighted linear combination of illuminated random masks. The method is so termed on account of its conceptual parallels with ghost imaging using randomly-patterned masks.

Two analytical cases of ghost projection were investigated, namely, a linearly weighted sum of random matrices and a filtered sum of random matrices. In either case, both weights and filtering were based on the pseudo-correlation coefficient between the desired image and a specified ensemble of random matrices. These two analytical cases, while providing both conceptual clarity and continuity with our previous work on ghost projection \cite{paganin2019writing}, have rather low inherent projection efficiency.  Accordingly, these two analytical cases of ghost projection were augmented by numerically-optimized filtration and weights, which were seen to significantly outperform the analytically tractable cases in terms of inherent projection efficiency.

To gain a more accurate impression of how ghost projection will perform in practice, two sources of noise were considered. Poisson noise was used to model shot noise in the photon counts. Gaussian noise was used to model errors in the exposure shutter-opening times which physically realize the weighting coefficients for each illuminated random mask. With these realistic sources of noise, ghost projection can still perform adequately for certain practical applications, e.g.~achieving an SNR on the order of 10 for a contrast on the order of 10,000 photon counts with a pedestal on the order of 100,000 counts. 

As well as these monochromatic cases of ghost projection, we introduced the concept of color ghost projection, whereby a desired spatial distribution of radiant exposure could be achieved for a number of different energy channels.  This could be achieved, for example, by illuminating a single thick spatially-random screen using a broadband source that is sufficiently small in spatial extent (i.e.~sufficiently spatially coherent) to create independent speckle images for each energy channel. 

In addition to its fundamental interest as a means to `build signals out of noise', several possible future applications provide broad motivation for this work.  The ghost-projection concept may be useful for constructing data projectors for radiation and matter wave fields (e.g.~hard x rays and neutrons) for which such devices either do not exist or have limited spatial resolution.  Lithography using e.g.~short-wavelength electromagnetic radiation, as well as atomic or molecular beams, is another potential application.  Three-dimensional printing or volumetric additive manufacturing via reverse tomography, particularly for electromagnetic radiation at extreme ultra-violet and shorter wavelengths, constitutes another possible avenue for future application.  One final potential application is ghost photocopying, which hybridizes ghost-imaging measurement of a sample and subsequent ghost-projection of the same sample.

A useful feature of ghost projection is its ability to employ a single spatially-random screen, as a universal mask for a wide variety of radiation and matter wave fields. Another advantage is its ability to utilize propagation-based phase contrast in a constructive manner, to sidestep the need for high aspect-ratio attenuation masks as well as being immune to the proximity-correction problem. These advantages of ghost projection should be balanced against the drawback of a significant pedestal of uniform exposure, upon which the desired contrast sits. While this pedestal may prove prohibitive for certain applications, it can be reduced and in many cases, adapted for. 

\section*{Acknowledgments}

We acknowledge useful discussions with Christian Dwyer, Scott Findlay, Kristian Helmerson, Andrew Kingston, Kaye Morgan and Daniele Pelliccia. D.~C.~acknowledges funding via an Australian Postgraduate Award. D.~M.~P.~acknowledges funding via Australian Research Council Discovery Project DP21010312.

\bibliography{Ghost_Projection_bib}

\begin{thebibliography}{56}%
\makeatletter
\providecommand \@ifxundefined [1]{%
 \@ifx{#1\undefined}
}%
\providecommand \@ifnum [1]{%
 \ifnum #1\expandafter \@firstoftwo
 \else \expandafter \@secondoftwo
 \fi
}%
\providecommand \@ifx [1]{%
 \ifx #1\expandafter \@firstoftwo
 \else \expandafter \@secondoftwo
 \fi
}%
\providecommand \natexlab [1]{#1}%
\providecommand \enquote  [1]{``#1''}%
\providecommand \bibnamefont  [1]{#1}%
\providecommand \bibfnamefont [1]{#1}%
\providecommand \citenamefont [1]{#1}%
\providecommand \href@noop [0]{\@secondoftwo}%
\providecommand \href [0]{\begingroup \@sanitize@url \@href}%
\providecommand \@href[1]{\@@startlink{#1}\@@href}%
\providecommand \@@href[1]{\endgroup#1\@@endlink}%
\providecommand \@sanitize@url [0]{\catcode `\\12\catcode `\$12\catcode
  `\&12\catcode `\#12\catcode `\^12\catcode `\_12\catcode `\%12\relax}%
\providecommand \@@startlink[1]{}%
\providecommand \@@endlink[0]{}%
\providecommand \url  [0]{\begingroup\@sanitize@url \@url }%
\providecommand \@url [1]{\endgroup\@href {#1}{\urlprefix }}%
\providecommand \urlprefix  [0]{URL }%
\providecommand \Eprint [0]{\href }%
\providecommand \doibase [0]{https://doi.org/}%
\providecommand \selectlanguage [0]{\@gobble}%
\providecommand \bibinfo  [0]{\@secondoftwo}%
\providecommand \bibfield  [0]{\@secondoftwo}%
\providecommand \translation [1]{[#1]}%
\providecommand \BibitemOpen [0]{}%
\providecommand \bibitemStop [0]{}%
\providecommand \bibitemNoStop [0]{.\EOS\space}%
\providecommand \EOS [0]{\spacefactor3000\relax}%
\providecommand \BibitemShut  [1]{\csname bibitem#1\endcsname}%
\let\auto@bib@innerbib\@empty
%</preamble>
\bibitem [{\citenamefont {Hecht}(2017)}]{HechtOpticsBook}%
  \BibitemOpen
  \bibfield  {author} {\bibinfo {author} {\bibfnamefont {E.}~\bibnamefont
  {Hecht}},\ }\href@noop {} {\emph {\bibinfo {title} {Optics}}},\ \bibinfo
  {edition} {5th}\ ed.\ (\bibinfo  {publisher} {Pearson, Boston},\ \bibinfo
  {year} {2017})\BibitemShut {NoStop}%
\bibitem [{\citenamefont {Johnson}(1960)}]{johnson1960}%
  \BibitemOpen
  \bibfield  {author} {\bibinfo {author} {\bibfnamefont {B.~K.}\ \bibnamefont
  {Johnson}},\ }\href@noop {} {\emph {\bibinfo {title} {Optics and Optical
  Instruments}}}\ (\bibinfo  {publisher} {Dover Publications, New York},\
  \bibinfo {year} {1960})\BibitemShut {NoStop}%
\bibitem [{\citenamefont {Pennycook}\ and\ \citenamefont
  {Nellist}(2011)}]{Pennycook2011}%
  \BibitemOpen
  \bibinfo {editor} {\bibfnamefont {S.~J.}\ \bibnamefont {Pennycook}}\ and\
  \bibinfo {editor} {\bibfnamefont {P.~D.}\ \bibnamefont {Nellist}},\ eds.,\
  \href@noop {} {\emph {\bibinfo {title} {Scanning Transmission Electron
  Microscopy: Imaging and Analysis}}}\ (\bibinfo  {publisher} {Springer, New
  York},\ \bibinfo {year} {2011})\BibitemShut {NoStop}%
\bibitem [{\citenamefont {Trundle}(2000)}]{Trundle2001}%
  \BibitemOpen
  \bibfield  {author} {\bibinfo {author} {\bibfnamefont {E.}~\bibnamefont
  {Trundle}},\ }\href@noop {} {\emph {\bibinfo {title} {Newnes Guide to
  Television and Video Technology}}},\ \bibinfo {edition} {3rd}\ ed.\ (\bibinfo
   {publisher} {Newnes, Oxford},\ \bibinfo {year} {2000})\BibitemShut {NoStop}%
\bibitem [{\citenamefont {Chen}(2015)}]{Chen2015}%
  \BibitemOpen
  \bibfield  {author} {\bibinfo {author} {\bibfnamefont {Y.}~\bibnamefont
  {Chen}},\ }\bibfield  {title} {\bibinfo {title} {Nanofabrication by electron
  beam lithography and its applications: A review},\ }\href
  {https://doi.org/https://doi.org/10.1016/j.mee.2015.02.042} {\bibfield
  {journal} {\bibinfo  {journal} {Microelectron. Eng.}\ }\textbf {\bibinfo
  {volume} {135}},\ \bibinfo {pages} {57} (\bibinfo {year} {2015})}\BibitemShut
  {NoStop}%
\bibitem [{\citenamefont {Bourdillon}\ \emph {et~al.}(2000)\citenamefont
  {Bourdillon}, \citenamefont {Boothroyd}, \citenamefont {Kong},\ and\
  \citenamefont {Vladimirsky}}]{Bourdillon2000}%
  \BibitemOpen
  \bibfield  {author} {\bibinfo {author} {\bibfnamefont {A.~J.}\ \bibnamefont
  {Bourdillon}}, \bibinfo {author} {\bibfnamefont {C.~B.}\ \bibnamefont
  {Boothroyd}}, \bibinfo {author} {\bibfnamefont {J.~R.}\ \bibnamefont
  {Kong}},\ and\ \bibinfo {author} {\bibfnamefont {Y.}~\bibnamefont
  {Vladimirsky}},\ }\bibfield  {title} {\bibinfo {title} {A critical condition
  in {F}resnel diffraction used for ultra-high resolution lithographic
  printing},\ }\href {https://doi.org/10.1088/0022-3727/33/17/307} {\bibfield
  {journal} {\bibinfo  {journal} {J. Phys. D: Appl. Phys.}\ }\textbf {\bibinfo
  {volume} {33}},\ \bibinfo {pages} {2133} (\bibinfo {year}
  {2000})}\BibitemShut {NoStop}%
\bibitem [{\citenamefont {Bourdillon}\ and\ \citenamefont
  {Boothroyd}(2001)}]{Bourdillon2001}%
  \BibitemOpen
  \bibfield  {author} {\bibinfo {author} {\bibfnamefont {A.~J.}\ \bibnamefont
  {Bourdillon}}\ and\ \bibinfo {author} {\bibfnamefont {C.~B.}\ \bibnamefont
  {Boothroyd}},\ }\bibfield  {title} {\bibinfo {title} {Proximity correction
  simulations in ultra-high resolution x-ray lithography},\ }\href
  {https://doi.org/10.1088/0022-3727/34/22/301} {\bibfield  {journal} {\bibinfo
   {journal} {J. Phys. D: Appl. Phys.}\ }\textbf {\bibinfo {volume} {34}},\
  \bibinfo {pages} {3209} (\bibinfo {year} {2001})}\BibitemShut {NoStop}%
\bibitem [{\citenamefont {Svalbe}\ \emph {et~al.}(2020)\citenamefont {Svalbe},
  \citenamefont {Paganin},\ and\ \citenamefont {Petersen}}]{Svalbe2020}%
  \BibitemOpen
  \bibfield  {author} {\bibinfo {author} {\bibfnamefont {I.~D.}\ \bibnamefont
  {Svalbe}}, \bibinfo {author} {\bibfnamefont {D.~M.}\ \bibnamefont
  {Paganin}},\ and\ \bibinfo {author} {\bibfnamefont {T.~C.}\ \bibnamefont
  {Petersen}},\ }\bibfield  {title} {\bibinfo {title} {Sharp computational
  images from diffuse beams: Factorization of the discrete delta function},\
  }\href {https://doi.org/10.1109/TCI.2020.3007549} {\bibfield  {journal}
  {\bibinfo  {journal} {IEEE Trans. Comput. Imaging}\ }\textbf {\bibinfo
  {volume} {6}},\ \bibinfo {pages} {1258} (\bibinfo {year} {2020})}\BibitemShut
  {NoStop}%
\bibitem [{\citenamefont {Bracewell}(1986)}]{Bracewellbook}%
  \BibitemOpen
  \bibfield  {author} {\bibinfo {author} {\bibfnamefont {R.~N.}\ \bibnamefont
  {Bracewell}},\ }\href@noop {} {\emph {\bibinfo {title} {The Fourier Transform
  and its Applications}}},\ \bibinfo {edition} {2nd}\ ed.\ (\bibinfo
  {publisher} {McGraw-Hill Book Company, New York},\ \bibinfo {year}
  {1986})\BibitemShut {NoStop}%
\bibitem [{\citenamefont {Mallat}(2009)}]{WaveletBook}%
  \BibitemOpen
  \bibfield  {author} {\bibinfo {author} {\bibfnamefont {S.}~\bibnamefont
  {Mallat}},\ }\href@noop {} {\emph {\bibinfo {title} {A Wavelet Tour of Signal
  Processing: The Sparse Way}}},\ \bibinfo {edition} {3rd}\ ed.\ (\bibinfo
  {publisher} {Academic Press, Burlington},\ \bibinfo {year}
  {2009})\BibitemShut {NoStop}%
\bibitem [{\citenamefont {Born}\ and\ \citenamefont {Wolf}(1999)}]{BornWolf}%
  \BibitemOpen
  \bibfield  {author} {\bibinfo {author} {\bibfnamefont {M.}~\bibnamefont
  {Born}}\ and\ \bibinfo {author} {\bibfnamefont {E.}~\bibnamefont {Wolf}},\
  }\href@noop {} {\emph {\bibinfo {title} {Principles of Optics}}},\ \bibinfo
  {edition} {7th}\ ed.\ (\bibinfo  {publisher} {Cambridge University Press,
  Cambridge},\ \bibinfo {year} {1999})\BibitemShut {NoStop}%
\bibitem [{\citenamefont {Harwit}\ and\ \citenamefont
  {Sloane}(1979)}]{HadamardBook}%
  \BibitemOpen
  \bibfield  {author} {\bibinfo {author} {\bibfnamefont {M.}~\bibnamefont
  {Harwit}}\ and\ \bibinfo {author} {\bibfnamefont {N.~J.~A.}\ \bibnamefont
  {Sloane}},\ }\href@noop {} {\emph {\bibinfo {title} {Hadamard Transform
  Optics}}}\ (\bibinfo  {publisher} {Academic Press, New York},\ \bibinfo
  {year} {1979})\BibitemShut {NoStop}%
\bibitem [{\citenamefont {Gorban}\ \emph {et~al.}(2016)\citenamefont {Gorban},
  \citenamefont {Tyukin}, \citenamefont {Prokhorov},\ and\ \citenamefont
  {Sofeikov}}]{Gorban2016}%
  \BibitemOpen
  \bibfield  {author} {\bibinfo {author} {\bibfnamefont {A.~N.}\ \bibnamefont
  {Gorban}}, \bibinfo {author} {\bibfnamefont {I.~Y.}\ \bibnamefont {Tyukin}},
  \bibinfo {author} {\bibfnamefont {D.~V.}\ \bibnamefont {Prokhorov}},\ and\
  \bibinfo {author} {\bibfnamefont {K.~I.}\ \bibnamefont {Sofeikov}},\
  }\bibfield  {title} {\bibinfo {title} {Approximation with random bases: {P}ro
  et contra},\ }\href@noop {} {\bibfield  {journal} {\bibinfo  {journal} {Inf.
  Sci.}\ }\textbf {\bibinfo {volume} {364--365}},\ \bibinfo {pages} {129}
  (\bibinfo {year} {2016})}\BibitemShut {NoStop}%
\bibitem [{\citenamefont {de~Beer}\ \emph {et~al.}(2019)\citenamefont
  {de~Beer}, \citenamefont {{van der Laan}}, \citenamefont {Cole},
  \citenamefont {Whelan}, \citenamefont {Burns},\ and\ \citenamefont
  {Scott}}]{Beer2019}%
  \BibitemOpen
  \bibfield  {author} {\bibinfo {author} {\bibfnamefont {M.~P.}\ \bibnamefont
  {de~Beer}}, \bibinfo {author} {\bibfnamefont {H.~L.}\ \bibnamefont {{van der
  Laan}}}, \bibinfo {author} {\bibfnamefont {M.~A.}\ \bibnamefont {Cole}},
  \bibinfo {author} {\bibfnamefont {R.~J.}\ \bibnamefont {Whelan}}, \bibinfo
  {author} {\bibfnamefont {M.~A.}\ \bibnamefont {Burns}},\ and\ \bibinfo
  {author} {\bibfnamefont {T.~F.}\ \bibnamefont {Scott}},\ }\bibfield  {title}
  {\bibinfo {title} {Rapid, continuous additive manufacturing by volumetric
  polymerization inhibition patterning},\ }\href@noop {} {\bibfield  {journal}
  {\bibinfo  {journal} {Sci. Adv.}\ }\textbf {\bibinfo {volume} {5}},\ \bibinfo
  {pages} {{eaau8723}} (\bibinfo {year} {2019})}\BibitemShut {NoStop}%
\bibitem [{\citenamefont {Kelly}\ \emph {et~al.}(2019)\citenamefont {Kelly},
  \citenamefont {Bhattacharya}, \citenamefont {Heidari}, \citenamefont
  {Shusteff}, \citenamefont {Spadaccini},\ and\ \citenamefont
  {Taylor}}]{TomographyInReverse2019}%
  \BibitemOpen
  \bibfield  {author} {\bibinfo {author} {\bibfnamefont {B.~E.}\ \bibnamefont
  {Kelly}}, \bibinfo {author} {\bibfnamefont {I.}~\bibnamefont {Bhattacharya}},
  \bibinfo {author} {\bibfnamefont {H.}~\bibnamefont {Heidari}}, \bibinfo
  {author} {\bibfnamefont {M.}~\bibnamefont {Shusteff}}, \bibinfo {author}
  {\bibfnamefont {C.~M.}\ \bibnamefont {Spadaccini}},\ and\ \bibinfo {author}
  {\bibfnamefont {H.~K.}\ \bibnamefont {Taylor}},\ }\bibfield  {title}
  {\bibinfo {title} {Volumetric additive manufacturing via tomographic
  reconstruction},\ }\href@noop {} {\bibfield  {journal} {\bibinfo  {journal}
  {Science}\ }\textbf {\bibinfo {volume} {363}},\ \bibinfo {pages} {1075}
  (\bibinfo {year} {2019})}\BibitemShut {NoStop}%
\bibitem [{\citenamefont {Shapiro}(2008)}]{shapiro2008computational}%
  \BibitemOpen
  \bibfield  {author} {\bibinfo {author} {\bibfnamefont {J.~H.}\ \bibnamefont
  {Shapiro}},\ }\bibfield  {title} {\bibinfo {title} {Computational ghost
  imaging},\ }\href@noop {} {\bibfield  {journal} {\bibinfo  {journal} {Phys.
  Rev. A}\ }\textbf {\bibinfo {volume} {78}},\ \bibinfo {pages} {061802}
  (\bibinfo {year} {2008})}\BibitemShut {NoStop}%
\bibitem [{\citenamefont {Erkmen}\ and\ \citenamefont
  {Shapiro}(2010)}]{erkmen2010ghost}%
  \BibitemOpen
  \bibfield  {author} {\bibinfo {author} {\bibfnamefont {B.~I.}\ \bibnamefont
  {Erkmen}}\ and\ \bibinfo {author} {\bibfnamefont {J.~H.}\ \bibnamefont
  {Shapiro}},\ }\bibfield  {title} {\bibinfo {title} {Ghost imaging: from
  quantum to classical to computational},\ }\href@noop {} {\bibfield  {journal}
  {\bibinfo  {journal} {Adv. Opt. Photonics}\ }\textbf {\bibinfo {volume}
  {2}},\ \bibinfo {pages} {405} (\bibinfo {year} {2010})}\BibitemShut {NoStop}%
\bibitem [{\citenamefont {Shapiro}\ and\ \citenamefont
  {Boyd}(2012)}]{Shapiro2012}%
  \BibitemOpen
  \bibfield  {author} {\bibinfo {author} {\bibfnamefont {J.~H.}\ \bibnamefont
  {Shapiro}}\ and\ \bibinfo {author} {\bibfnamefont {R.~W.}\ \bibnamefont
  {Boyd}},\ }\bibfield  {title} {\bibinfo {title} {The physics of ghost
  imaging},\ }\href@noop {} {\bibfield  {journal} {\bibinfo  {journal} {Quantum
  Inf. Process.}\ }\textbf {\bibinfo {volume} {11}},\ \bibinfo {pages} {949}
  (\bibinfo {year} {2012})}\BibitemShut {NoStop}%
\bibitem [{\citenamefont {Padgett}\ and\ \citenamefont
  {Boyd}(2017)}]{Padgett2017}%
  \BibitemOpen
  \bibfield  {author} {\bibinfo {author} {\bibfnamefont {M.~J.}\ \bibnamefont
  {Padgett}}\ and\ \bibinfo {author} {\bibfnamefont {R.~W.}\ \bibnamefont
  {Boyd}},\ }\bibfield  {title} {\bibinfo {title} {An introduction to ghost
  imaging: quantum and classical},\ }\href@noop {} {\bibfield  {journal}
  {\bibinfo  {journal} {Phil. Trans. R. Soc. A}\ }\textbf {\bibinfo {volume}
  {375}},\ \bibinfo {pages} {20160233} (\bibinfo {year} {2017})}\BibitemShut
  {NoStop}%
\bibitem [{\citenamefont {Donoho}(2006)}]{Donoho2006}%
  \BibitemOpen
  \bibfield  {author} {\bibinfo {author} {\bibfnamefont {D.~L.}\ \bibnamefont
  {Donoho}},\ }\bibfield  {title} {\bibinfo {title} {Compressed sensing},\
  }\href {https://doi.org/10.1109/TIT.2006.871582} {\bibfield  {journal}
  {\bibinfo  {journal} {IEEE Trans. Inf. Theory}\ }\textbf {\bibinfo {volume}
  {52}},\ \bibinfo {pages} {1289} (\bibinfo {year} {2006})}\BibitemShut
  {NoStop}%
\bibitem [{\citenamefont {{Rani}}\ \emph {et~al.}(2018)\citenamefont {{Rani}},
  \citenamefont {{Dhok}},\ and\ \citenamefont {{Deshmukh}}}]{Rani2017}%
  \BibitemOpen
  \bibfield  {author} {\bibinfo {author} {\bibfnamefont {M.}~\bibnamefont
  {{Rani}}}, \bibinfo {author} {\bibfnamefont {S.~B.}\ \bibnamefont {{Dhok}}},\
  and\ \bibinfo {author} {\bibfnamefont {R.~B.}\ \bibnamefont {{Deshmukh}}},\
  }\bibfield  {title} {\bibinfo {title} {A systematic review of compressive
  sensing: Concepts, implementations and applications},\ }\href@noop {}
  {\bibfield  {journal} {\bibinfo  {journal} {IEEE Access}\ }\textbf {\bibinfo
  {volume} {6}},\ \bibinfo {pages} {4875} (\bibinfo {year} {2018})}\BibitemShut
  {NoStop}%
\bibitem [{\citenamefont {Bromberg}\ \emph {et~al.}(2009)\citenamefont
  {Bromberg}, \citenamefont {Katz},\ and\ \citenamefont
  {Silberberg}}]{Bromberg2009}%
  \BibitemOpen
  \bibfield  {author} {\bibinfo {author} {\bibfnamefont {Y.}~\bibnamefont
  {Bromberg}}, \bibinfo {author} {\bibfnamefont {O.}~\bibnamefont {Katz}},\
  and\ \bibinfo {author} {\bibfnamefont {Y.}~\bibnamefont {Silberberg}},\
  }\bibfield  {title} {\bibinfo {title} {Ghost imaging with a single
  detector},\ }\href@noop {} {\bibfield  {journal} {\bibinfo  {journal} {Phys.
  Rev. A}\ }\textbf {\bibinfo {volume} {79}},\ \bibinfo {pages} {053840}
  (\bibinfo {year} {2009})}\BibitemShut {NoStop}%
\bibitem [{\citenamefont {Press}\ \emph {et~al.}(2007)\citenamefont {Press},
  \citenamefont {Teukolsky}, \citenamefont {Vetterling},\ and\ \citenamefont
  {Flannery}}]{NumericalRecipes}%
  \BibitemOpen
  \bibfield  {author} {\bibinfo {author} {\bibfnamefont {W.~H.}\ \bibnamefont
  {Press}}, \bibinfo {author} {\bibfnamefont {S.~A.}\ \bibnamefont
  {Teukolsky}}, \bibinfo {author} {\bibfnamefont {W.~T.}\ \bibnamefont
  {Vetterling}},\ and\ \bibinfo {author} {\bibfnamefont {B.~P.}\ \bibnamefont
  {Flannery}},\ }\href@noop {} {\emph {\bibinfo {title} {Numerical Recipes: The
  Art of Scientific Computing}}},\ \bibinfo {edition} {3rd}\ ed.\ (\bibinfo
  {publisher} {Cambridge University Press, Cambridge},\ \bibinfo {year}
  {2007})\BibitemShut {NoStop}%
\bibitem [{\citenamefont {Kingston}\ \emph {et~al.}(2019)\citenamefont
  {Kingston}, \citenamefont {Myers}, \citenamefont {Pelliccia}, \citenamefont
  {Svalbe},\ and\ \citenamefont {Paganin}}]{KingstonIEEE2019}%
  \BibitemOpen
  \bibfield  {author} {\bibinfo {author} {\bibfnamefont {A.~M.}\ \bibnamefont
  {Kingston}}, \bibinfo {author} {\bibfnamefont {G.~R.}\ \bibnamefont {Myers}},
  \bibinfo {author} {\bibfnamefont {D.}~\bibnamefont {Pelliccia}}, \bibinfo
  {author} {\bibfnamefont {I.~D.}\ \bibnamefont {Svalbe}},\ and\ \bibinfo
  {author} {\bibfnamefont {D.~M.}\ \bibnamefont {Paganin}},\ }\bibfield
  {title} {\bibinfo {title} {X-ray ghost-tomography: Artefacts, dose
  distribution, and mask considerations},\ }\href
  {https://doi.org/10.1109/TCI.2018.2880337} {\bibfield  {journal} {\bibinfo
  {journal} {IEEE Trans. Comput. Imaging}\ }\textbf {\bibinfo {volume} {5}},\
  \bibinfo {pages} {136} (\bibinfo {year} {2019})}\BibitemShut {NoStop}%
\bibitem [{\citenamefont {Paganin}(2019)}]{paganin2019writing}%
  \BibitemOpen
  \bibfield  {author} {\bibinfo {author} {\bibfnamefont {D.~M.}\ \bibnamefont
  {Paganin}},\ }\bibfield  {title} {\bibinfo {title} {Writing arbitrary
  distributions of radiant exposure by scanning a single illuminated spatially
  random screen},\ }\href@noop {} {\bibfield  {journal} {\bibinfo  {journal}
  {Phys. Rev. A}\ }\textbf {\bibinfo {volume} {100}},\ \bibinfo {pages}
  {063823} (\bibinfo {year} {2019})}\BibitemShut {NoStop}%
\bibitem [{\citenamefont {Paganin}(2006)}]{Paganin2006}%
  \BibitemOpen
  \bibfield  {author} {\bibinfo {author} {\bibfnamefont {D.~M.}\ \bibnamefont
  {Paganin}},\ }\href@noop {} {\emph {\bibinfo {title} {Coherent X-ray
  Optics}}}\ (\bibinfo  {publisher} {Oxford University Press, Oxford},\
  \bibinfo {year} {2006})\BibitemShut {NoStop}%
\bibitem [{\citenamefont {Cowley}(1995)}]{CowleyBook}%
  \BibitemOpen
  \bibfield  {author} {\bibinfo {author} {\bibfnamefont {J.~M.}\ \bibnamefont
  {Cowley}},\ }\href@noop {} {\emph {\bibinfo {title} {Diffraction Physics}}},\
  \bibinfo {edition} {3rd}\ ed.\ (\bibinfo  {publisher} {Elsevier, Amsterdam},\
  \bibinfo {year} {1995})\BibitemShut {NoStop}%
\bibitem [{\citenamefont {Utsuro}\ and\ \citenamefont
  {Ignatovich}(2010)}]{NeutronOpticsHandbook}%
  \BibitemOpen
  \bibfield  {author} {\bibinfo {author} {\bibfnamefont {M.}~\bibnamefont
  {Utsuro}}\ and\ \bibinfo {author} {\bibfnamefont {V.~K.}\ \bibnamefont
  {Ignatovich}},\ }\href@noop {} {\emph {\bibinfo {title} {Handbook of Neutron
  Optics}}}\ (\bibinfo  {publisher} {Wiley VCH Verlag GmbH, Weinheim},\
  \bibinfo {year} {2010})\BibitemShut {NoStop}%
\bibitem [{\citenamefont {Cimmino}\ \emph {et~al.}(2021)\citenamefont
  {Cimmino}, \citenamefont {Ambrosino}, \citenamefont {Anastasio},
  \citenamefont {{D'Errico}}, \citenamefont {Masone}, \citenamefont
  {Roscilli},\ and\ \citenamefont {Saracino}}]{MuonRadiography}%
  \BibitemOpen
  \bibfield  {author} {\bibinfo {author} {\bibfnamefont {L.}~\bibnamefont
  {Cimmino}}, \bibinfo {author} {\bibfnamefont {F.}~\bibnamefont {Ambrosino}},
  \bibinfo {author} {\bibfnamefont {A.}~\bibnamefont {Anastasio}}, \bibinfo
  {author} {\bibfnamefont {M.}~\bibnamefont {{D'Errico}}}, \bibinfo {author}
  {\bibfnamefont {V.}~\bibnamefont {Masone}}, \bibinfo {author} {\bibfnamefont
  {L.}~\bibnamefont {Roscilli}},\ and\ \bibinfo {author} {\bibfnamefont
  {G.}~\bibnamefont {Saracino}},\ }\bibfield  {title} {\bibinfo {title} {A new
  cylindrical borehole detector for radiographic imaging with muons},\
  }\href@noop {} {\bibfield  {journal} {\bibinfo  {journal} {Sci. Rep.}\
  }\textbf {\bibinfo {volume} {11}},\ \bibinfo {pages} {17425} (\bibinfo {year}
  {2021})}\BibitemShut {NoStop}%
\bibitem [{\citenamefont {Metcalf}\ and\ \citenamefont {{van der
  Straten}}(1999)}]{AtomBeamBook}%
  \BibitemOpen
  \bibfield  {author} {\bibinfo {author} {\bibfnamefont {H.~J.}\ \bibnamefont
  {Metcalf}}\ and\ \bibinfo {author} {\bibfnamefont {P.}~\bibnamefont {{van der
  Straten}}},\ }\href@noop {} {\emph {\bibinfo {title} {Laser Cooling and
  Trapping}}}\ (\bibinfo  {publisher} {Springer, New York},\ \bibinfo {year}
  {1999})\BibitemShut {NoStop}%
\bibitem [{\citenamefont {Ramsey}(1986)}]{MolcularBeamsBook}%
  \BibitemOpen
  \bibfield  {author} {\bibinfo {author} {\bibfnamefont {N.}~\bibnamefont
  {Ramsey}},\ }\href@noop {} {\emph {\bibinfo {title} {Molecular Beams}}}\
  (\bibinfo  {publisher} {Oxford University Press, Oxford},\ \bibinfo {year}
  {1986})\BibitemShut {NoStop}%
\bibitem [{\citenamefont {Joy}(2013)}]{IonBeamBook}%
  \BibitemOpen
  \bibfield  {author} {\bibinfo {author} {\bibfnamefont {D.~C.}\ \bibnamefont
  {Joy}},\ }\href@noop {} {\emph {\bibinfo {title} {Helium Ion Microscopy:
  Principles and Applications}}}\ (\bibinfo  {publisher} {Springer, New York},\
  \bibinfo {year} {2013})\BibitemShut {NoStop}%
\bibitem [{\citenamefont {Pelliccia}\ \emph {et~al.}(2018)\citenamefont
  {Pelliccia}, \citenamefont {Olbinado}, \citenamefont {Rack}, \citenamefont
  {Kingston}, \citenamefont {Myers},\ and\ \citenamefont
  {Paganin}}]{PellicciaIUCrJ}%
  \BibitemOpen
  \bibfield  {author} {\bibinfo {author} {\bibfnamefont {D.}~\bibnamefont
  {Pelliccia}}, \bibinfo {author} {\bibfnamefont {M.~P.}\ \bibnamefont
  {Olbinado}}, \bibinfo {author} {\bibfnamefont {A.}~\bibnamefont {Rack}},
  \bibinfo {author} {\bibfnamefont {A.~M.}\ \bibnamefont {Kingston}}, \bibinfo
  {author} {\bibfnamefont {G.~R.}\ \bibnamefont {Myers}},\ and\ \bibinfo
  {author} {\bibfnamefont {D.~M.}\ \bibnamefont {Paganin}},\ }\bibfield
  {title} {\bibinfo {title} {{Towards a practical implementation of X-ray ghost
  imaging with synchrotron light}},\ }\href
  {https://doi.org/10.1107/S205225251800711X} {\bibfield  {journal} {\bibinfo
  {journal} {IUCrJ}\ }\textbf {\bibinfo {volume} {5}},\ \bibinfo {pages} {428}
  (\bibinfo {year} {2018})}\BibitemShut {NoStop}%
\bibitem [{\citenamefont {Kingston}\ \emph {et~al.}(2020)\citenamefont
  {Kingston}, \citenamefont {Myers}, \citenamefont {Pelliccia}, \citenamefont
  {Salvemini}, \citenamefont {Bevitt}, \citenamefont {Garbe},\ and\
  \citenamefont {Paganin}}]{Kingston2020}%
  \BibitemOpen
  \bibfield  {author} {\bibinfo {author} {\bibfnamefont {A.~M.}\ \bibnamefont
  {Kingston}}, \bibinfo {author} {\bibfnamefont {G.~R.}\ \bibnamefont {Myers}},
  \bibinfo {author} {\bibfnamefont {D.}~\bibnamefont {Pelliccia}}, \bibinfo
  {author} {\bibfnamefont {F.}~\bibnamefont {Salvemini}}, \bibinfo {author}
  {\bibfnamefont {J.~J.}\ \bibnamefont {Bevitt}}, \bibinfo {author}
  {\bibfnamefont {U.}~\bibnamefont {Garbe}},\ and\ \bibinfo {author}
  {\bibfnamefont {D.~M.}\ \bibnamefont {Paganin}},\ }\bibfield  {title}
  {\bibinfo {title} {Neutron ghost imaging},\ }\href
  {https://doi.org/10.1103/PhysRevA.101.053844} {\bibfield  {journal} {\bibinfo
   {journal} {Phys. Rev. A}\ }\textbf {\bibinfo {volume} {101}},\ \bibinfo
  {pages} {053844} (\bibinfo {year} {2020})}\BibitemShut {NoStop}%
\bibitem [{\citenamefont {Ceddia}\ and\ \citenamefont
  {Paganin}(2018)}]{ceddia2018random}%
  \BibitemOpen
  \bibfield  {author} {\bibinfo {author} {\bibfnamefont {D.}~\bibnamefont
  {Ceddia}}\ and\ \bibinfo {author} {\bibfnamefont {D.~M.}\ \bibnamefont
  {Paganin}},\ }\bibfield  {title} {\bibinfo {title} {Random-matrix bases,
  ghost imaging, and x-ray phase contrast computational ghost imaging},\
  }\href@noop {} {\bibfield  {journal} {\bibinfo  {journal} {Phys. Rev. A}\
  }\textbf {\bibinfo {volume} {97}},\ \bibinfo {pages} {062119} (\bibinfo
  {year} {2018})}\BibitemShut {NoStop}%
\bibitem [{\citenamefont {Katz}\ \emph {et~al.}(2009)\citenamefont {Katz},
  \citenamefont {Bromberg},\ and\ \citenamefont {Silberberg}}]{Katz2009}%
  \BibitemOpen
  \bibfield  {author} {\bibinfo {author} {\bibfnamefont {O.}~\bibnamefont
  {Katz}}, \bibinfo {author} {\bibfnamefont {Y.}~\bibnamefont {Bromberg}},\
  and\ \bibinfo {author} {\bibfnamefont {Y.}~\bibnamefont {Silberberg}},\
  }\bibfield  {title} {\bibinfo {title} {Compressive ghost imaging},\
  }\href@noop {} {\bibfield  {journal} {\bibinfo  {journal} {Appl. Phys.
  Lett.}\ }\textbf {\bibinfo {volume} {95}},\ \bibinfo {pages} {131110}
  (\bibinfo {year} {2009})}\BibitemShut {NoStop}%
\bibitem [{\citenamefont {Gureyev}\ \emph {et~al.}(2014)\citenamefont
  {Gureyev}, \citenamefont {Nesterets}, \citenamefont {de~Hoog}, \citenamefont
  {Schmalz}, \citenamefont {Mayo}, \citenamefont {Mohammadi},\ and\
  \citenamefont {Tromba}}]{GureyevNRU}%
  \BibitemOpen
  \bibfield  {author} {\bibinfo {author} {\bibfnamefont {T.~E.}\ \bibnamefont
  {Gureyev}}, \bibinfo {author} {\bibfnamefont {{\relax{Ya}}.~I.}\ \bibnamefont
  {Nesterets}}, \bibinfo {author} {\bibfnamefont {F.}~\bibnamefont {de~Hoog}},
  \bibinfo {author} {\bibfnamefont {G.}~\bibnamefont {Schmalz}}, \bibinfo
  {author} {\bibfnamefont {S.~C.}\ \bibnamefont {Mayo}}, \bibinfo {author}
  {\bibfnamefont {S.}~\bibnamefont {Mohammadi}},\ and\ \bibinfo {author}
  {\bibfnamefont {G.}~\bibnamefont {Tromba}},\ }\bibfield  {title} {\bibinfo
  {title} {Duality between noise and spatial resolution in linear systems},\
  }\href@noop {} {\bibfield  {journal} {\bibinfo  {journal} {Opt. Express}\
  }\textbf {\bibinfo {volume} {22}},\ \bibinfo {pages} {9087} (\bibinfo {year}
  {2014})}\BibitemShut {NoStop}%
\bibitem [{\citenamefont {Gureyev}\ \emph {et~al.}(2020)\citenamefont
  {Gureyev}, \citenamefont {Kozlov}, \citenamefont {Paganin}, \citenamefont
  {Nesterets},\ and\ \citenamefont {Quiney}}]{GureyevNRU2020}%
  \BibitemOpen
  \bibfield  {author} {\bibinfo {author} {\bibfnamefont {T.~E.}\ \bibnamefont
  {Gureyev}}, \bibinfo {author} {\bibfnamefont {A.}~\bibnamefont {Kozlov}},
  \bibinfo {author} {\bibfnamefont {D.~M.}\ \bibnamefont {Paganin}}, \bibinfo
  {author} {\bibfnamefont {{\relax{Ya}}.~I.}\ \bibnamefont {Nesterets}},\ and\
  \bibinfo {author} {\bibfnamefont {H.~M.}\ \bibnamefont {Quiney}},\ }\bibfield
   {title} {\bibinfo {title} {Noise--resolution uncertainty principle in
  classical and quantum systems},\ }\href@noop {} {\bibfield  {journal}
  {\bibinfo  {journal} {Sci. Rep.}\ }\textbf {\bibinfo {volume} {10}},\
  \bibinfo {pages} {7890} (\bibinfo {year} {2020})}\BibitemShut {NoStop}%
\bibitem [{\citenamefont {{Baker Jr}}(1975)}]{PadeBook}%
  \BibitemOpen
  \bibfield  {author} {\bibinfo {author} {\bibfnamefont {G.~A.}\ \bibnamefont
  {{Baker Jr}}},\ }\href@noop {} {\emph {\bibinfo {title} {{Essentials of
  Pad\'{e} Approximants}}}}\ (\bibinfo  {publisher} {Academic Press, New
  York},\ \bibinfo {year} {1975})\BibitemShut {NoStop}%
\bibitem [{\citenamefont {Mandel}\ and\ \citenamefont
  {Wolf}(1995)}]{mandel1995optical}%
  \BibitemOpen
  \bibfield  {author} {\bibinfo {author} {\bibfnamefont {L.}~\bibnamefont
  {Mandel}}\ and\ \bibinfo {author} {\bibfnamefont {E.}~\bibnamefont {Wolf}},\
  }\href@noop {} {\emph {\bibinfo {title} {Optical Coherence and Quantum
  Optics}}}\ (\bibinfo  {publisher} {Cambridge University Press, Cambridge},\
  \bibinfo {year} {1995})\BibitemShut {NoStop}%
\bibitem [{\citenamefont {Blanter}\ and\ \citenamefont
  {B{\"u}ttiker}(2000)}]{blanter2000shot}%
  \BibitemOpen
  \bibfield  {author} {\bibinfo {author} {\bibfnamefont {Y.~M.}\ \bibnamefont
  {Blanter}}\ and\ \bibinfo {author} {\bibfnamefont {M.}~\bibnamefont
  {B{\"u}ttiker}},\ }\bibfield  {title} {\bibinfo {title} {Shot noise in
  mesoscopic conductors},\ }\href@noop {} {\bibfield  {journal} {\bibinfo
  {journal} {Phys. Rep.}\ }\textbf {\bibinfo {volume} {336}},\ \bibinfo {pages}
  {1} (\bibinfo {year} {2000})}\BibitemShut {NoStop}%
\bibitem [{\citenamefont {Cao}\ \emph {et~al.}(2015)\citenamefont {Cao},
  \citenamefont {Xu}, \citenamefont {Zhang},\ and\ \citenamefont
  {Wang}}]{colourGI}%
  \BibitemOpen
  \bibfield  {author} {\bibinfo {author} {\bibfnamefont {D.-Z.}\ \bibnamefont
  {Cao}}, \bibinfo {author} {\bibfnamefont {B.-L.}\ \bibnamefont {Xu}},
  \bibinfo {author} {\bibfnamefont {S.-H.}\ \bibnamefont {Zhang}},\ and\
  \bibinfo {author} {\bibfnamefont {K.-G.}\ \bibnamefont {Wang}},\ }\bibfield
  {title} {\bibinfo {title} {Color ghost imaging with pseudo-white-thermal
  light},\ }\href {https://doi.org/10.1088/0256-307x/32/11/114208} {\bibfield
  {journal} {\bibinfo  {journal} {Chin. Phys. Lett.}\ }\textbf {\bibinfo
  {volume} {32}},\ \bibinfo {pages} {114208} (\bibinfo {year}
  {2015})}\BibitemShut {NoStop}%
\bibitem [{\citenamefont {Mosk}\ \emph {et~al.}(2012)\citenamefont {Mosk},
  \citenamefont {Lagendijk}, \citenamefont {Lerosey},\ and\ \citenamefont
  {Fink}}]{FinkSpeckles2012}%
  \BibitemOpen
  \bibfield  {author} {\bibinfo {author} {\bibfnamefont {A.~P.}\ \bibnamefont
  {Mosk}}, \bibinfo {author} {\bibfnamefont {A.}~\bibnamefont {Lagendijk}},
  \bibinfo {author} {\bibfnamefont {G.}~\bibnamefont {Lerosey}},\ and\ \bibinfo
  {author} {\bibfnamefont {M.}~\bibnamefont {Fink}},\ }\bibfield  {title}
  {\bibinfo {title} {Controlling waves in space and time for imaging and
  focusing in complex media},\ }\href@noop {} {\bibfield  {journal} {\bibinfo
  {journal} {Nat. Photonics}\ }\textbf {\bibinfo {volume} {6}},\ \bibinfo
  {pages} {283} (\bibinfo {year} {2012})}\BibitemShut {NoStop}%
\bibitem [{\citenamefont {Gureyev}\ \emph {et~al.}(2018)\citenamefont
  {Gureyev}, \citenamefont {Paganin}, \citenamefont {Kozlov}, \citenamefont
  {Nesterets},\ and\ \citenamefont {Quiney}}]{Gureyev2018}%
  \BibitemOpen
  \bibfield  {author} {\bibinfo {author} {\bibfnamefont {T.~E.}\ \bibnamefont
  {Gureyev}}, \bibinfo {author} {\bibfnamefont {D.~M.}\ \bibnamefont
  {Paganin}}, \bibinfo {author} {\bibfnamefont {A.}~\bibnamefont {Kozlov}},
  \bibinfo {author} {\bibfnamefont {{\relax{Ya}}.~I.}\ \bibnamefont
  {Nesterets}},\ and\ \bibinfo {author} {\bibfnamefont {H.~M.}\ \bibnamefont
  {Quiney}},\ }\bibfield  {title} {\bibinfo {title} {On the efficiency of
  computational imaging with structured illumination},\ }\href@noop {}
  {\bibfield  {journal} {\bibinfo  {journal} {Phys. Rev. A}\ }\textbf {\bibinfo
  {volume} {97}},\ \bibinfo {pages} {053819} (\bibinfo {year}
  {2018})}\BibitemShut {NoStop}%
\bibitem [{\citenamefont {Lane}\ and\ \citenamefont
  {Ratner}(2020)}]{LaneRatner2020}%
  \BibitemOpen
  \bibfield  {author} {\bibinfo {author} {\bibfnamefont {T.~J.}\ \bibnamefont
  {Lane}}\ and\ \bibinfo {author} {\bibfnamefont {D.}~\bibnamefont {Ratner}},\
  }\bibfield  {title} {\bibinfo {title} {What are the advantages of ghost
  imaging? {M}ultiplexing for x-ray and electron imaging},\ }\href
  {https://doi.org/10.1364/OE.379503} {\bibfield  {journal} {\bibinfo
  {journal} {Opt. Express}\ }\textbf {\bibinfo {volume} {28}},\ \bibinfo
  {pages} {5898} (\bibinfo {year} {2020})}\BibitemShut {NoStop}%
\bibitem [{\citenamefont {Kingston}\ \emph {et~al.}(2021)\citenamefont
  {Kingston}, \citenamefont {Fullagar}, \citenamefont {Myers}, \citenamefont
  {Adams}, \citenamefont {Pelliccia},\ and\ \citenamefont
  {Paganin}}]{Kingston2021}%
  \BibitemOpen
  \bibfield  {author} {\bibinfo {author} {\bibfnamefont {A.~M.}\ \bibnamefont
  {Kingston}}, \bibinfo {author} {\bibfnamefont {W.~K.}\ \bibnamefont
  {Fullagar}}, \bibinfo {author} {\bibfnamefont {G.~R.}\ \bibnamefont {Myers}},
  \bibinfo {author} {\bibfnamefont {D.}~\bibnamefont {Adams}}, \bibinfo
  {author} {\bibfnamefont {D.}~\bibnamefont {Pelliccia}},\ and\ \bibinfo
  {author} {\bibfnamefont {D.~M.}\ \bibnamefont {Paganin}},\ }\bibfield
  {title} {\bibinfo {title} {Inherent dose-reduction potential of classical
  ghost imaging},\ }\href {https://doi.org/10.1103/PhysRevA.103.033503}
  {\bibfield  {journal} {\bibinfo  {journal} {Phys. Rev. A}\ }\textbf {\bibinfo
  {volume} {103}},\ \bibinfo {pages} {033503} (\bibinfo {year}
  {2021})}\BibitemShut {NoStop}%
\bibitem [{\citenamefont {Vladimirsky}\ \emph {et~al.}(1999)\citenamefont
  {Vladimirsky}, \citenamefont {Bourdillon}, \citenamefont {Vladimirsky},
  \citenamefont {Jiang},\ and\ \citenamefont {Leonard}}]{Vladimirsky1999}%
  \BibitemOpen
  \bibfield  {author} {\bibinfo {author} {\bibfnamefont {Y.}~\bibnamefont
  {Vladimirsky}}, \bibinfo {author} {\bibfnamefont {A.}~\bibnamefont
  {Bourdillon}}, \bibinfo {author} {\bibfnamefont {O.}~\bibnamefont
  {Vladimirsky}}, \bibinfo {author} {\bibfnamefont {W.}~\bibnamefont {Jiang}},\
  and\ \bibinfo {author} {\bibfnamefont {Q.}~\bibnamefont {Leonard}},\
  }\bibfield  {title} {\bibinfo {title} {Demagnification in proximity x-ray
  lithography and extensibility to 25 nm by optimizing {F}resnel diffraction},\
  }\href {https://doi.org/10.1088/0022-3727/32/22/102} {\bibfield  {journal}
  {\bibinfo  {journal} {J. Phys. D: Appl. Phys.}\ }\textbf {\bibinfo {volume}
  {32}},\ \bibinfo {pages} {L114} (\bibinfo {year} {1999})}\BibitemShut
  {NoStop}%
\bibitem [{\citenamefont {Bremmer}(1952)}]{Bremmer1952}%
  \BibitemOpen
  \bibfield  {author} {\bibinfo {author} {\bibfnamefont {H.}~\bibnamefont
  {Bremmer}},\ }\bibfield  {title} {\bibinfo {title} {On the asymptotic
  evaluation of diffraction integrals with a special view to the theory of
  defocusing and optical contrast},\ }\href@noop {} {\bibfield  {journal}
  {\bibinfo  {journal} {Physica}\ }\textbf {\bibinfo {volume} {18}},\ \bibinfo
  {pages} {469} (\bibinfo {year} {1952})}\BibitemShut {NoStop}%
\bibitem [{\citenamefont {Snigirev}\ \emph {et~al.}(1995)\citenamefont
  {Snigirev}, \citenamefont {Snigireva}, \citenamefont {Kohn}, \citenamefont
  {Kuznetsov},\ and\ \citenamefont {Schelokov}}]{Snigirev1995}%
  \BibitemOpen
  \bibfield  {author} {\bibinfo {author} {\bibfnamefont {A.}~\bibnamefont
  {Snigirev}}, \bibinfo {author} {\bibfnamefont {I.}~\bibnamefont {Snigireva}},
  \bibinfo {author} {\bibfnamefont {V.}~\bibnamefont {Kohn}}, \bibinfo {author}
  {\bibfnamefont {S.}~\bibnamefont {Kuznetsov}},\ and\ \bibinfo {author}
  {\bibfnamefont {I.}~\bibnamefont {Schelokov}},\ }\bibfield  {title} {\bibinfo
  {title} {On the possibilities of x‐ray phase contrast microimaging by
  coherent high‐energy synchrotron radiation},\ }\href
  {https://doi.org/10.1063/1.1146073} {\bibfield  {journal} {\bibinfo
  {journal} {Rev. Sci. Instrum.}\ }\textbf {\bibinfo {volume} {66}},\ \bibinfo
  {pages} {5486} (\bibinfo {year} {1995})}\BibitemShut {NoStop}%
\bibitem [{\citenamefont {Wilkins}\ \emph {et~al.}(2014)\citenamefont
  {Wilkins}, \citenamefont {Nesterets}, \citenamefont {Gureyev}, \citenamefont
  {Mayo}, \citenamefont {Pogany},\ and\ \citenamefont
  {Stevenson}}]{Wilkins2014}%
  \BibitemOpen
  \bibfield  {author} {\bibinfo {author} {\bibfnamefont {S.~W.}\ \bibnamefont
  {Wilkins}}, \bibinfo {author} {\bibfnamefont {{\relax{Ya}}.~I.}\ \bibnamefont
  {Nesterets}}, \bibinfo {author} {\bibfnamefont {T.~E.}\ \bibnamefont
  {Gureyev}}, \bibinfo {author} {\bibfnamefont {S.~C.}\ \bibnamefont {Mayo}},
  \bibinfo {author} {\bibfnamefont {A.}~\bibnamefont {Pogany}},\ and\ \bibinfo
  {author} {\bibfnamefont {A.~W.}\ \bibnamefont {Stevenson}},\ }\bibfield
  {title} {\bibinfo {title} {On the evolution and relative merits of hard x-ray
  phase-contrast imaging methods},\ }\href@noop {} {\bibfield  {journal}
  {\bibinfo  {journal} {Phil. Trans. R. Soc. A}\ }\textbf {\bibinfo {volume}
  {372}},\ \bibinfo {pages} {20130021} (\bibinfo {year} {2014})}\BibitemShut
  {NoStop}%
\bibitem [{\citenamefont {Nie}\ \emph {et~al.}(2021)\citenamefont {Nie},
  \citenamefont {Yang}, \citenamefont {Liu}, \citenamefont {Zhao},
  \citenamefont {Nessler}, \citenamefont {Peng}, \citenamefont {Zubairy},\ and\
  \citenamefont {Scully}}]{PinkNoiseGI2021}%
  \BibitemOpen
  \bibfield  {author} {\bibinfo {author} {\bibfnamefont {X.}~\bibnamefont
  {Nie}}, \bibinfo {author} {\bibfnamefont {F.}~\bibnamefont {Yang}}, \bibinfo
  {author} {\bibfnamefont {X.}~\bibnamefont {Liu}}, \bibinfo {author}
  {\bibfnamefont {X.}~\bibnamefont {Zhao}}, \bibinfo {author} {\bibfnamefont
  {R.}~\bibnamefont {Nessler}}, \bibinfo {author} {\bibfnamefont
  {T.}~\bibnamefont {Peng}}, \bibinfo {author} {\bibfnamefont {M.~S.}\
  \bibnamefont {Zubairy}},\ and\ \bibinfo {author} {\bibfnamefont {M.~O.}\
  \bibnamefont {Scully}},\ }\bibfield  {title} {\bibinfo {title} {Noise-robust
  computational ghost imaging with pink noise speckle patterns},\ }\href
  {https://doi.org/10.1103/PhysRevA.104.013513} {\bibfield  {journal} {\bibinfo
   {journal} {Phys. Rev. A}\ }\textbf {\bibinfo {volume} {104}},\ \bibinfo
  {pages} {013513} (\bibinfo {year} {2021})}\BibitemShut {NoStop}%
\bibitem [{\citenamefont {Alperin}\ \emph {et~al.}(2019)\citenamefont
  {Alperin}, \citenamefont {Grotelueschen},\ and\ \citenamefont
  {Siemens}}]{Alperin2019}%
  \BibitemOpen
  \bibfield  {author} {\bibinfo {author} {\bibfnamefont {S.~N.}\ \bibnamefont
  {Alperin}}, \bibinfo {author} {\bibfnamefont {A.~L.}\ \bibnamefont
  {Grotelueschen}},\ and\ \bibinfo {author} {\bibfnamefont {M.~E.}\
  \bibnamefont {Siemens}},\ }\bibfield  {title} {\bibinfo {title} {Quantum
  turbulent structure in light},\ }\href
  {https://doi.org/10.1103/PhysRevLett.122.044301} {\bibfield  {journal}
  {\bibinfo  {journal} {Phys. Rev. Lett.}\ }\textbf {\bibinfo {volume} {122}},\
  \bibinfo {pages} {044301} (\bibinfo {year} {2019})}\BibitemShut {NoStop}%
\bibitem [{\citenamefont {Paganin}\ and\ \citenamefont {{S\'anchez del
  R\'{\i}o}}(2019)}]{PaganinSanchezDelRio2019}%
  \BibitemOpen
  \bibfield  {author} {\bibinfo {author} {\bibfnamefont {D.~M.}\ \bibnamefont
  {Paganin}}\ and\ \bibinfo {author} {\bibfnamefont {M.}~\bibnamefont
  {{S\'anchez del R\'{\i}o}}},\ }\bibfield  {title} {\bibinfo {title} {Speckled
  cross-spectral densities and their associated correlation singularities for a
  modern source of partially coherent x rays},\ }\href
  {https://doi.org/10.1103/PhysRevA.100.043813} {\bibfield  {journal} {\bibinfo
   {journal} {Phys. Rev. A}\ }\textbf {\bibinfo {volume} {100}},\ \bibinfo
  {pages} {043813} (\bibinfo {year} {2019})}\BibitemShut {NoStop}%
\bibitem [{\citenamefont {Paganin}\ and\ \citenamefont
  {Pelliccia}(2021)}]{PaganinPelliccia2021}%
  \BibitemOpen
  \bibfield  {author} {\bibinfo {author} {\bibfnamefont {D.~M.}\ \bibnamefont
  {Paganin}}\ and\ \bibinfo {author} {\bibfnamefont {D.}~\bibnamefont
  {Pelliccia}},\ }\bibfield  {title} {\bibinfo {title} {X-ray phase-contrast
  imaging: a broad overview of some fundamentals},\ }in\ \href
  {https://doi.org/https://doi.org/10.1016/bs.aiep.2021.04.002} {\emph
  {\bibinfo {booktitle} {Advances in Imaging and Electron Physics}}},\ Vol.\
  \bibinfo {volume} {218},\ \bibinfo {editor} {edited by\ \bibinfo {editor}
  {\bibfnamefont {M.}~\bibnamefont {H\"{y}tch}}\ and\ \bibinfo {editor}
  {\bibfnamefont {P.~W.}\ \bibnamefont {Hawkes}}}\ (\bibinfo  {publisher}
  {Academic Press, London},\ \bibinfo {year} {2021})\ pp.\ \bibinfo {pages}
  {63--158}\BibitemShut {NoStop}%
\bibitem [{\citenamefont {Kingston}\ \emph {et~al.}(2018)\citenamefont
  {Kingston}, \citenamefont {Pelliccia}, \citenamefont {Rack}, \citenamefont
  {Olbinado}, \citenamefont {Cheng}, \citenamefont {Myers},\ and\ \citenamefont
  {Paganin}}]{KingstonOptica2018}%
  \BibitemOpen
  \bibfield  {author} {\bibinfo {author} {\bibfnamefont {A.~M.}\ \bibnamefont
  {Kingston}}, \bibinfo {author} {\bibfnamefont {D.}~\bibnamefont {Pelliccia}},
  \bibinfo {author} {\bibfnamefont {A.}~\bibnamefont {Rack}}, \bibinfo {author}
  {\bibfnamefont {M.~P.}\ \bibnamefont {Olbinado}}, \bibinfo {author}
  {\bibfnamefont {Y.}~\bibnamefont {Cheng}}, \bibinfo {author} {\bibfnamefont
  {G.~R.}\ \bibnamefont {Myers}},\ and\ \bibinfo {author} {\bibfnamefont
  {D.~M.}\ \bibnamefont {Paganin}},\ }\bibfield  {title} {\bibinfo {title}
  {Ghost tomography},\ }\href {https://doi.org/10.1364/OPTICA.5.001516}
  {\bibfield  {journal} {\bibinfo  {journal} {Optica}\ }\textbf {\bibinfo
  {volume} {5}},\ \bibinfo {pages} {1516} (\bibinfo {year} {2018})}\BibitemShut
  {NoStop}%
\bibitem [{\citenamefont {Yu}(2017)}]{Yu2017}%
  \BibitemOpen
  \bibfield  {author} {\bibinfo {author} {\bibfnamefont {F.~T.~S.}\
  \bibnamefont {Yu}},\ }\href@noop {} {\emph {\bibinfo {title} {Entropy and
  information optics: connecting information and time}}},\ \bibinfo {edition}
  {2nd}\ ed.\ (\bibinfo  {publisher} {Taylor \& Francis and CRC Press, Boca
  Raton},\ \bibinfo {year} {2017})\BibitemShut {NoStop}%
\end{thebibliography}%

\appendix

\section{Correlation Coefficient} \label{AppA}

\begin{figure*}[ht!]
     \centering
     \begin{subfigure}{0.329\textwidth}
         \centering
         \includegraphics[width=\textwidth]{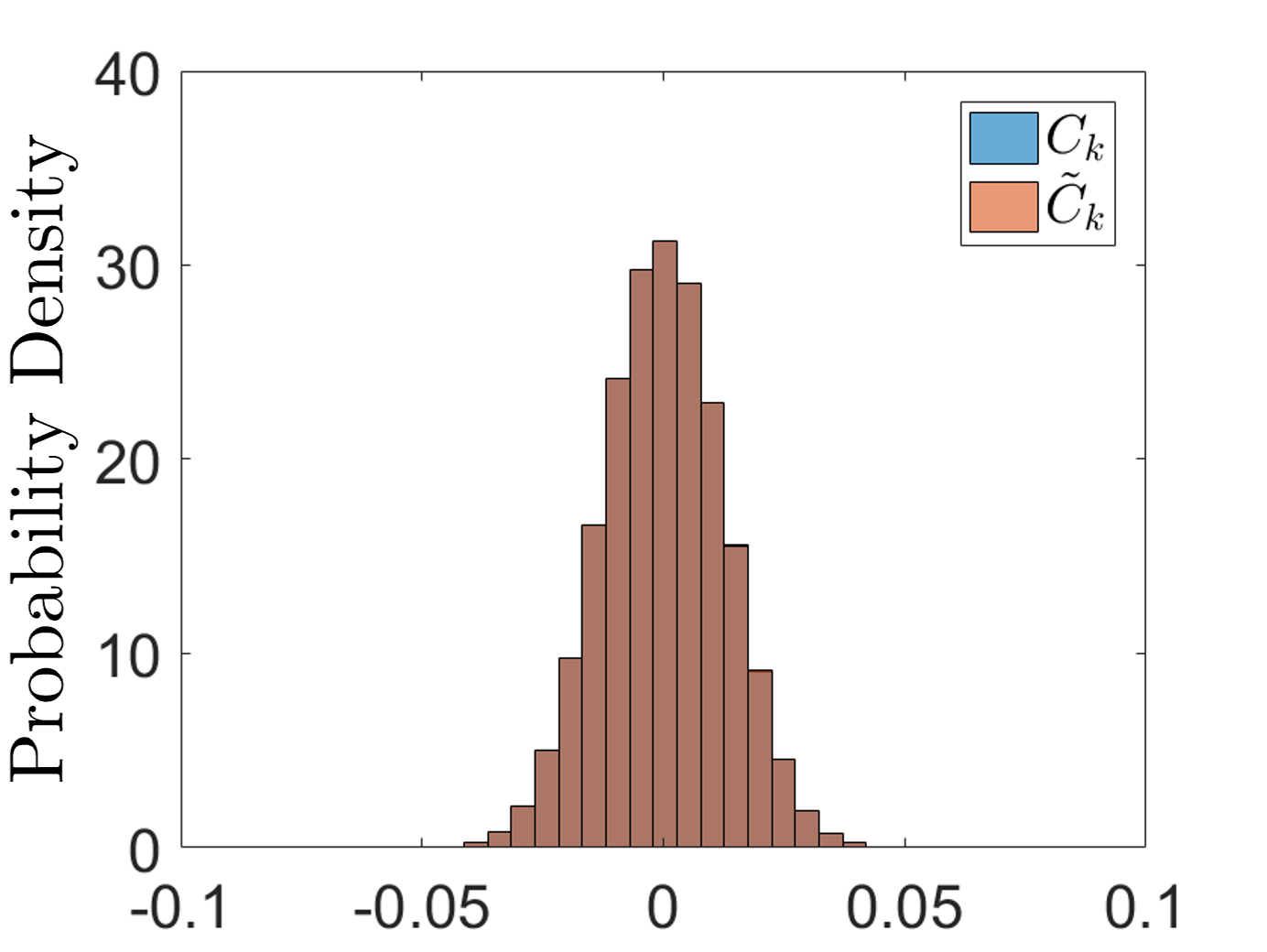}
         \caption{ }
         \label{subfig:Correlation C1}
     \end{subfigure}
     \begin{subfigure}{0.329\textwidth}
         \centering
         \includegraphics[width=\textwidth]{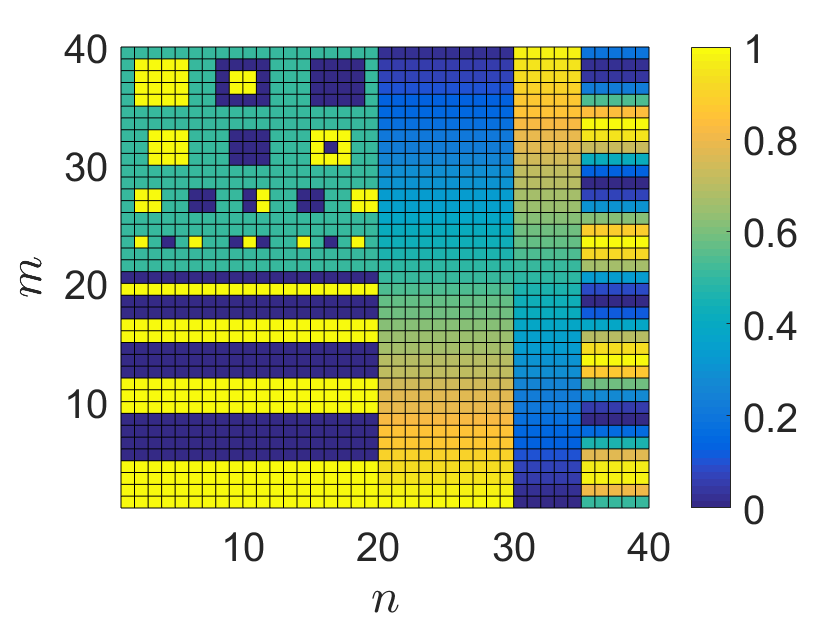}
         \caption{ }
         \label{subfig:Correlation C2}
     \end{subfigure}
     \begin{subfigure}{0.329\textwidth}
         \centering
         \includegraphics[width=\textwidth]{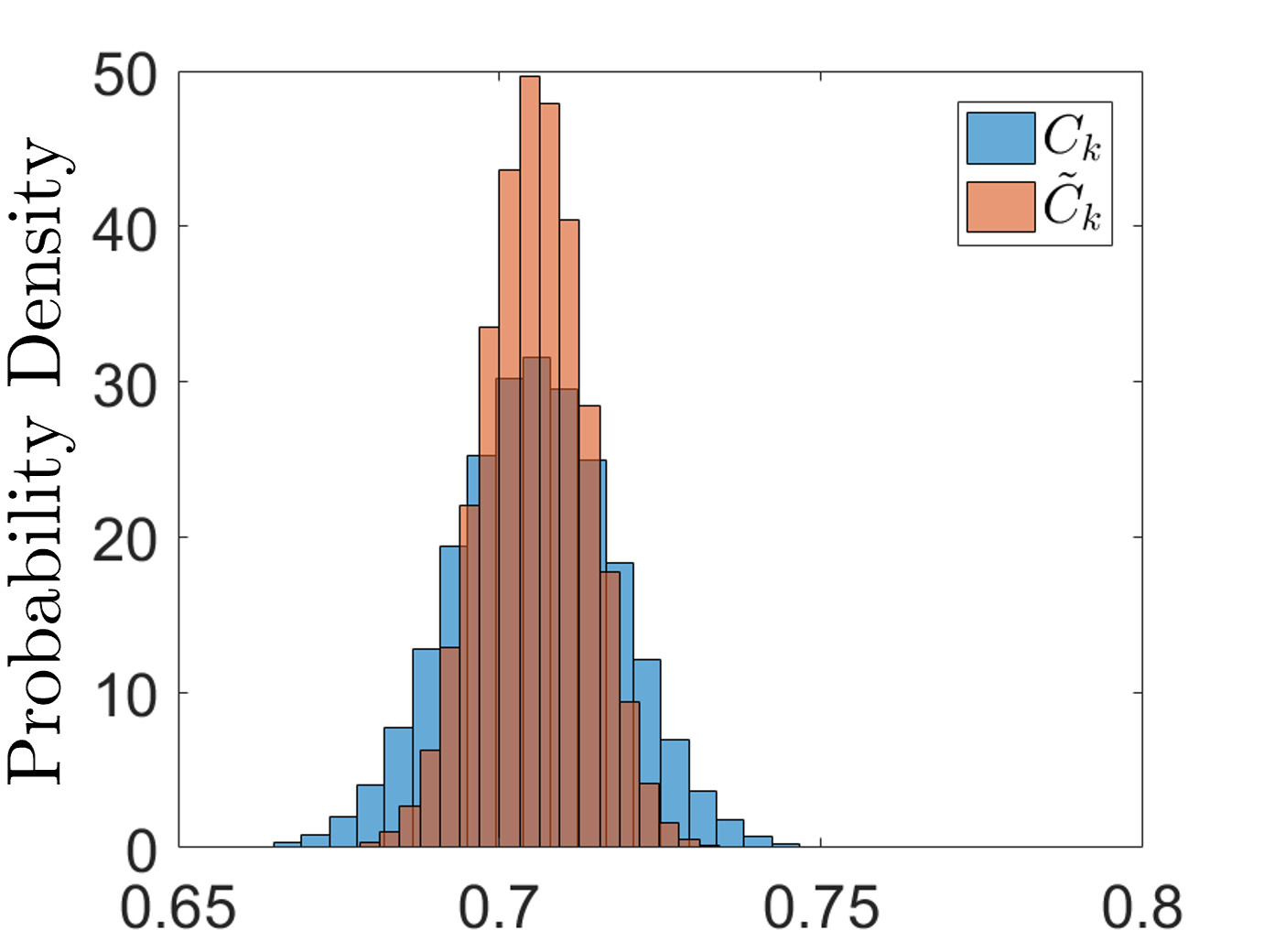}
         \caption{ }
         \label{subfig:Correlation C3}
     \end{subfigure}
        \caption{(a) Overlay of the pseudo-correlation coefficient $C_k$ (Eq.~(\ref{eq:DefinitionForPseudoCorrelation})) and correlation coefficient $\tilde{C}_k$ (Eq.~(\ref{eq:GenuineCorrelationCoefficient})) for a zero-centered image, for 200,000 realizations. (b) Non-zero-centered test image for the comparison of the statistical properties of the pseudo-correlation coefficient and correlation coefficient. (c) Overlay of $C_k$ and  $\tilde{C}_k$ for the non-zero-centered image expressed in panel (b), for 200,000 realizations.}
        \label{fig:Correlation Vs Pseudo-Correlation}
\end{figure*}

Having introduced and explored the simpler pseudo-correlation coefficient defined in Eq.~(\ref{eq:DefinitionForPseudoCorrelation}), it might appear natural to extend consideration to the correlation coefficient itself. The difference between the two is that in the pseudo-correlation case, we normalize by the expected magnitude of a basis member rather than the particular magnitude of each basis member. Doing this has the drawback of suggesting that a random-basis member with a higher average value is more correlated with the desired image than it in fact is. The reason for having introduced the pseudo-correlation coefficient first (i.e.~in Eq.~(\ref{eq:DefinitionForPseudoCorrelation}) of the main text) is that the analysis is considerably simpler and yet still yields the dominant descriptive power that we are seeking. To justify these claims, consider the genuine correlation coefficient:
\begin{align}
\tilde{C}_k = \frac{R_{ijk}I^{ij}}{\sqrt{(R_{\alpha  \beta k}R^{\alpha  \beta}_{\ \ k})(I_{\mu \nu} I^{\mu \nu})}} =  \frac{R_{ijk}I^{ij}}{\sqrt{nm \text{E}[I^2] (R_{\alpha  \beta k}R^{\alpha  \beta}_{\ \ k})}}.
\label{eq:GenuineCorrelationCoefficient}
\end{align}
This correlation coefficient can be approximated by taking a Taylor series expansion of the denominator $R_{\alpha  \beta k}^2 J^{\alpha  \beta}/(nm)$ about its expected value $\text{E}[R^2]$, truncating to first order:
\begin{widetext}
\begin{align}
\tilde{C}_k & \approx \frac{R_{ijk}I^{ij}}{\sqrt{ nm \text{E}[I^2]}}  \left( \frac{1}{\sqrt{nm\text{E}[R^2]}} - \left( \frac{R_{\alpha \beta k}^2J^{\alpha \beta}}{nm} -  \text{E}[R^2] \right) \frac{1}{2 \sqrt{nm \text{E}[R^2]^3}} \right)  \approx  \frac{3}{2} C_k - \frac{R_{ijk}I_{ij}R_{\alpha \beta k}^2 J^{ij} J^{\alpha \beta}}{2(nm)^2 \sqrt{ \text{E}[I^2] \text{E}[R^2]^3}}. \label{eq:Correlation Taylor}
\end{align}
\end{widetext}
We can justify truncating this series at first order in the limit that the product $nm$ is large (i.e.~a reasonable resolution image) as the terms decrease by powers of $1/(nm)$. Taking the expectation value of our truncated series yields:
\begin{align*}
\text{E}[\tilde{C}_k] \approx \text{E}[C_k] - \frac{\text{E}[I] (\text{E}[R^3] - \text{E}[R^2]\text{E}[R])}{2nm \sqrt{\text{E}[I^2]\text{E}[R^2]^3}}.
\end{align*}
Observe that for a zero-centered image, the expected correlation and pseudo-correlation is the same to first order. To determine if it is worthwhile to  calculate the variance of Eq.~(\ref{eq:Correlation Taylor}), consider the simulations in Fig.~\ref{fig:Correlation Vs Pseudo-Correlation}.  For a zero-centered image (i.e. $\text{E}[I] =0$), there is little functional difference between the statistical properties of the correlation coefficient and pseudo-correlation coefficient. For the non-zero-centered image case, we do see a noticeable reduction in variance for the correlation coefficient as compared to the pseudo-correlation coefficient. How this impacts the reconstruction of a ghost projection is, at this stage, unknown. Owing to the negligible difference between the pseudo-correlation coefficient and correlation coefficient for the zero-centered image case (which can often be arranged), and owing to the gains in simplicity, we will continue using the pseudo-correlation coefficient over the correlation coefficient for the purposes of this work.

\section{Pseudo-correlation Filtered Ghost Projection with Linear Weights}\label{AppB:Pseudo-correlation Filtered Ghost Projection with Linear Weights}

Suppose we wish to construct a ghost projection that filters out a portion of the random-matrix basis and then linearly weights the remaining elements according to the pseudo-correlation coefficient in some optimum fashion. That is, we consider the scheme:
\begin{align}
P_{ij}(C_{\text{min}},\alpha,\beta) = (\alpha C_k - \beta J_k) {R'}_{ij}^{\ \ k},
\end{align}
where $C_{\text{min}}$, $\alpha$ and $\beta$ are parameters to be determined via optimization of SNR. Our aim in doing this is to produce a scheme that achieves better SNR than just weighting or filtering alone. Note that, in this Appendix, we restrict ourselves to the case that our desired image is zero-centered, i.e.~$\text{E}[I] = 0$. 

\begin{figure*}[ht!]
     \centering
     \begin{subfigure}{0.329\textwidth}
         \centering
         \includegraphics[width=\textwidth]{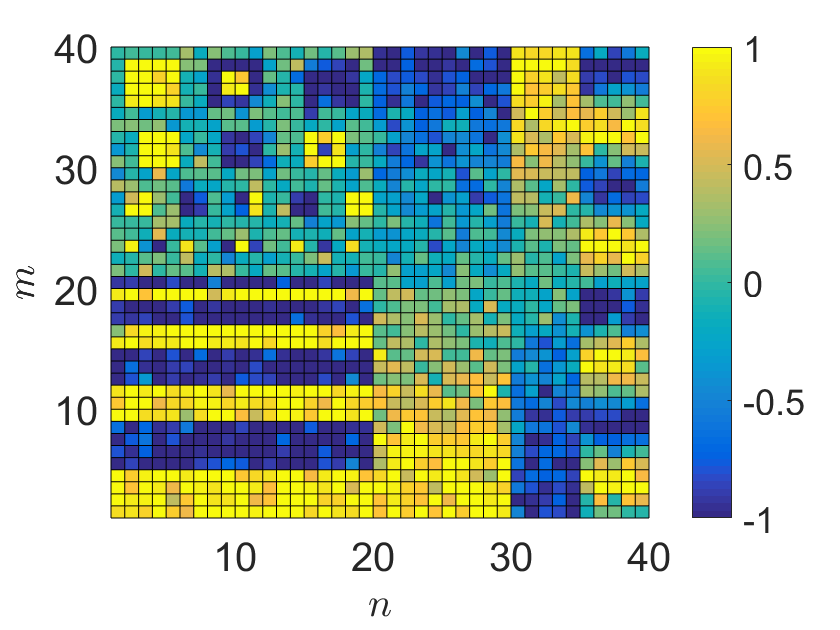}
         \caption{ }
         \label{subfig: GP1 Pseudo Linear}
     \end{subfigure}
     \begin{subfigure}{0.329\textwidth}
         \centering
         \includegraphics[width=\textwidth]{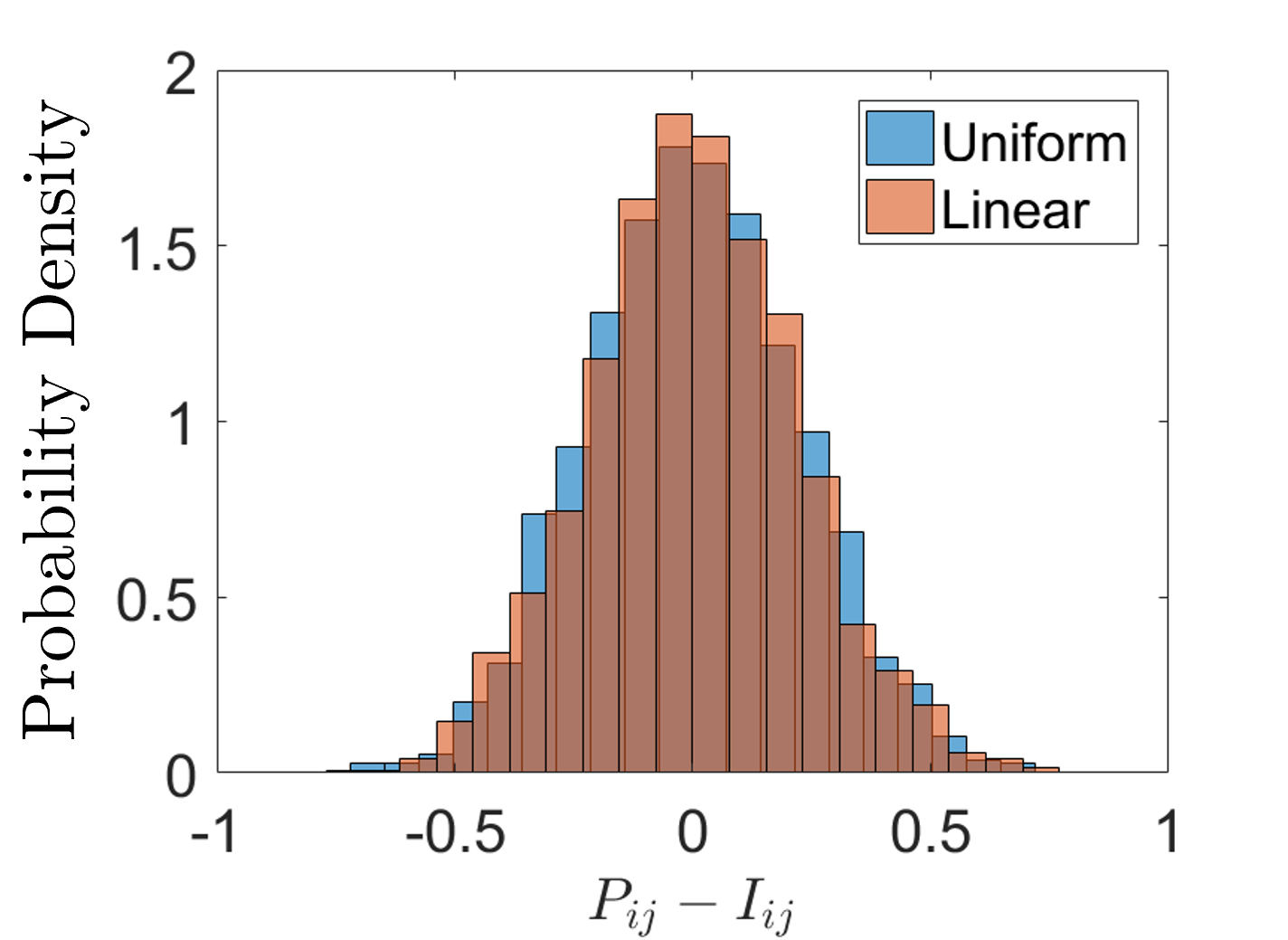}
         \caption{ }
         \label{subfig: GP2 Pseudo Linear}
     \end{subfigure}
     \begin{subfigure}{0.329\textwidth}
         \centering
         \includegraphics[width=\textwidth]{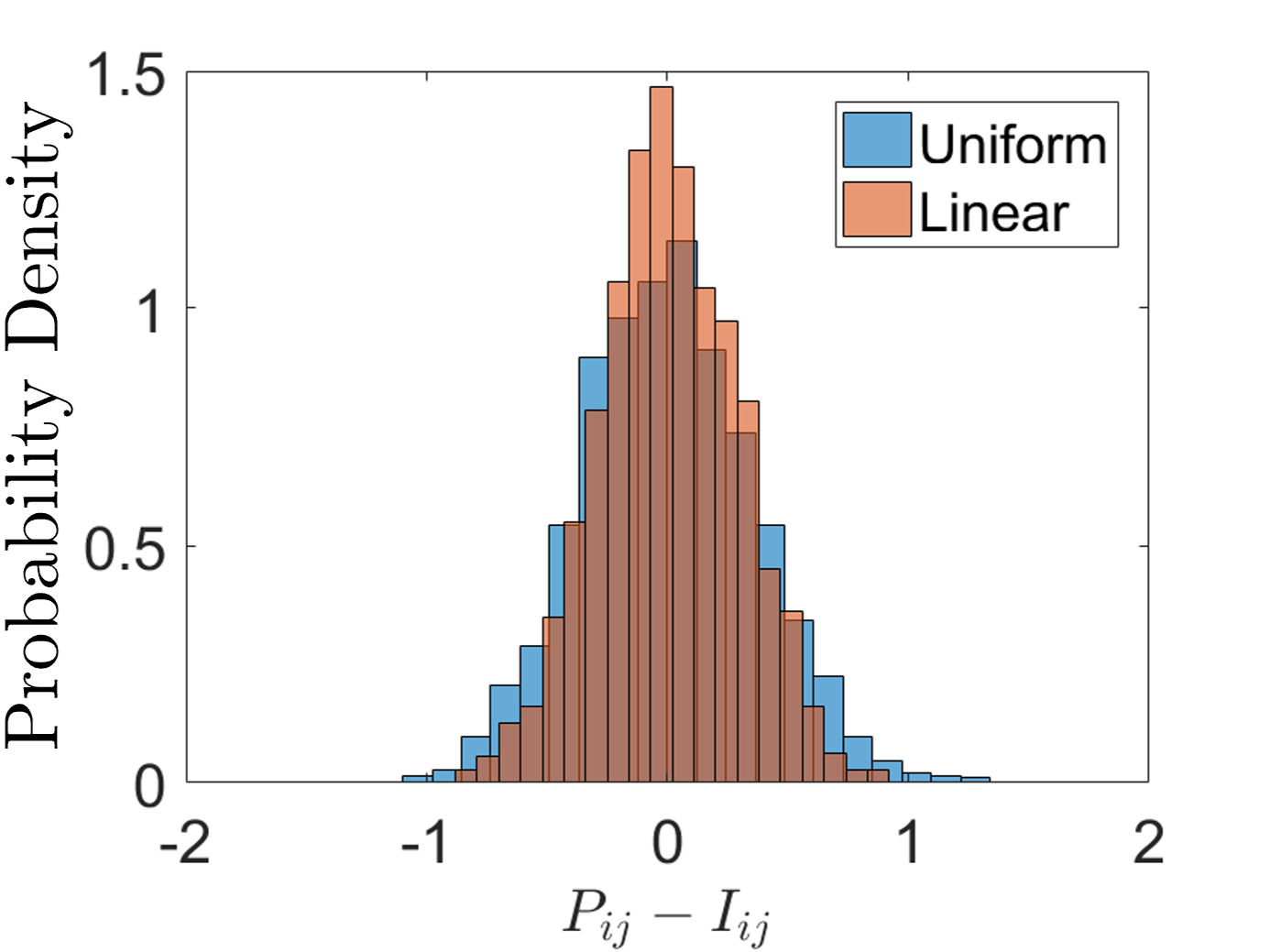}
         \caption{ }
         \label{subfig: GP3 Pseudo Linear}
     \end{subfigure}
        \caption{(a) Pseudo-correlation filtered ghost projection with linear weights for the same conditions as Fig.~\ref{fig:pseudo-correlation result graphs} in Sec.~\ref{subsec:Pseudo-correlation Filtered Random Matrix Ghost Projection Simulation}, with $\frac{\beta}{\alpha} = C_{\text{min}}$. (b) The noise obtained in the pseudo-correlation filtered ghost projection with linear weights, as compared to without linear weights. (c) The noise obtained in the pseudo-correlation filtered ghost projection with linear weights, as compared to without linear weights for sub-optimal pseudo-correlation cut-off $C_{\text{min}} = \text{E}[C]$. Note that, despite the fact that the linear weights indeed out performs the uniform weights, the variance of this case is still higher than for the optimal pseudo-correlation cut-off of $C_{\text{min}} = \text{E}[C]+ 0.612 \sqrt{\text{Var}[C]}$. }
        \label{fig:pseudo-correlation with linear weights result graphs}
\end{figure*}

From filtering, we know that the expectation value of each basis member is skewed according to:
\begin{equation}
\nonumber
\text{E}[R'_{ij}] = \text{E}[R] J_{ij} + I_{ij} \gamma ,
\end{equation}
where 
\begin{equation}
\nonumber
\gamma = (\text{E}[C'] - \text{E}[C]) \sqrt{\text{E}[R^2]/\text{E}[I^2]}. 
\end{equation}
To make the above scheme viable, we need to determine the scaling factor of our desired image, together with the offset that this produces in expectation value. Motivated by what we have seen with the previous case of pseudo-correlation filtering, we obtained a scaling factor of the expected correlation, normalized to the appropriate length (i.e.~we divide by the magnitude of the basis member and multiply by the magnitude of the image, $\sqrt{\text{E}[I^2]/\text{E}[R^2]}$). In this case, we expect the scaling constant that we need to divide by, to be $ \text{E}[ \alpha{C'}^2 - \beta C'] \sqrt{\text{E}[R^2]/\text{E}[I^2]} $. Moreover, for the offset value inherent to this scheme, we can deduce this for the non-filtered, zero image $I_{ij} = 0$ case to be $\text{E}[R](\alpha \text{E}[C']-\beta)$. Making these appropriate substitutions, we obtain the expected result of the scheme to be:
\begin{align}
\text{E} [P_{ij}] = &   \left( \alpha \text{E}[{C'}^2] - \beta \text{E}[ C'] \right) \sqrt{\frac{\text{E}[R^2]}{\text{E}[I^2]}}I_{ij} \\ \nonumber & + \text{E}[R]( \alpha \text{E}[C']-\beta).
\end{align}
This implies that for our scheme to produce our desired image, we need to normalize the coefficient of $I_{ij}$. Redefining our scheme then, we have:
\begin{align}
P_{ij} = t_k {R'}_{ij}^{\ \ k} \approx I_{ij} +  \frac{\text{E}[R] \left( \text{E}[C'] - \frac{\beta}{\alpha} \right)}{\text{E}[{C'}^2] - \frac{\beta}{\alpha} \text{E}[C']} \sqrt{\frac{\text{E}[I^2]}{\text{E}[R^2]}} J_{ij},
\end{align}
where 
\begin{equation}
t_k = \frac{C_k - \frac{\beta}{\alpha} J_k}{N' \left(\text{E}[{C'}^2] - \frac{\beta}{\alpha} \text{E}[C'] \right)} \sqrt{\frac{\text{E}[I^2]}{\text{E}[R^2]}}. 
\end{equation}
We performed two simulations for the same conditions as the pseudo-correlation filtered ghost projection in Sec.~\ref{subsec:Pseudo-correlation Filtered Random Matrix Ghost Projection Simulation}, one with the optimal pseudo-correlation cut-off $C_{\text{min}} = \text{E}[C]+ 0.612 \sqrt{\text{Var}[C]}$ (Fig.~\ref{subfig: GP1 Pseudo Linear}) and another with the reduced cut-off value of $C_{\text{min}} = \text{E}[C]$, both with $\frac{\beta}{\alpha} = C_{\text{min}}$. Based on the very mild reduction in variance exhibited in  Fig.~\ref{subfig: GP2 Pseudo Linear}, we will not pursue analytically determining the variance of this scheme, nor calculating an SNR uncertainty principle. There is a greater relative improvement between the filtered-and-weighted scheme as compared to the just-filtered scheme for the sub-optimal cut-off criterion of $C_{\text{min}} = \text{E}[C]$, however, this relative improvement is still short of the former case. The reason we suspect the gains are so mild is that, although we are privileging those basis members with a higher pseudo-correlation value, the variance within pseudo-correlation values is low. That is, those that are relatively more correlated are still not correlated to any high degree in an absolute sense, and the differentiation of these pseudo-correlation values does not produce significant results for the conditions explored above. Moreover, it is worth noting that, even for the very mild gains in reduced variance, this comes at the significant complexity cost of having to expose each mask for its own unique period of time, as opposed to exposing all masks for the same period of time.  

% These figures were generated using the MATLAB code GP_filteredBasis

\section{Minimum Number of Basis Vectors to Guarantee Non-negative Coefficients} \label{App:C Minimum number of basis vectors}

Ghost projection can conceptually be thought as a non-negative basis representation. One naturally arising question is then, what is the minimum number of basis vectors one needs to guarantee non-negative coefficients exists? In an unrestricted-weights context, such a question would have the answer of an orthogonal basis. 

We remark that non-negative coefficients will always exist if one has an orthogonal basis paired with a complementary orthogonal basis that is a mirrored (i.e.~multiplied by $-1$) version of the original, e.g. in two dimensions, $\{(1,0), (0,1)\}$ paired with $\{(-1,0), (0,-1)\}$. However, this is not the minimal number of basis vectors required to ensure a non-negative solution exists.

\begin{theorem}
For an $n$-dimensional linear vector space, $(n+1)$ basis vectors can be arranged to guarantee non-negative coefficients exists.
\end{theorem}

\begin{proof}
Base case: starting with the 1D case, we can see that the basis set $\{(1), (-1)\}$ ensures non-negative coefficients can be found to express every real number. 

To increase by a dimension, we propose the algorithm:
\begin{enumerate}
    \item Indexing the basis vectors chronologically according to $(1)$ as the first, $(-1)$ as the second, and so on for subsequent basis vectors that are introduced.
    \item Adding a zero column to the end of all basis vectors. 
    \item Taking the last basis vector, creating two copies, and replacing the 0 in the last column with $\sqrt{3}$ and $-\sqrt{3}$, respectively.
\end{enumerate}

In going from 1D to 2D, this amounts to taking $\{(1),(-1)\}$ and transforming it to $\{(1,0), (-1,\sqrt{3}), (-1,-\sqrt{3})\}$. Graphically speaking, these three basis vectors are such that the same angle is obtained between any two vectors. Note, they need not necessarily be equally angularly-spaced but, for the sake of argument, we assume them to be. In general, the basis vector being `split' could take any angle between $(0,90)$ degrees from its previous orientation. 

To prove these basis vectors guarantee non-negative coefficients, we will show that a non-negative solution exists for the vectors $\{(1,0), (0,1), (-1,0), (0,-1), (0,0)\}$ -- which are an orthonormal set, the complementary mirrored orthonormal set and a zero vector (note, constructing the zero vector is an important property for the induction step). Assigning our basis vectors a name, let's call $\vec{a} = (1,0)$, $\vec{b} = (-1,\sqrt{3})$ and $\vec{c} = (-1,-\sqrt{3})$. From here, we can see that $\vec{a} =(1,0)$, $(\vec{a}+\vec{b})/\sqrt{3} = (0,1)$, $\vec{a}+\vec{b}+\vec{c} = (-1,0)$, $(\vec{a}+\vec{c})/\sqrt{3} = (0,-1)$ and $2\vec{a} + \vec{b} + \vec{c} = (0,0)$. 

Induction step: assume that we have $n+1$ basis vectors for an $n$-dimensional space that have a non-trivial, non-negative solution for: (i) the orthonormal basis, (ii) the mirrored complementary orthonormal basis and (iii) the zero vector. We notate this basis set by the matrix $B_{ij}$ where the rows $i \in [1,n+1]$ index the different basis vectors and the columns $j \in [1,n]$ index the elements of each basis vector. For this matrix, let us also explicitly define the non-trivial, non-negative row-vector $T_{n}$ with dimensions [1,$n+1$] that, when acting from the left on $B$, achieves a vector of all zeros, i.e.~$ T_n = (n,n-1,...,4,3,2,1,1)$.

Moving onto the $(n+1)$-dimensional space, we have the basis set $\{ (B_{ij},0) \ \forall \ i \in[1,n], (B_{(n+1)j},\sqrt{3}), (B_{(n+1)j},-\sqrt{3})\}$. To prove this has non-negative coefficients for all possible vectors in the $(n+1)$-dimensional space, we need to be able to create $(0,...,0,1), (0,...,0,-1)$ and $(0,...,0,0)$. To achieve the first case, take the matrix $(B_{ij},0) \forall i \in [1,n]$ and augment it with $(B_{(n+1)j},\sqrt{3})$ in the final row, then apply $T_{n}$ and divide by $\sqrt{3}$. To achieve the second case, we take the matrix $(B_{ij},0) \forall i \in [1,n]$ and augment it with $(B_{(n+1)j},-\sqrt{3})$ in the final row, then apply $T_{n}$ and divide by $\sqrt{3}$. To achieve the final case, we can augment the matrix $(B_{ij},0) \forall i \in [1,n]$ with both $(B_{(n+1)j},\sqrt{3})$ and $(B_{(n+1)j},-\sqrt{3})$ in the final two rows, and define the new non-trivial, non-negative row-vector $T_{n+1}$ with dimensions [1,$n+2$] that, when acting from the left on $B$, achieves a vector of all zeros. Explicitly, this transformation is $T_{(n+1)} = (n+1,n,...,4,3,2,1,1)$.

Conclusion: we have shown that with three vectors, we can ensure a non-negative solution exists for the coefficients of a vector in 2D, as well as a non-trivial, non-negative solution to create an all-zeros vector. Assuming these properties hold for $n$-dimensions, we showed they must hold for $(n+1)$-dimensions. So, since they hold for 2D, they must hold for 3D. Since they hold for 3D, they must hold for 4D, and so on.
\end{proof}

\end{document}